\documentclass[usenatbib]{mnras}

\usepackage{newtxtext,newtxmath}
\usepackage[T1]{fontenc}
\usepackage{ae,aecompl}

\usepackage{graphicx}	% Including figure files
\usepackage{amsmath}	% Advanced maths commands
\usepackage{hyperref}
\title[Star Formation in Nearby Galaxies]{A Multi-Wavelength Study of Star Formation in 15 Local Star-Forming Galaxies}

\author[M. V. Smith et al.]{
Madison V. Smith,$^{1}$\thanks{E-mail: madvsmit@iu.edu}
L. van Zee$^{1}$,
S. Salim$^{1}$,
D. Dale$^{2}$,
S. Staudaher $^{3}$, \newauthor
T. Wrock $^{1}$,
A. Maben $^{1}$
\\
$^{1}$Department of Astronomy, Indiana University, Bloomington IN, USA\\
$^{2}$Department of Physics \& Astronomy, University of Wyoming, Laramie WY, USA\\
$^{3}$College of Osteopathic Medicine, Sam Houston State University, Conroe TX, USA \\
}
 
\date{Accepted XXX. Received YYY; in original form ZZZ}

\pubyear{2021}

\begin{document}
\label{firstpage}
\pagerange{\pageref{firstpage}--\pageref{lastpage}}
\maketitle


\begin{abstract}
 
We have fit the far-ultraviolet (FUV) to mid-infrared (MIR) spectral energy distributions (SEDs) for several nearby galaxies ($<$ 20 Mpc). 
Global, radial, and local photometric measurements are explored to better understand how SED-derived star formation histories (SFHs) and classic star formation rate (SFR) tracers manifest at different scales. Surface brightness profiles and radial SED fitting provide insight into stellar population gradients in stellar discs and haloes. A double exponential SFH model is used in the SED fitting to better understand the distributions of young vs. old populations throughout these galaxies. Different regions of a galaxy often have undergone very different SFHs, either in strength, rate, timing, or some combination of all these factors. An analysis of individual stellar complexes within these galaxies shows a relationship between the ages of stellar clusters and how these clusters are distributed throughout the galaxy. These star formation properties are presented alongside previously published HI observations to provide a holistic picture of a small sample of nearby star-forming galaxies. The results presented here show that there is a wide variety of star formation gradients and average stellar age distributions that can manifest in a $\Lambda$CDM universe.
\end{abstract}

\begin{keywords}
galaxies: star formation -- galaxies: evolution -- galaxies: fundamental parameters
\end{keywords}

%%%%%%%%%%%%%%%%%%%%%%%%%%%%%%%%%%%%%%%%%%%%%%%%%%

%%%%%%%%%%%%%%%%% BODY OF PAPER %%%%%%%%%%%%%%%%%%
 
\section{  Introduction}

The way a galaxy evolves is inherently dependent on the laws of structure formation in the universe. In particular, $\Lambda$CDM models support the theory that galaxies form "inside-out", meaning that there is an expectation that the edges of stellar discs are to be richer in young, metal-poor stars than the inner regions \citep[e.g.,][]{abadi2003,aumer2013}. The color gradients of many spiral galaxies have corroborated these models by showing a trend of red to blue colors as a function of galactocentric distance \citep[e.g.,][]{dejong1996,macarthur2004}. Beyond the stellar disc, the stellar halo can be simulated as a build up of past major/minor merger and accretion events \citep[e.g.,][]{alberts2011,dsouza2014} or as stars undergoing stellar migration \citep{sellwood2002}, both allowing the edges of a galaxy to contain older populations. While dwarf galaxies also show older stellar populations at larger radii, this population gradient may be better explained by stellar migration \citep[e.g.,][]{zhang2012,elbadry2016}. Simulations wishing to recreate accurate evolutionary models should be validated by an extensive catalogue of the star formation properties and the stellar population distributions of real galaxies. This project is the first step in providing those details for several nearby galaxies, so as to enumerate the possible outcomes of the complex process of galaxy evolution. 

A popular technique in constraining a galaxy's components and physical properties is spectral energy distribution (SED) fitting. Multi-wavelength photometry is matched to libraries of SEDs made from convolutions of stellar synthesis models, dust attenuation curves, and more to infer details about the physical properties of a galaxy \citep[e.g.;][]{larson_tinsley1978,hunt2019}. A wide variety of physical properties, such as total mass, dust attenuation, and metallicity, can be inferred from the best-fitting modelled SED. Of interest for this project, SED fitting can give insight into a galaxy's star formation history (SFH) both on global and spatially-resolved scales \citep[e.g.,][]{reddy2012,cooper2012,Dale2016,dale2020}. 
Combining the results of the model fitting with commonly used SFR tracers, such as the FUV+24$\mu$m flux, can provide a more comprehensive look at the variety of current and past star formation histories possible for nearby star-forming galaxies.
Previous HI observations are also presented alongside these photometric results to enhance the discussion on the current and past evolutionary steps these galaxies are experiencing, even though the HI is not used in the SED fitting. In particular, HI surface density maps and velocity fields deepen our understanding of the immediate environment around and the varying kinematics within these galaxies \citep[e.g.,][]{swaters2002,holwerda2012}.

In addition to the multiwavelength approach, this project also looks at star formation properties on different spatial scales because the interpretation of a galaxy's properties may be methodology- and scale-dependent \citep{kennicutt2012}. Measuring a galaxy's global star formation properties is a method for classifying whether that galaxy is star-forming or quiescent \citep{noeske2007}. While the global properties of galaxies are important to know, nearby galaxies, especially those in the Local Volume, can be observed with more detail. Especially in the case of spirals, a radial analysis of the star formation properties in a galaxy can give more detailed information about where individual stellar populations tend to reside \citep[e.g.,][]{munozmateos2011,monachesi2013}. The EDGES Survey \citep{vanzee2012,staudaher2019}, the source of the NIR measurements used here, was able to detect the low-surface brightness, extended stellar haloes that serve as evidence for hierarchical evolution \citep{bullock2005,cooper2013}.

On a smaller scale than simply looking at radial trends, tracing and analyzing individual stellar complexes throughout the disc can be enlightening as to how star formation persists in different environments. \citet{thilker2005} showed the existence of moderately young stellar clusters in the outer, low-density regions of M83, furthering the argument that star formation thresholds fall short of describing the ability of star formation to take place in specific areas of a galaxy. Continuing the studies of \citet{thilker2005} and others \citep[e.g.,][]{alberts2011,barnes2014}, this project will look at the properties of UV-selected stellar clusters in our sample of local galaxies. A combination of SED-fitting and clustering algorithms is used to look at the characteristic separation of young stellar complexes depending on their age. With global, radial, and local star formation properties presented in one place, this project aims to catalogue the abundance of star formation properties that can be produced in a $\Lambda$CDM universe. 

Section \ref{sec:sample} details the sources of the measured images and gives an overview of the sample. Section \ref{sec:methods} explains the methods used in this analysis. Section \ref{sec:results} presents the main results. Section \ref{sec:discussion} provides more context for these results, including a comparison of these SED-derived SFH results to past CMD-derived results for several galaxies. Section \ref{sec:conclusion} summarizes the major conclusions for this project, and discusses possible future avenues of research. Appendix \ref{appen:measurements} presents the individual analysis of the photometric measurements and radial SFHs for each galaxy, including information about each galaxy in the literature. Appendix \ref{appen:measurements} also presents the HI distributions of these galaxies, making it possible in some cases to draw more distinct conclusions about the evolutionary path the galaxy may be on.

\section{  Observations and Sample}
\label{sec:sample}
This analysis of 15 galaxies relies on multiwavelength observations gathered from archival GALEX and \textit{Spitzer} data, as well as optical observations published here for the first time.  Because this subsample is selected from the EDGES sample \citep{vanzee2012}, exceptionally deep 3.6 and 4.5$\mu$m \textit{Spitzer} Infrared Array Camera (IRAC) \citep{fazio2004} images are available. For more coverage in the mid-IR, 5.8, 8.0, and 24 $\mu$m images from either SINGS  \citep{kennicutt2003} or the LVL Survey \citep{dale2009} are used in our measurements. The observing strategy outlined for both surveys made it such that there was a net integration time of 240s (160s) for each pixel in the mosaicked IRAC (MIPS) images. Table \ref{tab:observations} shows the integration times for the GALEX images and the sources of these supplementary IR observations used in this analysis. 

The campaign to observe the EDGES galaxies in the optical, including narrowband H$\alpha$ observations, at the WIYN 0.9m on Kitt Peak gives this project the opportunity to incorporate more multiwavelength measurements in the SED fitting. The narrowband H$\alpha$ measurements were included in the SED fitting. It is important to note, however, not all of the H$\alpha$ flux can be attributed to the ionizing radiation of O stars. All H$\alpha$ flux within an annulus is averaged for the radial analysis component of this project. This makes the EW(H$\alpha$) profiles more difficult to interpret in the context of star formation, but they are still included in the data presentation.
In all cases, the B and R observations have been used in the methods outlined in Section \ref{sec:methods}. For some galaxies, the full UVBR coverage is available and was included in the methodology. These optical images were observed with either S2KB or HDI on the WIYN 0.9m telescope between May 25, 2009 and April 19, 2018. Typical integration times for the U, B, V, R, and H$\alpha$ single exposures are 1200, 900, 600, 300, and 1200s respectively. In all cases, the final science images utilized for this project were combinations of at least two individual exposures for each filter. The integration times and NCOMBINE procedure ensured that the observations at optical wavelengths were not the limiting factor when measuring low surface brightness regions of the galaxies. The surface brightness limit for the B measurements is \textasciitilde26.5 mag arcsec$^{-2}$.

Some properties of our sample are given in Table \ref{tab:distances}, including morphology, distances, and Galactic reddening ($A_{V}$). All 15 galaxies are within 20 Mpc, but span a wide range of morphologies, inclination, and luminosity. This sample is the first in a series of papers to look at the star formation in local galaxies using the methods outlined in Section \ref{sec:methods}. 
Another important aspect in understanding these galaxies can be studied by looking at their HI distributions.  HI gas plays an important role in the regulation of star formation in galaxies \citep[e.g.,][]{vanzee1997}; therefore, the HI surface density and velocity fields are presented in conjunction with the SED fitting and photometric measurements. The HI observations come from a variety of sources. These sources and cursory analysis of the HI distributions in the galaxies are discussed in Appendix \ref{appen:measurements}.

\begin{table}
\caption{UV integration times (in seconds)  and the survey where the UV and IR images were published. Sources for the IR images : (1) \citet{kennicutt2003}, (2) \citet{dale2009}. \textsuperscript{a} Optical images for NGC 7793 were also from \citet{kennicutt2003}.}
\centering
\begin{tabular}{cccccccc}\hline
Galaxy    & FUV  & NUV                                & Source \\ \hline
NGC 0024  & 1577 & 1577                                 & (1)     \\
NGC 3344  & 895  & 1400             & (2)     \\
NGC 3486  & 1488 & 3072                  & (2)  \\
NGC 3938  & -    & 3045                           & (1)    \\
NGC  4068 & 1350 & 2281            & (2)     \\
NGC 4096  & 1650 & 1650      & (2)     \\
NGC 4214  & 1066 & 2006          &  (2)     \\
NGC 4242  & 1683 & 3282	  &  (2)     \\
NGC 4618  & 1630 & 3259                    & (2)    \\
NGC 4625  & 1630 & 3259                       &  (2)  \\
NGC 7793 \textsuperscript{a}  & 1553 & 1553  & (1)      \\
UGC 07408 & 1450 & 2691           & (2)  \\
UGC 07577 & 1684 & 1684                  & (2)  \\
UGC 07608 & 1684 & 1684              &(2)     \\
UGC 08320 & 1584 & 3116                       & (2) \\ \hline
\end{tabular}
\label{tab:observations}
\end{table}

\begin{table}
     \caption{   Sources for distances: (1) \citet{tully2013}, (2) \citet{tully2016}, (3)  \citet{poznanski2009}, (4)  \citet{jacobs2009}, (5) \citet{sorce2014}, (6) \citet{mcquinn2017}, (7) \citet{zgirski2017}, (8) \citet{tully1988}. All information is taken from the NASA/IPAC Extragalactic Database (NED).
    }
\centering
    \begin{tabular}{c|c|c|c|c}
    \hline
        Galaxy & Morphology & $A_{V}$ & $cz$ (km s$^{-1}$) & d (Mpc)   \\ \hline
        NGC 0024   &    SA(s)c  &   0.053   &   555   &   7.7 (1)  \\
        NGC 3344   &    (R)SAB(r)bc  &  0.091   &   582   &   9.8 (1) \\
        NGC 3486   &    SAB(r)c  &  0.059   &   681 &   12.6 (2)  \\
        NGC 3938   &    SA(s)c HII  &   0.058   &   810 &   17.9 (3)  \\
        NGC 4068    &   IAm &   0.059   &   210 &   4.4 (4)  \\
        NGC 4096    &   SAB(rs)c    &   0.050    &   567 &   12.4 (1) \\
        NGC 4214    &   IAB(s)m &   0.060   &   291 &   2.9 (1) \\
        NGC 4242    &   SAB(s)dm    &   0.033   &   507 &   5.2 (5) \\
        NGC 4618    &   SB(rs)m HII &   0.058   &   543 &   6.5 (2) \\
        NGC 4625    &   SAB(rs)m pec    &   0.50    &   621 &   11.8 (6) \\
        NGC 7793    &   SA(s)d HII  &   0.053   &   231 &   3.4 (7)  \\
        UGC 07408   &   IAm &   0.032   &   462 &   7.3 (1) \\
        UGC 07577   &   Im  &   0.056   &   195 &   2.6 (1) \\
        UGC 07608   &   Im  &   0.047   &   537 &   7.6 (8) \\
        UGC 08320   &   IBm &    0.042   &   192 &      4.3 (1) \\ \hline 
    \end{tabular}
    \label{tab:distances}
\end{table}

\section{  Methods}
\label{sec:methods}

The surface brightness profiles are measured using the \texttt{ellipse} package in IRAF. Images were cleaned such that any foreground or background sources were removed. Any of these sources directly on a galaxy were interpolated over, instead of having the pixel values simply set to zero. Interpolating over the sources helped to preserve the flux of these galaxies especially in the case of bright foreground stars. 
 Families of ellipses were grown inward and outward with characteristics (ellipticity, position angle, etc.) drawn from the properties of the isophotal ellipse at $R_{25}$, which is the radius at which the average surface brightness in the B band is 25 mag arcsec$^{-2}$. 
  The growth of these ellipses is logarithmic with a multiplicative factor of 1.1 to ensure adequate signal to noise even in the far edges of the galaxy.  The smallest of the ellipses (the inner annulus) was never smaller than the highest resolution limit of all the available images, which is set by the 24$\mu$m MIPS images at 6".
The annulus-averaged surface magnitude in each ellipse is recorded to create the surface brightness profile according to the best-practice methods of \citet{barnes2014}. These surface magnitudes are inclination corrected, assuming the ellipticity of a galaxy corresponds directly to its inclination. Next, the surface magnitudes are corrected for Galactic extinction \citep{schlaflyfinkbeiner2011} assuming $A_{V}/E(B-V) \approx 3.1$ and the reddening curve from \cite{fitzpatrick1999}. 

It is important to note that the inclination corrections are based on the photometric properties of the isophote at $R_{25}$, and do not include effects based on the thickness of the disc. For this correction, cos$i$ = $b/a$, where $b/a$ is the axial ratio of the minor to the major axis. This interpretation is used even in the case of dwarf galaxies where inclination angle may not be well-defined, for consistency. On a similar consistency note, often concentric ellipses are not the most intuitive method of exploring a surface brightness profile, but this method is consistently used across the sample even in the case of highly-irregular or clumpy galaxies. While there are more precise formulations of the inclination correction, the simple cos$i$ correction can be used consistently across many morphologies. The simplicity of using the ellipse parameters ($b/a$) to define the inclination of a galaxy does not affect the results presented in the project.

While this logarithmic growth of the ellipses allows for finely-spaced, consistent coverage of each galaxy, it creates a large amount of concentric regions (\textasciitilde30 per galaxy) to be fed into CIGALE\footnote{Version 2018.0, \url{https://cigale.lam.fr}} \citep{noll2009,boquien2019} that then need to be analysed independently. In the interest of being able to effectively analyse the trends across a galaxy without being overwhelmed by the regions of interest, the above method is used only to show the surface brightness and color profiles. For the SED-fitting, a slightly different method is used for clarity in the analysis. The preliminary photometry provides the details needed to calculate the total light from a galaxy using its asymptotic magnitude derived from the curve of growth \citep{cairos2001}. The total magnitude of the galaxy is then used to find the half-light radius.  

Using the same centre, ellipticity, and position angle from the $R_{25}$ analysis, but with the half light radius as the parent ellipse's semi-major axis, a new family of ellipses is grown. These ellipses are grown by a factor of 1.8 such that there are three regions interior and two to four regions exterior to the half light radius. The number of exterior regions is dependent on the spatial extent of the disc in all of the bands, allowing for a maximum of seven concentric regions to be measured for any given galaxy.
This coarser definition of annuli provides another advantage, besides having fewer regions to analyse. We opted not to smooth the higher resolution images (the optical, 3.6 $\mu$m, and 4.5 $\mu$m images) to match the other, lower resolution images. \citet{Dale2016} showed that coarsely placed concentric ellipses lead to consistent results whether or not the higher resolution images are smoothed. 

The surface magnitudes created from this method are then inclination corrected and corrected for Galactic extinction (same as before), but are then converted to fluxes (mJy) to be fed into the SED-fitting code, CIGALE. Galactic extinction and redshifts were obtained through the NASA/IPAC Extragalactic Database (NED). This methodology effectively assumes independent evolution of the concentric rings being given to CIGALE, as the SED-fitting is done on each ring without knowledge about the neighboring rings. Although this simplifying assumption does not take into account effects of stellar migration, modeling galaxies as independent rings has been shown to reproduce realistic stellar properties \citep{boissier1999,boissier2000,munozmateos2011}.

CIGALE allows for a large parameter space to be explored, but the parameters used for this project are similar to those of \cite{Dale2016} and \cite{dale2020} in order to model realistic parameters for nearby galaxies.
We use the \citet{bruzual_charlot2003} library of single stellar populations (SSPs) to model the stellar part of the spectrum.  We use the SSPs modelled with the \citet{salpeter1955} stellar IMF (lower and upper mass cut offs are $m_{L} = 0.1 M_{\odot}$ and $m_{U} = 100 M_{\odot}$) and a range of metallicities (see Table \ref{tab:cigale_parameters}). The single parameter dust emission model with no AGN component from \citet{dale2014} is used as well.
 
 The \texttt{nebular} module, an optional module in CIGALE, was included in the parameter space because the emission-line flux can contribute a significant amount to the broadband fluxes of galaxies with high equivalent widths and sSFRs. Without accounting for emission lines, the SED-derived specific star formation rates (sSFR = SFR / stellar mass) of these already high-sSFR galaxies are overestimated \citep{salim2016}. Because the galaxies in this sample are already designated as star forming, including the nebular emission in the SEDs produce better fits at UV and optical wavelengths. 
 It has also been shown that gaseous nebular emission can become a major source of flux for young, low-metallicity stellar populations \citep{anders2003}. 
 The dust attenuation curve based on \citet{calzetti2000} and \citet{leitherer2002} was used. The exact ranges of values allowed for the SED parameters are given in Table \ref{tab:cigale_parameters}.

We opt to use a double exponential SFH for this project. This functional form of a SFH involves five parameters. There are two onsets of star formation, starting at times $t_{o}$ and $t_{y}$ for the older and younger stellar populations, respectively. Both epochs of star formation have exponential decreases in the SFR, which is described by the e-folding times, $\tau_{o}$ and $\tau_{y}$. The final parameter related to the double exponential SFH is $f_{burst}$, the fraction of stars formed in the younger burst of star formation relative to the total stellar mass formed.
We set the age of the older stellar population, $t_{o}$ to be 10 Gyr, but allow the e-folding time, $\tau_{o}$ to vary, as in \citet{noll2009}. For the more recent burst of star formation, we allow the age $t_{y}$, $\tau_{y}$, and burst fraction, $f_{burst}$ to vary.  A double exponential SFH is favored over the single exponential because the double exponential has been shown to produce better fits, especially in the case of more complicated, bursty SFHs often associated with low-mass dwarfs \citep{weisz2011,salim2016}. 

Setting the older stellar population's age to 10 Gyr and only allowing the newer population to be at most 4 Gyr old forces the stars to be formed in one of two possible formation mechanisms. An older population is expected to be made during the formation and assembly of the galaxy, while a newer population forms during the evolution and current times of the galaxy \citep{mcquinn2010}. Keeping these time-scales separate from one another is also important because any measure of SFH has systematically larger uncertainties on the more ancient SFH than on the recent SFH, so allowing only a single $t_{o}$ still captures the essence of ancient star formation without needlessly complicating the modelling process. This SFH template allows for a clear division between older and younger populations, with the focus of this research being on the younger population.

This type of SED fitting requires us to limit our SFH models to a library of functional forms, even though the true SFH of a galaxy is likely to be more complicated. Parametric models have been shown to produce biases and incorrect uncertainties for output physical parameters. The use of nonparametric models have been used recently with success in reducing biases and producing accurate error bars on physical parameters \citep{leja2019,iyer2019}.  However, a parametric SFH model is used for this project because it confines the parameter space to a reasonable number of dimensions. Some of the shortcomings of choosing a single SFH model for such a wide variety of galaxies is discussed in Section \ref{ssec:cmd sfh}.

In addition to the radial analysis of these galaxies, a similar SED-fitting process was done on individual star forming complexes in the galaxies. These sources were found using SExtractor \citet{bertin1996} in order to automate the process. For this analysis the images were smoothed to a common PSF and the NUV image was used as the basis for finding sources. The average size of these automatically identified sources was 14.3" and there were approximately 200 sources per galaxy on average. The distribution of physical sizes (in parsecs) of the sources changed depending on the galaxy because the minimum resolvable source depends on the distance to that galaxy. 
 For the closest galaxy, UGC 07577, the smallest regions detectable can be \textasciitilde 30 pc in radius. NGC 3938 is the furthest away at 17.90 Mpc, making the smallest sources possible \textasciitilde300 pc in radius. A maximum size of 1.5 kpc was allowed, so sources larger than this were split up into smaller regions. A recent study of the NGC 7793 found UV clumps sized 12-70 pc \citep{mondal2021}. In comparison, the smallest size permitted for NGC 7793 in this project was \textasciitilde40 pc.
The fluxes of the UV-selected sources were measured at each available wavelength and were fit to SEDs to recover their physical properties. Although the parameter space for the radial analysis allowed for the recent onset of star formation to be as old as 4 Gyr, the UV-selection of these individual star-forming complexes gives a younger upper limit for the age of these sources. Because of this, the parameter space was changed such that $t_{ySP}$ could be no older than 1 Gyr. Limiting the oldest $t_{ySP}$ allowed for a finer spacing of younger values to be explored in the models. 

To describe the way star formation manifests in a galaxy based on this small-scale analysis, we wanted to look at the characteristic separation between sources of similar star formation properties. The \textit{scikit-learn} Python package \citep{scikit-learn} contains numerous clustering algorithms that would be applicable to finding groups of stellar complexes with similar star formation parameters. The \textit{k}-means clustering algorithm, where \textit{k} is the number of groupings permitted, was chosen for this particular project. 
The optimal \textit{k} for each galaxy was chosen after reviewing three classic diagnostics: the elbow method, the silhouette method, and using the approximation $k \approx \sqrt{n/2}$ \citep{kodinariya2013review}. The relevant features were chosen to be three SFH parameters from the SED-fitting results ($f_{burst}$, $age_{burst}$, and $\tau_{burst}$) as well as the NUV magnitude of the UV-selected sources. Membership to each group depends on an individual source's proximity to all of the groups in the feature space, not the physical proximity to other sources in the galaxy. 
 After the group membership of the stellar complexes were determined, the average Euclidean distance between sources within an assigned group was calculated and then normalized by the measured $R_{25}$ for easier comparison between galaxies.

\begin{table*}
\centering
	\caption{     Module and parameter values used for SED fitting.}
	\label{tab:cigale_parameters}
	\begin{tabular}{lll} 
		\hline
		Module & Parameter & Value \\
		\hline
	    \texttt{sfh2exp} & \texttt{tau\_main} & 0.2, 0.5, 1, 5, 10 Gyr \\
	    
		 & \texttt{tau\_burst} & 0.05, 0.1, 0.2, 1, 2 Gyr \\
		 & \texttt{f\_burst} & 0.005, 0.01, 0.05, 0.1, 0.2, 0.5, 0.75 \\
		 & \texttt{age} & 10 Gyr \\
		 & \texttt{age\_burst} & 0.01, 0.1, 0.5, 1.0, 2.0, 3.0, 4.0 Gyr \\
		 \hline
		\texttt{bc03} & \texttt{imf} & Salpeter\\
		 & \texttt{metallicity} & 0.008, 0.02, 0.05 \\
		 \hline 
		 \texttt{nebular} & \texttt{logU} & -3.0, -2.0 \\
		 \hline
		 \texttt{dustatt\_modified\_starburst} & \texttt{E\_BV\_lines} &  0.0, 0.025, 0.05, 0.1, 0.15, 0.2, 0.25, 0.3, 0.4 mag\\
		 & \texttt{powerlaw\_slope} & -0.5, -0.4, -0.3, -0.2, -0.1, 0.0 \\
		 \hline
		 \texttt{dale2014} & \texttt{fracAGN} & 0 \\
		 & \texttt{alpha} &  0.5, 1.0, 1.25, 1.50, 1.75, 2.0, 2.25, 2.50, 3.00 \\
		\hline
	\end{tabular}
\end{table*}

\section{  Results}
\label{sec:results}

\subsection{  Surface Brightness and Color Profiles}
\label{ssec:results_sb}

The ellipse properties from the IRAF-based method for deriving $R_{25}$ are shown in Table \ref{table:measurevalues}.
The surface brightness profile, B-R color gradient, log specific star formation rate (sSFR) and log equivalent width (EW) profiles are given for each galaxy next to the cumulative star formation histories in Appendix \ref{appen:measurements}. An example of the results are shown for NGC 4242 in Figures \ref{fig:ngc4242_sfh} and \ref{fig:ngc4242_fulldata}. The EW(H$\alpha$) has often been used as an indicator for the current normalized SFR compared to the past average SFR rate \citep{kennicutt1983}. The EW(H$\alpha$) profiles are used in a limited capacity for the aforementioned issues with different H$\alpha$ sources. However, the profiles can be looked at in a relative sense to get a better understanding of where the most star formation is occurring in a galaxy.

To calculate the sSFR profile for these galaxies, broadband indicators for SFR (FUV+24 $\mu$m flux) and stellar mass (3.6 $\mu$m flux) were used. 
The FUV+24 $\mu$m SFR indicator was used to ensure re-radiation of the ultraviolet star formation tracer was taken into account \citep{hao2011}. When the FUV data were unavailable, the NUV data were substituted instead as the SFR indicator and the appropriate NUV coefficient was used in the conversion. However, this was only the case for NGC 3938. As more galaxies are added to this study, more galaxies may require this slightly alternative method of deriving SFR. The mass of the stellar disc was calculated from the 3.6 $\mu$m data, assuming a mass-to-light ratio of $\Upsilon_*^{3.6\micron} = 0.5 \pm 0.1$ \citep{Cook2014,meidt2014}. The log(sSFR) panels show an inverse relationship with the (B-R) color gradients, as expected with the understanding that bluer populations are more actively forming stars than redder regions. 

\begin{table*}
 
\centering
\caption{    Ellipse properties. The centre of each ellipse may not correspond to the canonical centre of each galaxy, but is derived from the centre of the ellipse at $R_{25}$. Position angle (PA) is measured East of North.}
\label{table:measurevalues}
\begin{tabular}{c|c|c|c|c|c|c} \hline
 Galaxy &   RA	&	Dec &  $R_{25}$ (") &     $r_{e}$ (") &      b/a &      PA (deg) \\ \hline
 NGC 0024 &  00:09:56.65 & -24:57:44.1 &  114.7 &   55.20 &  0.256 &   44.52 \\
 NGC 3344 &  10:43:31.14 & 24:55:20.40 & 206.9 &   59.00 &  0.970 &  -40.79 \\
 NGC 3486 &  11:00:23.88 & 28:58:32.23 & 178.3 &   77.58 &  0.792 &   82.69 \\
 NGC 3938 &  11:52:49.65 & 44:07:14.86 & 154.5 &   48.60 &  0.930 &    10.0 \\
 NGC 4068 &   12:04:02.48 & 52:35:29.50 & 73.47 &   51.30 &  0.602 &   30.49 \\
 NGC 4096 &  12:06:00.10 & 47:28:28.47 & 155.3 &   69.00 &  0.267 &    18.8 \\
 NGC 4214 &  12:15:39.07 & 36:19:44.15 & 206.9 &   61.30 &  0.967 &  -172.2 \\
 NGC 4242 &  12:17:29.96 & 45:37:07.64 & 137.0 &   84.60 &  0.695 &    30.7 \\
 NGC 4618 &  12:41:32.95 & 41:08:41.92 & 117.1 &   51.95 &  0.832 &   22.71 \\
 NGC 4625 &   12:41:52.52 & 41:16:19.12 & 49.34 &   22.35 &  0.913 &  -60.38 \\
 NGC 7793 &  23:57:49.52 & -32:35:28.2  & 283.7 &  122.6 &  0.595 &  -80.88 \\
 UGC 07408 &   12:21:15.32 & 45:48:51.93 & 53.03 &   45.45 &  0.627 &  -83.45 \\
 UGC 07577 &  12:27:42.48 & 43:29:32.75 & 85.25 &   68.15 &  0.699 &   56.73 \\
 UGC 07608 &   12:28:44.95 & 43:13:29.93 & 46.53 &   54.30 &  0.497 &  -50.89 \\
 UGC 08320 &   13:14:26.56 & 45:55:30.60 & 80.54 &   80.75 &  0.424 &  -28.88 \\ \hline
\end{tabular}
\end{table*}

\subsection{Global Properties}
\begin{figure}
    \centering
    \includegraphics[width=\columnwidth]{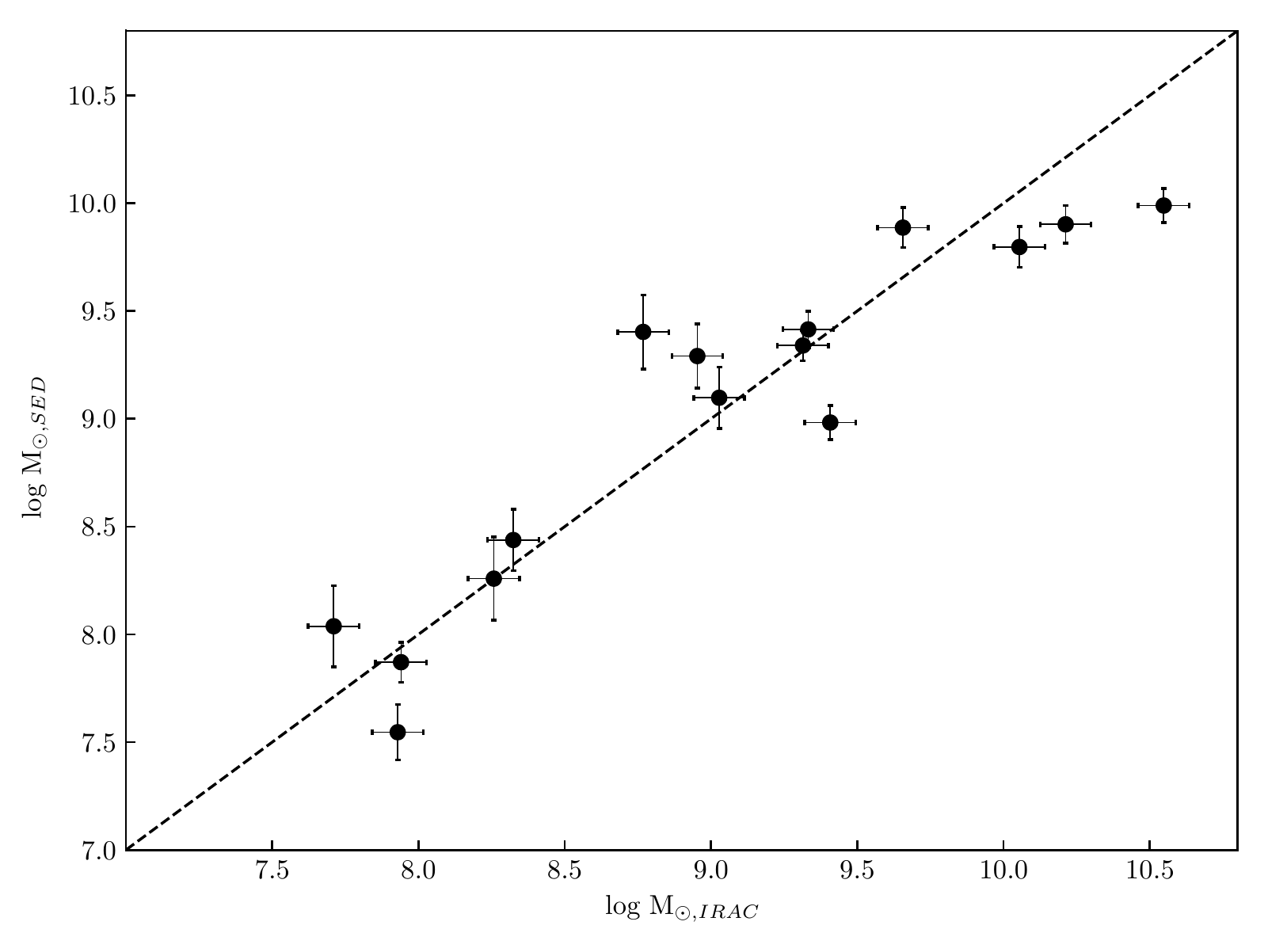}
    \caption{     Comparison of the total stellar mass estimates from SED fitting and from 3.6$\mu$m-to-mass conversion \citep{hao2011}. The dashed line shows a one-to-one correspondence.}.
    \label{fig:mass_comparison}
\end{figure}

Comparing two values for each galaxy's total stellar mass, calculated through both the SED fitting of the measured integrated magnitudes and by using the M/L ratio at 3.6$\mu$m, was used as a first pass at determining whether the SED fitting process was producing reasonable results. Although the focus of this project is on the trends at radial and local scales, the results from the globally measured properties are given here. Figure \ref{fig:mass_comparison} displays the global SED-estimated log $M_{*}$ compared to the log $M_{*}$ derived from the 3.6$\mu$m data for the sample. Although several galaxies fall on the one-to-one correspondence, some galaxies deviate a significant amount. A possible systematic affecting the higher mass galaxies may be the contamination of PAH emission in the 3.6$\mu$m measurements, leading to an overestimation in log $M_{*,IRAC}$ \citep{meidt2012}. A more formal statistical analysis of the differences in these global measurements should be done when more of the EDGES sample is covered in future papers.

The general agreement between the two values suggests that the parameter space used here is probing the relevant values to estimate the physical properties for this sample of galaxies. 

\subsection{  Radial Trends}

\begin{figure*}
    \centering
    \includegraphics[width=\textwidth]{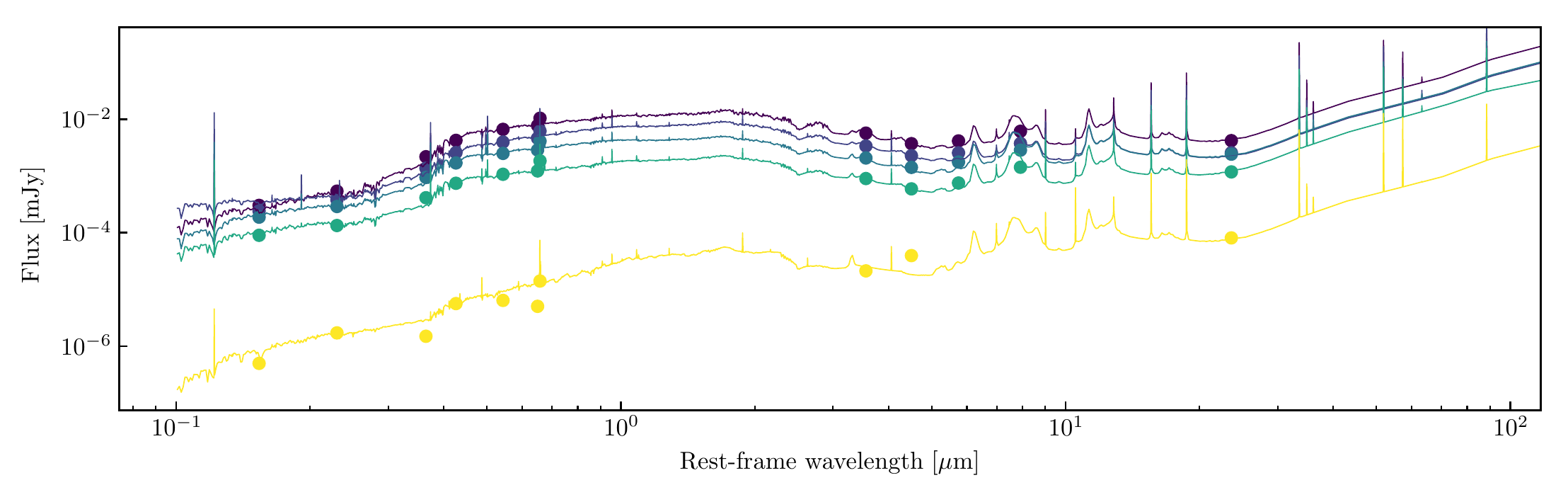}
    \caption{The best-fitting SEDs for five of the annuli covering NGC 4242. The innermost annulus is the brightest, with each consecutive annulus getting dimmer. In the final annulus (yellow line), the IRAC 3 \& 4 filters did not have signal, so they were left out of the fitting.}
    \label{fig:ngc4242_sed_colored}
\end{figure*}

\begin{figure}
    \centering
    \includegraphics[width=\columnwidth]{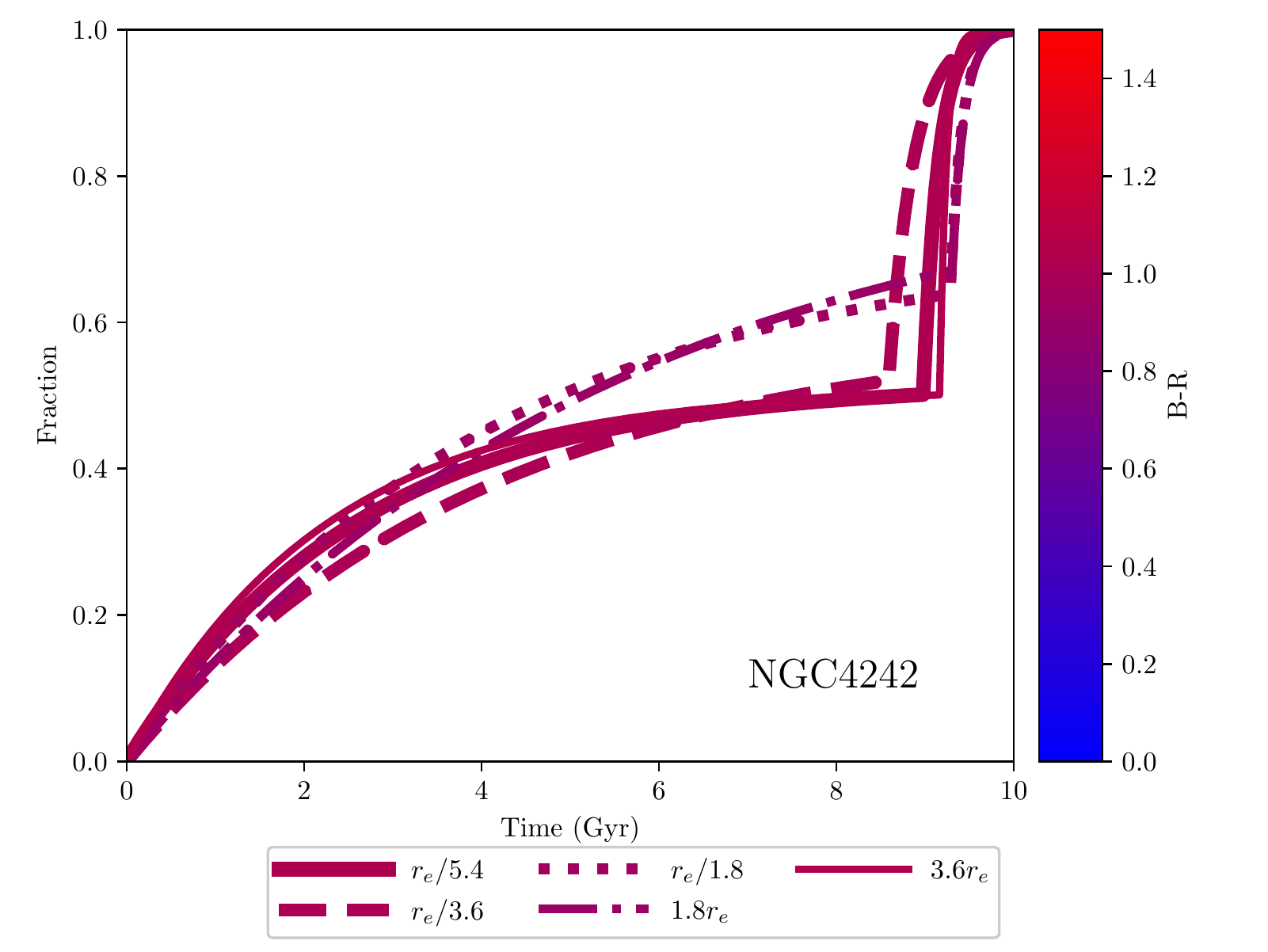}
    \caption{     Cumulative SFH for the five well-fit annuli of NGC 4242. The stellar mass fraction is equal to one at current times. As stated in Section \ref{sec:methods}, the older population is set to start forming 10 Gyr ago, while the younger population has a flexible start time, $t_{ySP}$. With this framework, the x-axis shows time since the oldest stars were allowed to form.}
    \label{fig:ngc4242_sfh}
\end{figure}

The parameters outlined in Section \ref{sec:methods} and Table \ref{tab:cigale_parameters} create a grid of $\sim$4 million models.
The best-fitting models are found by computing $\chi^{2}$ from the residuals and photometric uncertainty, and then the likelihood-weighted parameters are recorded. The full statistical process of finding the output parameters for the SED fitting process are outlined in \citet{boquien2019}. The best-fit SED for each annulus in NGC 4242, along with the observed fluxes, are shown in Figure \ref{fig:ngc4242_sed_colored}.
Cumulative star formation histories were constructed from the Bayesian-estimated SFH parameters for each annulus of each galaxy. Figure \ref{fig:ngc4242_sfh} shows the five annuli that cover the extent of NGC 4242's disc, including information on the B-R color and the form of the cumulative SFH for each annulus. The line thickness depends on the annulus' distance from the galaxy's photometrically-determined centre, such that thicker lines are closer to the centre. As mentioned in Section \ref{sec:methods}, each galaxy was evaluated individually to determine how many annuli would be appropriate to fully explore the farthest extents of each galaxy. In the case of NGC 4242, five concentric ellipses were used to capture the relevant parts of the galaxy, with all five annuli having their goodness of fit below the set threshold of reduced $\chi^{2}$, $\chi_{\nu}^{2}$ = 15. In the majority of cases where a region's SED had a reduced $\chi^{2}$ above the threshold, the annulus was on a low surface brightness part of the galaxy, away from the often bright nucleus.

\begin{figure}
     \centering
     \includegraphics[width=\columnwidth]{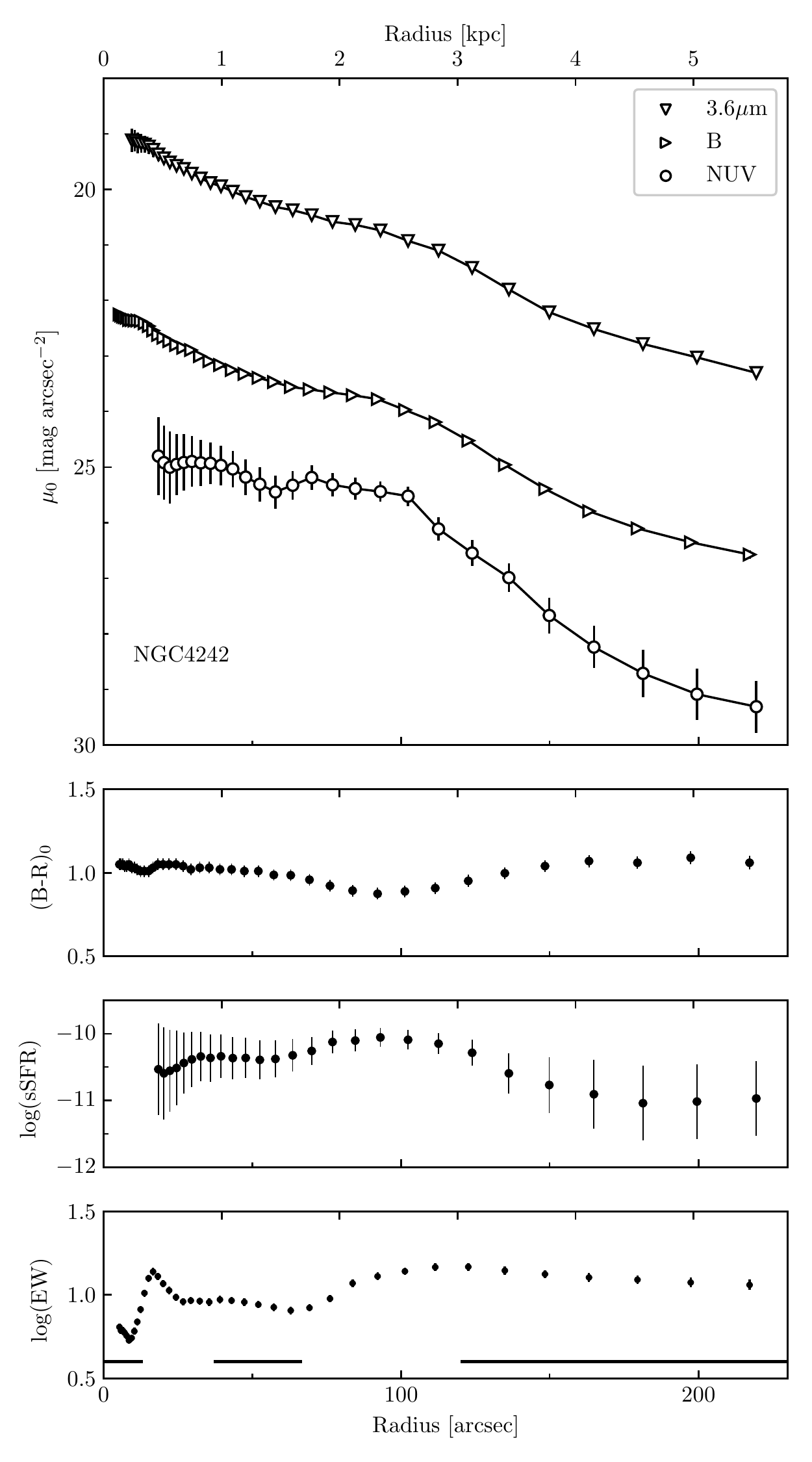}
     \caption{     The results from the photometric measurements of NGC 4242. The innermost annulus size is set by the resolution of the data. The variations in the colour gradient are mirrored in the log(sSFR) profile, showing a connection between B-R colour and SFR. The bluest region in NGC 4242 is coincident with the highest sSFR and EW(H$\alpha$) values. All of these radial profiles are measured from the images using finely-spaced, azimuthally-averaged annuli. In the bottom panel, solid horizontal lines are used to show the larger radial bins used for SED fitting.}
     \label{fig:ngc4242_fulldata}
 \end{figure}

Figure \ref{fig:ngc4242_sfh} matches the expected trends in SF, based on the photometric measurements seen in the two middle panels of Figure \ref{fig:ngc4242_fulldata}. Between the centre and farthest reaches of this galaxy, there is a region of elevated star formation. The innermost and outermost regions of the galaxy underwent bursts at earlier times (\textasciitilde 1--2 Gyr ago) than the middle region, which experienced elevated star formation starting 500 Myr ago. 
The (B-R) coloring of each region is not corrected for internal extinction, so the correlation between bluer colors and more recent star formation cannot be immediately drawn without exploring the dust attenuation. The SED-derived dust parameters (such as shape of the attenuation curve) have been shown to be the least accurate parameters returned in the fitting process \citep{giovannoli2011,boquien2012}, so the dust results are mentioned here only briefly. 
 In most cases, the attenuation ($A_{V}$) showed smooth transitions as a function of radius, suggesting there were not many instances where degeneracies brought a chaotic nature to the results, or affected other SED-derived parameters.

\subsection{Stellar Complexes}\label{ssec:sc_analysis}

Looking at star formation on a local scale helps alleviate some of the shortcomings of the radial analysis. The azimuthal averaging needed to create surface brightness profiles smoothed over many of the important structural features in these galaxies, specifically the spiral arms of the more massive galaxies. As a result, areas of recent star formation were averaged with adjacent areas of older stellar populations, biasing some results towards older SFHs than what may be truly characteristic of parts of a galaxy. In a similar vein, dwarf irregulars undergoing clumpy star formation may not be the ideal candidates for the radial treatment. The stellar complex analysis presented here is a morphology-independent way to look at the most recent areas of star formation in a wide range of galaxies. 

\begin{figure}
	\includegraphics[width=\columnwidth]{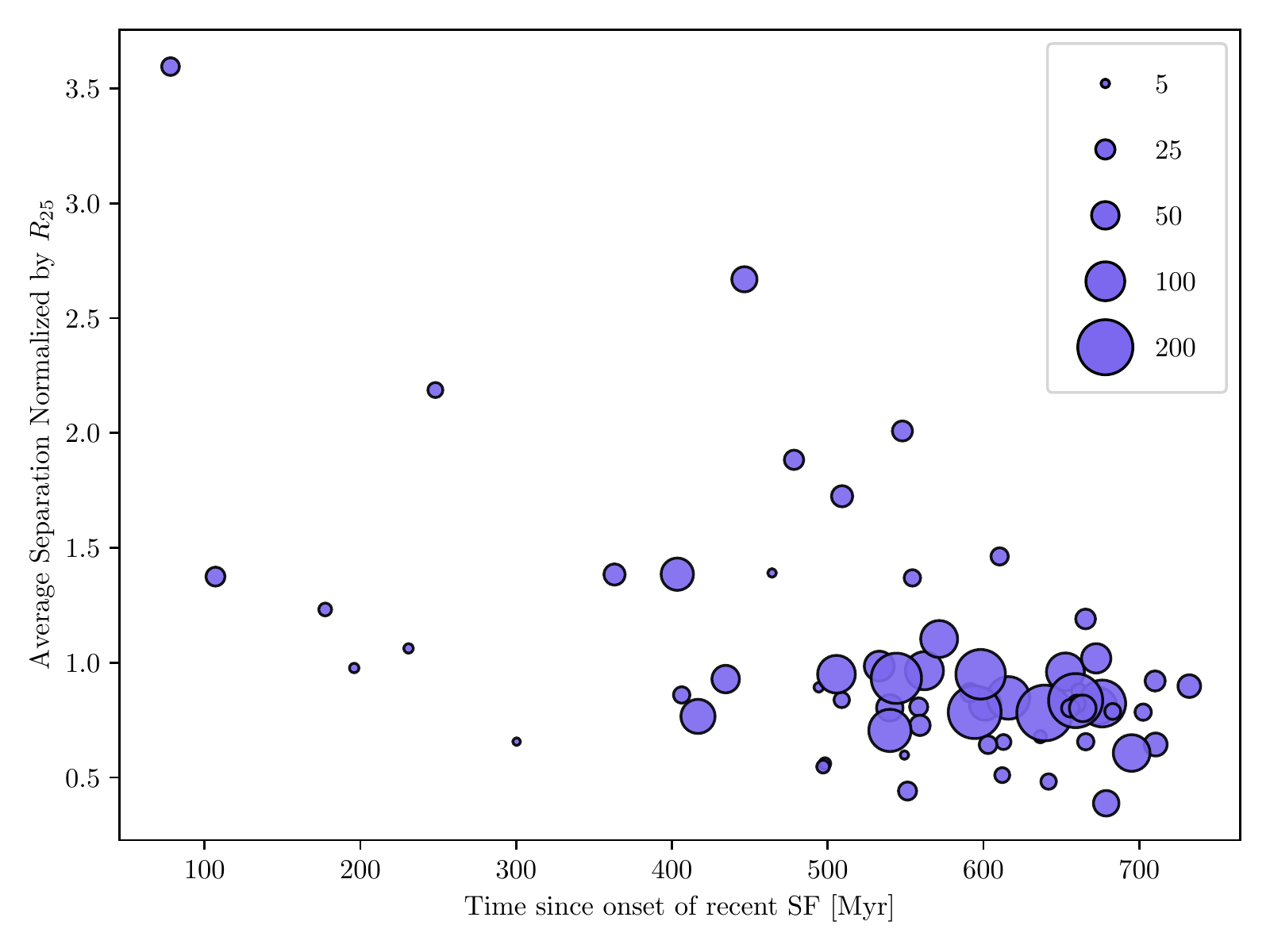}
	\caption{The average physical separation between stellar complexes within a group assigned by the k-means algorithm versus the average time since the last onset of star formation ($age_{burst}$). The size of each point is determined by the number of stellar compxlexes in each group. The smallest point has only four sources (a cluster in NGC 4242) and the largest has 205 sources (a cluster in NGC 7793). Because the sources were UV-selected, the lack of older clusters is expected. The legend shows how many sources are in each grouping. The uncertainty on the time of the burst is large, with an average of 200 Myrs.}
	\label{fig:sep_vs_age_clustering}
\end{figure}

The SED-fitting results of the UV-selected sources were used to examine the way different stellar populations exist in a galaxy. Are younger populations often grouped together or more spread throughout the disc? In the $\Lambda$CDM framework, it should be expected to find older populations more clumped together (often near the centre of the galaxy) than the areas of recent/current star formation. As an initial test of this concept, the scatter within groups of similar SFH properties was measured. Figure \ref{fig:sep_vs_age_clustering} shows the results of the clustered sources from all of the galaxies. The older (but still relatively young) groups of star formation were found to be closer together (lower average separation) than the groups of extremely recent onsets of star formation, as expected.

Although there may be an inverse relationship between the average separation between star-forming sources within a group of similar SFHs and the $t_{ySP}$ of that group's common SFH, the results seen in Figure \ref{fig:sep_vs_age_clustering} may be influenced by other things. Although the distances between sources within a common cluster were normalized by the size of the galaxy they are contained within so that the results could be compared between galaxies without biasing results, other biases arise. For example, the clusters with the seven highest values of average separation ($>1.4R_{25}$ in separation) come from just two galaxies, NGC 4625 and UGC 07608. NGC 4625 has an extended XUV disc with UV emission extending out to 4$\times$ its optical radius \citep{gildepaz2005}; therefore, normalizing the separation to $R_{25}$ biases the normalized, average separation to higher values than the rest of the galaxies that do not have the extended UV emission. In a similar vein, UGC 7608 is a low surface brightness galaxy where R$_{25}$ encompasses less of the total emission of the galaxy. This led to more UV sources being well-beyond R$_{25}$ and, therefore, led to more clustered sources appearing to have higher average separation in Figure \ref{fig:sep_vs_age_clustering}.

Another hidden variable in this stellar complex analysis is that it lacks information on the intrinsic nature of these sources. Their sizes are not encoded into these results, but may contain information regarding the more detailed stage of star formation these regions are experiencing. Hierarchical star formation on scales  similar to those discussed here have been studied extensively \citep[e.g.,][]{efremov1995,elmegreen1996,elmegreen1999}. Although the detailed analysis of the scale and nature of the UV-selected sources are beyond the scope of this paper, the information is available to be studied in a future project. Adding more galaxies to this sample may provide the information needed to draw more quantitative conclusions about the distribution of young star forming regions in local star forming galaxies.

\section{  Discussion}
\label{sec:discussion}

\subsection{  Comparing to Resolved Stellar Populations}
\label{ssec:cmd sfh}

Some of the galaxies in this sample are close enough to have \textit{HST} resolved photometry, making it possible to use CMDs to recover detailed SFHs. In particular, \citet{mcquinn2010} studied NGC 4068 and NGC 4214, exploring their recent/current starbursting characteristics. Their SFH's based strictly on stellar evolutionary models suggest that the most recent burst in NGC 4068 started at about 500 Myr, with which the three inner regions of our radial analysis agree (Figure \ref{fig:ngc4068_sfh}). The SFH from the resolved stellar population of NGC 4214 \citep{mcquinn2010} suggests that there were elevations in SFR 1-2 Gyr and $<$500 Myr ago (see their Figures 8 \& 9). The results of our radial analysis show that different radial components of the galaxy may have experienced these two episodes of elevated SF. The inner regions have more recent star formation, while the outer regions increasingly see older $t_{ySP}$. In these ways, the radial analysis of SED-derived SFHs agree with past studies of resolved stellar populations in these two galaxies. 

\begin{figure}
	\centering
	\includegraphics[width=\columnwidth]{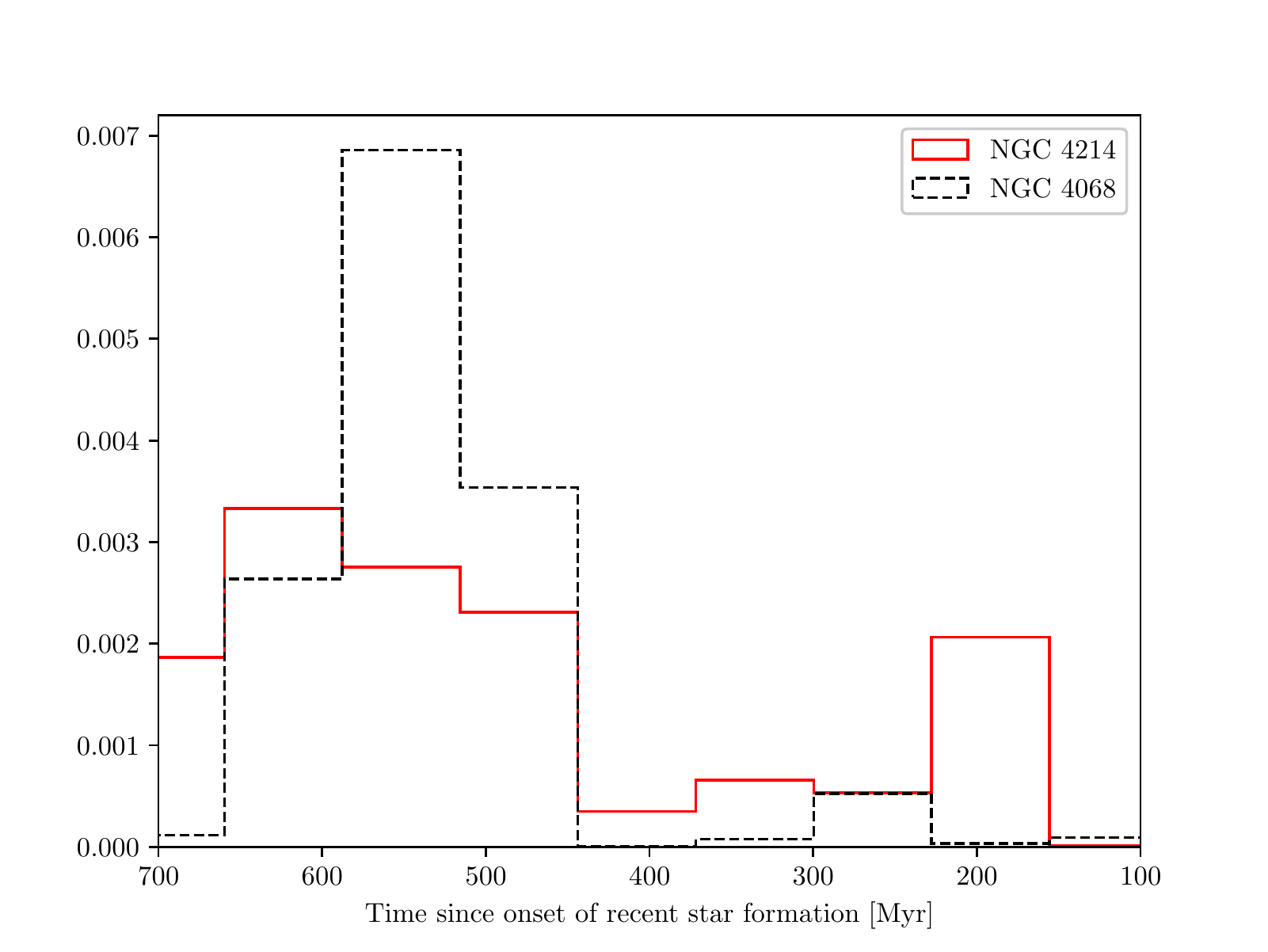}
	\caption{Normalized histogram of the SED-derived burst age for the UV-selected stellar complexes. In order to account for the differences in other properties, like size of the source and the strength of the current SFR, the times are weighted by the total luminosity of the source. Comparing these results to those in \citet{mcquinn2010}, our results show increased onsets of SF at the most recent times ($<$1 Gyr).}
	\label{fig:resolved_stellar_pop_lumweighted_comparison}
\end{figure}

The results of the individual stellar complex SED-fitting for NGC 4068 (and NGC 4214) are shown in Figure \ref{fig:resolved_stellar_pop_lumweighted_comparison}. The luminosity-weighted, normalized histograms of the $t_{ySP}$ values for NGC 4068 and NGC 4214 can be compared to the recent SFHs of \citet{mcquinn2010}. The CMD-derived SFHs are expected to have more resolution and less uncertainty than the currently presented method, but the general trends should be comparable. The common value of $t_{ySP} \approx$ 500 Myr among many of the UV-selected sources in NGC 4068 align with the idea that NGC 4068 began a large onset of star formation at about that time. There are not many star-forming complexes with $t_{ySP}$ < 500 Myr for NGC 4068. This result combined with the H$\alpha$ equivalent width profile with consistently high values (provided in Figure \ref{fig:ngc4068_fulldata}), suggests that the elevated levels of star formation that began \textasciitilde500 Myr years ago has persisted at a consistent level from then until present times. These conclusions agree with the results of the CMD-derived SFH for NGC 4068 in \citet{mcquinn2010}. 

On the other hand, NGC 4214's individual sources appear to have a more bimodal distribution of $t_{ySP}$ (Figure \ref{fig:resolved_stellar_pop_lumweighted_comparison}). While this shape is similar to the CMD-derived SFH presented in \citet{mcquinn2010}, the peaks (450 \& 175 Myr for \citet{mcquinn2010} and 600 \& 200 Myr for this project) do not directly align. This difference in times could be due to either binning or SED-based uncertainty issues. With more testing and validation, extending this individual UV source analysis to galaxies more distant than \textasciitilde 8 Mpc may allow for SFHs to be studied on smaller scales when resolved stellar populations are harder to observe. Presenting the SED-dervied SFHs in this way may also alleviate some of the difficulties associated with having to choose between strictly parametric forms of SFHs, since the distribution of SED results across a galaxy may be more insightful than using single $\tau$ value for the entire galaxy.

Another galaxy from our sample that has had its CMD-derived SFH computed was NGC 7793 in \citet{radburnsmith2012} and \citet{sacchi2019}. Both of their results suggest that about 80\% of the galaxy's stars were formed within the last 8 Gyr and 40\% within the last 4 Gyr. Our results (Figure \ref{fig:ngc7793_sfh}) agree with their cumulative SFH's, although the shapes from our analysis are different. As mentioned before, a single parametric SFH (the double exponential form) was used for the entire sample, so it is expected that some galaxies are more well-fit than others. In the case of NGC 7793, a delayed SFH might be the more appropriate functional form to have chosen, but the flexibility of the available parameters produced results that appear quite similar to the single parameter $\tau$ models in the case of NGC 7793.

\subsection{  Stellar Migration}

Numerical simulations such as those in \citet{radburnsmith2012} have been utilized to explain the breaks in surface brightness profiles via radial migration. In the case of \citet{radburnsmith2012}, NGC 7793, one of the galaxies in this sample, showed evidence that the outer region of the galaxy was older than the interior because of strong radial stellar migration. Type I and III surface brightness profiles (those with pure exponential profiles and those with upbending profiles after the break radius) are linked with high radial migration efficiency \citep{ruizlara2017} and are not uncommon, so these dynamical effects should not be ignored in the context of stellar populations. The deep 3.6 and 4.5 $\mu$m images from EDGES can help determine which galaxies in our sample may be exhibiting the observational evidence of stellar migration from their surface brightness profiles \citep{staudaher2019}. If stellar migration is a major factor in these galaxies, our initial assumption that the concentric ellipses defined from the photometry evolve independently of each other may falter. In this case, it would be important to create a system in which the SED-fitting process would have knowledge of neighboring annuli as it fits each annulus. 

\section{  Conclusion}
\label{sec:conclusion}

We have presented new UBVR observations from the WIYN 0.9m telescope for several nearby galaxies. These observations combined with archival \textit{GALEX} and Spitzer imaging allow us to take a deeper look into the radial trends of galaxies in the local volume. Surface brightness profiles were presented for each galaxy, based on fixed-ellipse photometry. Classic star formation rate tracers, 
UV $+$ mid-IR and H$\alpha$ fluxes, were examined as a function of radius to explore regions of recent star formation, on two different time-scales (100 and 10 Myrs respectively). In conjunction with these classic star formation indicators, SED fitting was utilized to derive more star formation properties, with a focus on star formation histories. While there is a plethora of parameters that SED fitting can infer, this analysis focused on the most recent onset of star formation (t$_{ySP}$) in a double-exponential SFH model.  

Radial SFHs often showed the same trends seen in the photometric profiles, with small incremental changes in star formation of adjacent ellipses. Larger changes in the SED-inferred SFHs between radial bins often signified a turnover from one structural feature to another (the difference between an old bulge and a younger stellar disc, for example). For the spiral galaxies in this sample, the intermediate regions showed the youngest stellar populations which suggests a growing stellar disc within a larger, older stellar halo, agreeing with the results of \citet{dsouza2014} and \citet{Dale2016}.

The UV-selected stellar complex analysis provided local insight - more than was possible with the azimuthally-averaged radial analysis - about the regions of youngest star formation in each galaxy. We showed a tentative relationship between the ages of clusters of UV-sources and the way they are distributed throughout a galaxy. Stellar complexes that have not recently begun star formation are more likely to be clustered together. Future studies similar to this may provide direct constraints on this relationship. There was no discernible difference in this relationship when morphologies were taken into consideration. This conclusion was counterintuitive due to the differences in color and SFH gradients seen in the radial analysis of the same galaxies. 

In the future, we hope to utilize the methods presented here on a larger sample of nearby galaxies to better understand the variety of stellar population distributions possible in a $\Lambda$CDM universe. With a larger sample comes the possibility of placing better physical constraints on possible star formation properties on different scales (global, radial, or local). These constraints can better inform the realistic simulations of galaxy evolution.

\section*{Acknowledgements}
The authors thank the referee for taking the time to provide thoughtful feedback on this paper.
The authors would also like to thank the CIGALE team for making their SED-fitting software publically available. This project is funded by NASA award NNX17AF17G. MVS worked on this project as a recipient of the Indiana Space Grant Consortium Doctoral Fellowship. This research made use of Astropy,\footnote{http://www.astropy.org} a community-developed core Python package for Astronomy \citep{astropy2018}. This research has made use of the NASA/IPAC Extragalactic Database, which is funded by the National Aeronautics and Space Administration and operated by the California Institute of Technology.
% The Acknowledgements section is not numbered. Here you can thank helpful
% colleagues, acknowledge funding agencies, telescopes and facilities used etc.

%%%%%%%%%%%%%%%%%%%%%%%%%%%%%%%%%%%%%%%%%%%%%%%%%%

\section*{Data availability}
The GALEX and Spitzer data underlying this article were accessed from public sources. The derived data generated in this research will be shared on reasonable request to the corresponding author.


\bibliographystyle{mnras}
\bibliography{edges}

%\begin{thebibliography}{99}
%\end{thebibliography}

%%%%%%%%%%%%%%%%%%%%%%%%%%%%%%%%%%%%%%%%%%%%%%%%%%

%%%%%%%%%%%%%%%%% APPENDICES %%%%%%%%%%%%%%%%%%%%%
\appendix
\section{  Photometric Measurements \& HI Observations}
\label{appen:measurements}
For the 14 galaxies not shown in the body of this paper (see Figures \ref{fig:ngc4242_sfh} and \ref{fig:ngc4242_fulldata} for NGC 4242), the photometric measurements and radial cumulative star formation histories are described here. As described in Section \ref{ssec:results_sb}, the surface brightness profiles of each galaxy at three wavelengths (NUV, B, and IRAC Channel 1) are measured. The (B-R) color gradient is shown because a change in optical colors can be linked to a change in stellar populations \citep{larson_tinsley1978}. The sSFR profile, from the FUV+24$\mu$ SFR indicator \citep{hao2011}, and the H$\alpha$ EW profile trace the recent star formation on different time scales (100 Myr and 10 Myr time-scales respectively)

This appendix also presents the HI measurements and color images of the galaxies in the sample. The HI data comes from a variety of sources, shown in Table \ref{table:hi}. The HI data is presented as additional observations that can be drawn from to form conclusions about the past, present, and future ability of a galaxy to form stars and grow its stellar disc. The top left panel shows the IRAC 3.6$\mu$m image from EDGES, with an HI surface density contour superimposed. Unless stated otherwise, the contour represents 1 $\times$ 10$^{20}$ atoms cm$^{-2}$. The top left panel shows an RGB color image where red is 3.6$\mu$m from Spitzer, green is R from WIYN 0.9m, and blue is B + FUV from WIYN 0.9m/GALEX, unless noted otherwise. Although the surface brightness profile analysis and SED fitting relied on the azimuthal averaging of concentric radial bins, there are often features that can be lost in this averaging process. The color images are designed to bring attention to some of the features that may be important in further understanding the growth of the stellar disc. The bottom left image is the HI surface density contours starting at 0.5 $\times$ 10$^{20}$ atoms cm$^{-2}$. The bottom right image is the velocity field, including some annotations to indicate the contour step sizes. In all panels that include HI data, the beam size is represented in the bottom left corner.

\setlength\tabcolsep{1.5pt}
\begin{table*}
\caption{Published HI data references: (1) \citet{vanderhulst2001}; (2) \citet{richards2016}; (3) \citet{richards2018}; (4) \citet{walter2008}; (5) \citet{swaters2002}; (6) AW618 C config.: \citet{bush2004}, AW618
B,C config.: \citet{kaczmarek2012}; (7) \citet{hunter1998}; (8) \citet{patterson1996}; (9) \citet{broeils1992}.  $^{*}$ only one polarization}

\begin{tabular}{lcccccccc}
\hline
Galaxy        & \begin{tabular}[c]{@{}c@{}}Telescope;\\  array\end{tabular} & \begin{tabular}[c]{@{}c@{}}Project\\ code\end{tabular} & \begin{tabular}[c]{@{}c@{}}Time on\\  source\\ (h)\end{tabular} & \begin{tabular}[c]{@{}c@{}}Chan.\\ sep.\\ (km s$^{-1}$)\end{tabular} & \begin{tabular}[c]{@{}c@{}}Beam\\ size\\ (arcsec)\end{tabular} & \begin{tabular}[c]{@{}c@{}}Beam\\ PA\\ (deg)\end{tabular} & \begin{tabular}[c]{@{}c@{}}Noise\\ (mJy beam$^{-1}$)\end{tabular} & \begin{tabular}[c]{@{}c@{}}Pub.\\ ref\end{tabular} \\ \hline
NGC 0024      & VLA; DnC, CnB   & AF219, AC343          & 9.5         & 2.6    & 30.0 x 21.6     & 11.2      & 1.0   & --       \\
NGC 3344      & WSRT               	   & WHISP       & 12           & 4.1     & 62.0 x 48.8     & 0.0   & 1.9   & (1) 		  \\
NGC 3486      & VLA; C      & 13A-107      & 6.6     		       & 5.0   	  & 25.5 x 17.7     & 55.8     & 0.47  & (2)       \\
NGC 3938      &   VLA; C, D  &  AW358           &     6.0$^{*}$           &       2.6            &      27.5 x 24.6           &         9.50     &   1.1    &           --                \\
NGC 4068      & VLA; C         & 16A-013            & 6.8                    & 5.0       & 18.7 x 16.7       & 43.7      & 0.46  & (3)          \\
NGC 4096      & VLA; C       & 16A-013         & 7.4        & 5.0         & 20.1 x 16.6       & -75.9       & 0.47  & (3)       \\
NGC 4214      & VLA; B, C, D          & AM418      & 11             & 5.2             & 21.7 x 19.4        & -23.7        & 0.65  & (4)             \\
NGC 4242      & WSRT              & WHISP       & 12                  & 4.1          & 30.9 x 30.8             & 0.0           & 4.0   & (5)        \\
NGC 4618/4625 & VLA; B, C, D    & AW618, AO108, AO101        & 38              & 5.2            & 19.7 x 18.2         & 84.4         & 0.25  & (6)     \\
NGC 7793      & VLA; BnA,CnB,DnC       & AW605       & 9.5               & 2.6                & 15.6 x 10.9              & 10.7          & 0.92  & (4)                                        \\
UGC 07408     &   WSRT    &   WHISP            &         12       &             2.1          &       17.5 x 12.7               &           0.0     &   4.3    &      (5)             \\
UGC 07577     & VLA; D      & AH540        & 0.7 $^{*}$                & 5.2                   & 60.6 x 55.5      & 81.2       & 1.5   & (7)                                 \\
UGC 07608     & VLA    & AP198          & 1.7                  & 2.58                   & 18.8 x 18.1          & 67.4       & 1.6     &  (8)      \\
UGC 08320     & WSRT         & --            & 48         & 2.1       & 14.0 x 19.0        & 0.0       & 1.1   & (9)              \\ \hline                      
\end{tabular}
\label{table:hi}
\end{table*}
\newpage
\subsection{NGC 0024}
NGC 0024 has only a moderate B-R gradient, and not a much variation in sSFR outside of the bulge, suggesting a fairly uniform SFH.
The cumulative SFH reflects this. This galaxy's high inclination has been noted in observations of its HI and H$\alpha$ distributions \citep{chemin2006,dicaire2008}. As it is one of the more highly inclined galaxies in the sample, NGC 0024's lack of variation across the disc and red (B-R) values may be caused by inclination effects. The consistently elevated values of log(sSFR) and log(EW) observed in this radial analysis agrees with the even distribution of H$\alpha$ noted in \citet{rossa2003}. The SED-derived SFHs show a similar consistency across the disc, even with more coarsely-spaced bins. A large fraction of star formation began across the entire galaxy approximately 2 Gyr ago.

\subsection{NGC 3344}
 NGC 3344 has a well-established strong B-R color gradient \citep{prugniel1998}. NGC 3344 has also been identified as having an XUV disc that fades more slowly than the optical disc \citep{thilker2007}.  The analysis of the observations done here show both of these attributes of  NGC 3344. As is seen in the colour image of Figure \ref{fig:ngc3344_4panel}, as well as the profiles of Figure \ref{fig:ngc3344_fulldata}, NGC 3344 has a very red (bright in the NIR) centre, but a strong young, blue (particularly in the B+NUV) stellar component in the outskirts of the galaxy.  The SED-fitted, cumulative SFH results show a variation in time of the onset of recent star formation across the disc. A recent, strong burst of star formation seen in the annulus reaching out to 3.6$r_{e}$ represents a region that coincides with the highest sSFR values ($r_{e}$ = 59"). 
 
 The inner and outer rings, defined in \citet{verdesmontenegro2000}, are most clearly seen in the HI (Figure \ref{fig:ngc3344_4panel}). The HI data for NGC 3344 was taken from the Westerbork H I Survey of Spiral and Irregular Galaxies (WHISP) project \citep{swaters2002}. The extended HI on the eastern side of the galaxy could be indicative of interaction. Although the evidence for interaction is axisymmetric, the extended B+UV emission (shown in blue in the RGB image) appears to show azimuthal symmetry. The bluest regions with the highest sSFR values are outside of $R_{25}$ but are well within the HI disc, suggesting that the stellar disc is actively growing.

\subsection{NGC 3486}
 With the exception with the outer most ring, NGC 3486 has one of the most obvious examples of $t_{ySP}$ decreasing and $f_{burst}$ increasing with galactocentric radius. The central region is host to a possible Type 2 Seyfert \citep{ho1997,annuar2020}. The outermost region, although slightly bluer than the intermediate regions, has a similar $t_{ySP}$ as the intermediate regions. This combination of information may be a metallicity effect, caused by a small low-metallicity stellar halo.
 
 There are two HI arcs extending from the southeast and northwest parts of NGC 3486 (Figure \ref{fig:ngc3486_4panel}). However, these features appear to be rotating with the rest of the galaxy as expected by the velocity field contours. The wiggle seen just outside D$_{25}$ in the velocity contours roughly coincides with the downbend seen in the 3.6$\mu$m surface brightness profile of Figure \ref{fig:ngc3486_fulldata}. The color image shows a bright, red bulge, coinciding with a region of decreased HI surface density. 
 
 \subsection{NGC 3938}
NGC 3938 is a nearly face-on and symmetric Sc galaxy. This galaxy has also been observed as part of the SINGS sample and its multiwavelength surface brightness profiles have been documented \citep{munozmateos2009}. Similar to NGC 3344, NGC 3938 has prominent B+NUV emission at its edges in the colour image of \ref{fig:ngc3938_4panel}. The SED-fitting results in Figure \ref{fig:ngc3938_sfh} show that NGC 3938 does not have much radial variation in SFH. Prior analysis of neutral gas distribution and kinematics is presented in \citet{vanderkruit1982}. The HI (Figure \ref{fig:ngc3938_4panel}) does not extend much past the stellar disk and is rather symmetric in its velocity.

\subsection{NGC 4068}
NGC 4068 is a starbursting dwarf galaxy, identified by its extremely blue colors \citep{gallagher1986}. SED-derived SFHs, shown in Figure \ref{fig:ngc4068_sfh}, agree with the CMD-derived SFH of \citet{mcquinn2010}. Because NGC 4068 has a moderate blue to red gradient, its SED results were expected to have a proportional relationship between $t_{ySP}$ and radius. Even with a moderate color gradient, the color profile shows consistently bluer (B-R) values ((B-R) $\leq$ 1 everywhere) than most other galaxies in the sample. The SED-derived results of the three innermost regions suggest that about 50\% of the stars in this galaxy formed within the past 1 Gyr, where the burst of star formation has been largely located. 

Figure \ref{fig:ngc4068_4panel} shows how the HI extends to the south much farther than the stellar disc in NGC 4068. There is an apparent offset between the photometric and HI position angles. This twisting effect is also seen in the velocity field contours. The highest HI surface density is located in the southwest corner of the galaxy. Two other clumps of high HI surface density are seen to the northeast of this point. All three of these knots are coincident with clumps of H$\alpha$.

\subsection{NGC 4096}
NGC 4096 is an SABc galaxy with a bright nucleus \citep{devaucouleurs1964}. Well-defined spiral arms combined with the effect of the high inclination (b/a = 0.267) results in quickly-varying, bumpy radial profiles (Figure \ref{fig:ngc4096_fulldata}). A lopsidedness has been noted in the HI density profile and kinematics as well \citep{garcia2002,vaneymeren2011}. The HI data presented in Figure \ref{fig:ngc4096_4panel} was originally published in \citet{richards2018}.
NGC 4096 is one of the redder galaxies in the sample, and it correspondingly has one of the highest median values of $t_{ySP}$ from the radial component analysis at 2.3 Gyr.

\subsection{NGC 4214}
NGC 4214 is another starbursting dwarf irregular galaxy. Previously studied multiwavelength observations revealed regions of the galaxy that are currently experiencing high levels of star formation \citep{huchra1983,sargent_filippenko1991}. The bright UV and H$\alpha$ emission, noted in the literature, can also be seen in the radial profiles of Figure \ref{fig:ngc4214_fulldata}. The close proximity of this galaxy has also allowed resolved stellar population studies and CMD-derived SFHs \citep{mcquinn2010,weisz2011,williams2011}. The comparison between these SFH results and those of \citet{mcquinn2010} are discussed more in Section \ref{ssec:cmd sfh}. The decreasing EW(H$\alpha$) profile in Figure \ref{fig:ngc4214_fulldata} matches the expectations of starburst dwarf galaxies, but a B-R profile that grows more blue with radius matches the trend seen in larger spirals. Although this project does not quantify the steepness of the measured color gradients, doing this in the future may lead to exploring the connection between slope of color gradients, the age of stellar populations, and the masses of galaxies. In the case of NGC 4214, there is not a strong radial color gradient out to the low surface brightness and the galaxy appears to be dominated by an extremely young stellar population.

The HI distribution of NGC 4214 extends past the reach of the stellar component by a considerable amount (upper left panel of Figure \ref{fig:ngc4214_4panel}). Clumps of UV emission are coincident with higher HI surface densities in the centre as well as along a weak bar structure. \citet{bagetakos2011} documented 51 HI holes in NGC 4214, suggesting the scale height of this galaxy is low which would result in holes breaking out of the disc more easily than in other galaxies. 
Although there has been detailed studies of the extremely young ($<$ 10 Myr) stars in NGC 4214, there is still a prevalent old stellar population throughout the disc \citep{williams2011}. The more evenly distributed older stars may be the reason that the radial SFHs do not show extremely young $t_{ySP}$'s everywhere ($\langle t_{ySP} \rangle = 940$ Myr).

\subsection{NGC 4242}
NGC 4242 is classified as an SAB(s)dm (RC3), with a low surface brightness disc \citep{eskridge2002}. The B-R color and sSFR profiles for NGC 4242, shown in Figure \ref{fig:ngc4242_fulldata}, suggest a younger stellar population with heightened star formation exists at mid-galactic regions. The signature "U-shaped" B-R colour profile (i.e., with a minimum blue colour in the middle region) indicates a galaxy that may have undergone inside-out growth in addition to the substantial building up of an older stellar halo through accretion/mergers \citep{bakos2008,Dale2016}. 

Most of the HI is coincident with the stars, although there are detections of low surface density HI outside of the immediate region of the stellar disc (see Figure \ref{fig:ngc4242_4panel}). The area of highest current star formation, as indicated by both the EW(H$\alpha$) profile and the SED-fitted SFH, is just within D$_{25}$ (as shown as the white ellipse in the velocity field panel of Figure \ref{fig:ngc4242_4panel}). The HI is mostly symmetric with only a slight extension to the southwest of the galaxy where the HI, as well as the H$\alpha$, extends further than the stellar population traced at 3.6$\mu$m.

\subsection{NGC 4618}
NGC 4618 is an SBm galaxy that forms a physical pair with NGC 4625. \citet{bertola1967} likened its morphology to the LMC because of its strong bar and its main arm that encompasses most of the central region. NGC 4618 has a bar that is not centered on the nucleus \citep{eskridge2002}, and this offset can be seen in the photometric profiles in Figure \ref{fig:ngc4618_fulldata}. The tidal interaction between NGC 4618 and NGC 4625 is seen in the morphology and HI distributions of both (Figure \ref{fig:ngc4618_4panel}). The asymmetry of NGC 4618 makes the radial analysis difficult to interpret, but was a good candidate for the individual stellar complex analysis. %I wish i could include the B-R stellar complex image here because it reveals blue areas in the SE and the NW was more red 
\citet{staudaher2019} used the deep 3.6$\mu$m image of NGC 4618 to quantify the breaks in the stellar disc, finding a bulge and possible four disc components. 
A complicated evolutionary history, whether through stellar migration, tidal interactions, mergers, or some mixture of all these factors, may be an explanation for the changes in NGC 4618's stellar disc.  

\subsection{NGC 4625}
NGC 4625 was one of the first galaxies to be classified as having an extended UV (XUV) disc \citep{gildepaz2005,thilker2007}. The UV and H$\alpha$ emission that persists far outside of D$_{25}$ combined with elevated HI surface densities (Figure \ref{fig:ngc4618_4panel}) and young $t_{ySP}$'s (Figure \ref{fig:ngc4625_sfh}) in the same region points to late stage disc growth \citep{elmegreen_hunter2006,bush2014}.  According to the color profile (Figure \ref{fig:ngc4625_fulldata}) and the radial SFHs (Figure \ref{fig:ngc4625_sfh}), the oldest stellar population in NGC4625 resides near the D$_{25}$ isophote. Because of the clear difference between its inner and outer regions, NGC 4625 is a prime example of a galaxy with an "inside-out" growth signature. 

\subsection{NGC 7793}
 NGC 7793 is an SAd galaxy with very little structure in its old stellar disc and is part of the Sculptor group \cite{elmegreen1984}. Because of its close proximity to the Milky Way, NGC 7793 has been extensively studied in many wavelengths. The photometry in \citet{carignan1985} as well as the measurements in Figure \ref{fig:ngc7793_fulldata} show little to no color gradient across the galaxy. The SED-derived radial SFH of NGC 7793 (Figure \ref{fig:ngc7793_sfh}) shows how NGC 7793 has a roughly constant SFH over its lifetime with the exception of the small bulge region, as mentioned in Section \ref{ssec:cmd sfh} and in \citet{sacchi2019}. The similar results among all annuli included in the radial analysis suggests that the continuous SFH was experienced across the galaxy, and not limited to either the inner or outer regions. 
 The HI velocity field of NGC 7793 is relatively symmetric, with the exception of the low density southeast region (Figure \ref{fig:ngc7793_4panel}). The HI disc does not extend much more than the stellar disc at 3.6$\mu$, agreeing with the assessment of \citet{deblok2008}.

\subsection{UGC 07408}
UGC 07408 has weakly detected H$\alpha$ emission, has very little HI content \citep{swaters2002}, and has only a moderate blue-to-red color gradient \citep{hunter2006}. The inner regions do not change rapidly in B-R, which is reflected in the cumulative SFH. The two inner most regions are identical in their SFHs. Between the two inner regions and the next annulus, there is a larger change in $t_{ySP}$ (Figure \ref{fig:ugc07408_sfh}) than would be suggested by the photometric diagnostics of Figure \ref{fig:ugc07408_fulldata}. The strong, recent burst suggested by the results in the central regions would most likely lead to some current star formation, which goes undetected by the H$\alpha$ image. This contradiction leads us to conclude that the results for UGC 07408 may be overestimating the burst fraction, $f_{burst}$. 
The HI observations come from WHISP and are shown in Figure \ref{fig:ugc07408_4panel}.

\subsection{UGC 07577}
UGC 07577 (DDO 125) is a low mass dwarf irregular galaxy that is often associated with NGC 4449 (physical separation of 40 kpc) \citep{barnes_hernquist1992}. UGC 07577 lacks a dominant dark matter component and could have formed in the tidal streams of NGC 4449 \citep{swaters1999,hunter2000}. Figures \ref{fig:ugc07577_fulldata} and \ref{fig:ugc07577_sfh} show a galaxy experiencing elevated star formation in the central regions. An analysis of its global star formation history \citep{weisz2011} suggested that the majority (\textasciitilde 90\%) of the current fraction of stars in the galaxy was formed more than 6 Gyrs ago. The rest of the stars were formed within the past 1 Gyr. The SED-derived SFHs in Figure \ref{fig:ugc07577_sfh}, therefore, may have overestimated $f_{burst}$. However, the SED-derived $t_{ySP}$ at every radii is less than 1 Gyr. A galaxy that may be experiencing its first large star formation episode in several Gyr may be especially suited for the double exponential SFH parametrization. The HI observations shown in Figure \ref{fig:ugc07577_4panel} were previously published in \citet{hunter1998}. The velocity field shows solid body rotation that is undisturbed. 

\subsection{UGC 07608}
UGC 07608 (DDO 129) is an irregular galaxy with bright UV sources in the north west region of the galaxy (upper right panel of Figure \ref{fig:ugc07608_4panel}). Its irregular morphology made UGC 07608's azimuthally averaged photometry appear bumpy, seen in Figure \ref{fig:ugc07608_fulldata} and previously seen with the photometry in \citet{patterson1996}. However, the coarser placement of ellipses for the SED fitting analysis makes the knots and clumps of H${\alpha}$ and UV emission become less apparent (Figure \ref{fig:ugc07608_sfh}). The high EW(H$\alpha$) coupled with the rather blue colors seen in UGC 07608 are correlated with the high $f_{burst}$ and low $t_{ySP}$ in the cumulative SFHs in all annuli.

\subsection{UGC 08320}
UGC 08320 (DDO 168) is a well-studied dwarf irregular in the Canes Venatici I group \citep{karachentsev2003}. \citet{bremnes1999} notes a luminosity excess above a pure exponential disc at large radii in the surface brightness profile of this dwarf. Although there are clear breaks in the surface brightness profiles of Figure \ref{fig:ugc08320_fulldata}, a closer look at how the profiles deviate from an exponential fit would be needed to confirm that the methods used in this project produce the same result as \citet{bremnes1999}. 

The SFH results for UGC 08320 are similar to those of \citet{Dale2016}, even with slightly differing SED parameters spaces. A large portion of the stellar population formed more recently than most of the other galaxies in this sample, according to the SFH results in Figure \ref{fig:ugc08320_sfh}. The oldest stellar population, residing in the outermost annulus, still has undergone recent star formation. This display of an extremely recent, large onset of star formation, coinciding with the extended HI disc \citep{ghosh2018} and a slow HI bar \citep{patra2019}, points to a galaxy going through an important evolutionary phase.

The HI extends past the stellar component by many arcseconds (Figure \ref{fig:ugc08320_4panel}). The HI surface density contours and velocity field appears twisted, suggesting a warped disc. The low density HI coincides with the reddest stellar populations in the galaxy measured by the color gradients, log(sSFR) profile, EW(H$\alpha$) profile, SED-derived SFH (Figures \ref{fig:ugc08320_fulldata} and \ref{fig:ugc08320_sfh}). There is a misalignment of the stellar and HI position angles.

\twocolumn
  \begin{figure}
     \centering
     \includegraphics[width=\columnwidth]{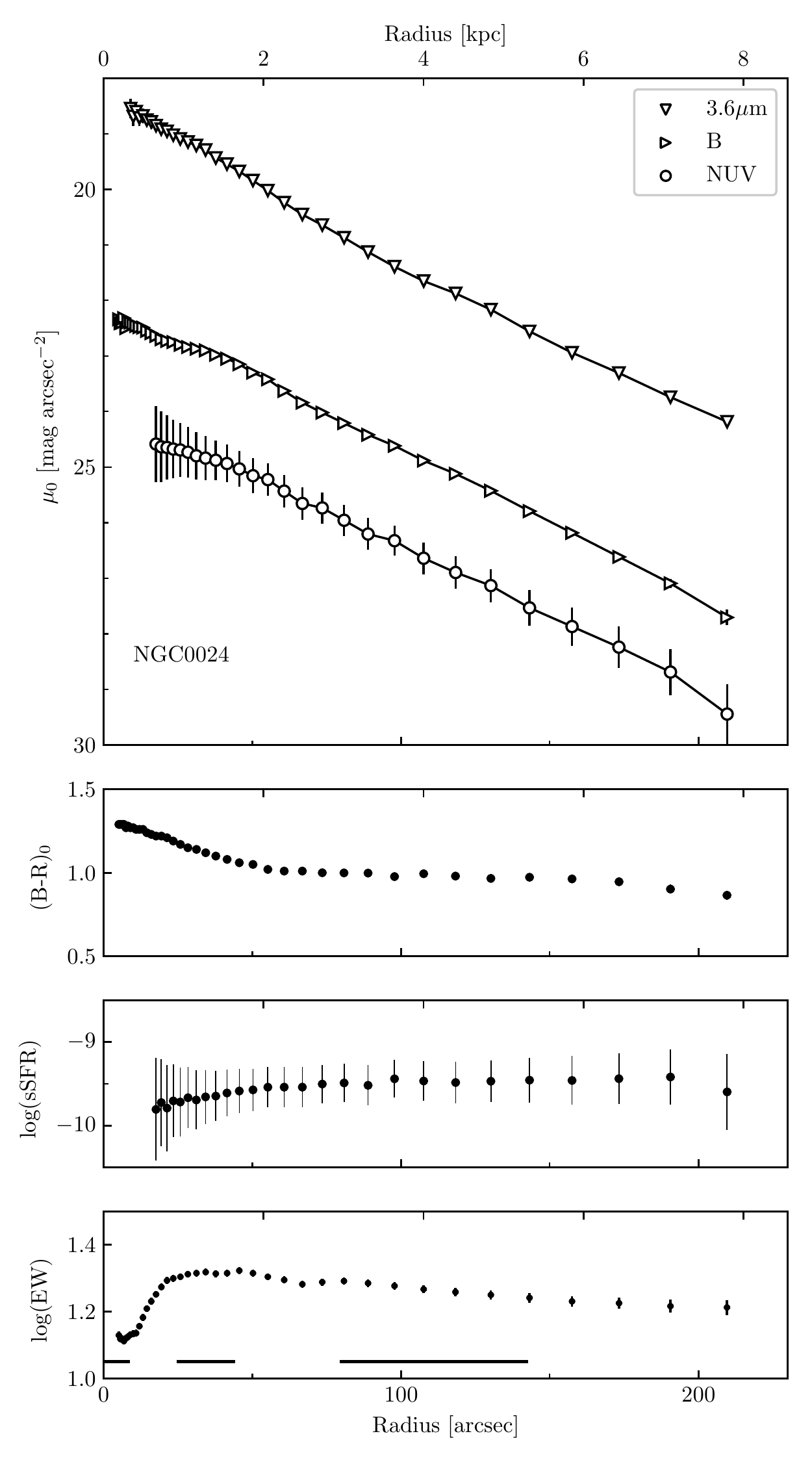}
     \caption{     The photometric and classical SFR tracer profiles for NGC 0024. Profiles are corrected for foreground extinction and inclination. In the bottom panel, solid horizontal lines are used to show the larger radial bins used for SED fitting.}
     \label{fig:ngc0024_fulldata}
 \end{figure}
 
 \begin{figure}
     \centering
     \includegraphics[width=\columnwidth]{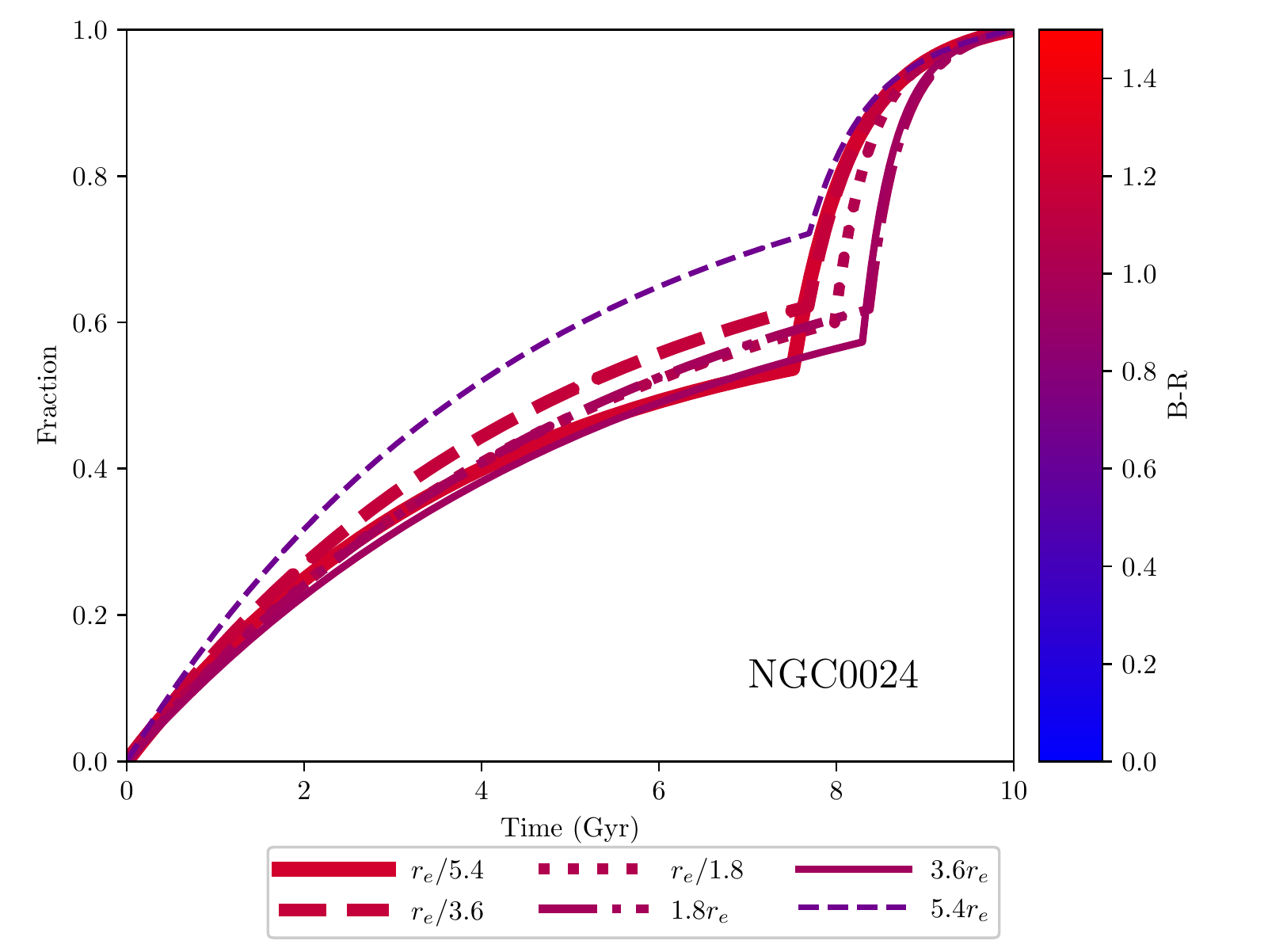}
     \caption{     The cumulative SFH for radial bins in NGC 0024. The central region is a 6" aperture. All radii are measured along the major axis. Any radial bin that is present in the legend but not in the figure means that the SED fitting resulted in a fit with $\chi^{2}_{\nu} > $ 15 for that bin.  }
     \label{fig:ngc0024_sfh}
 \end{figure}

\begin{figure*}
	\centering
	\includegraphics[width=2\columnwidth]{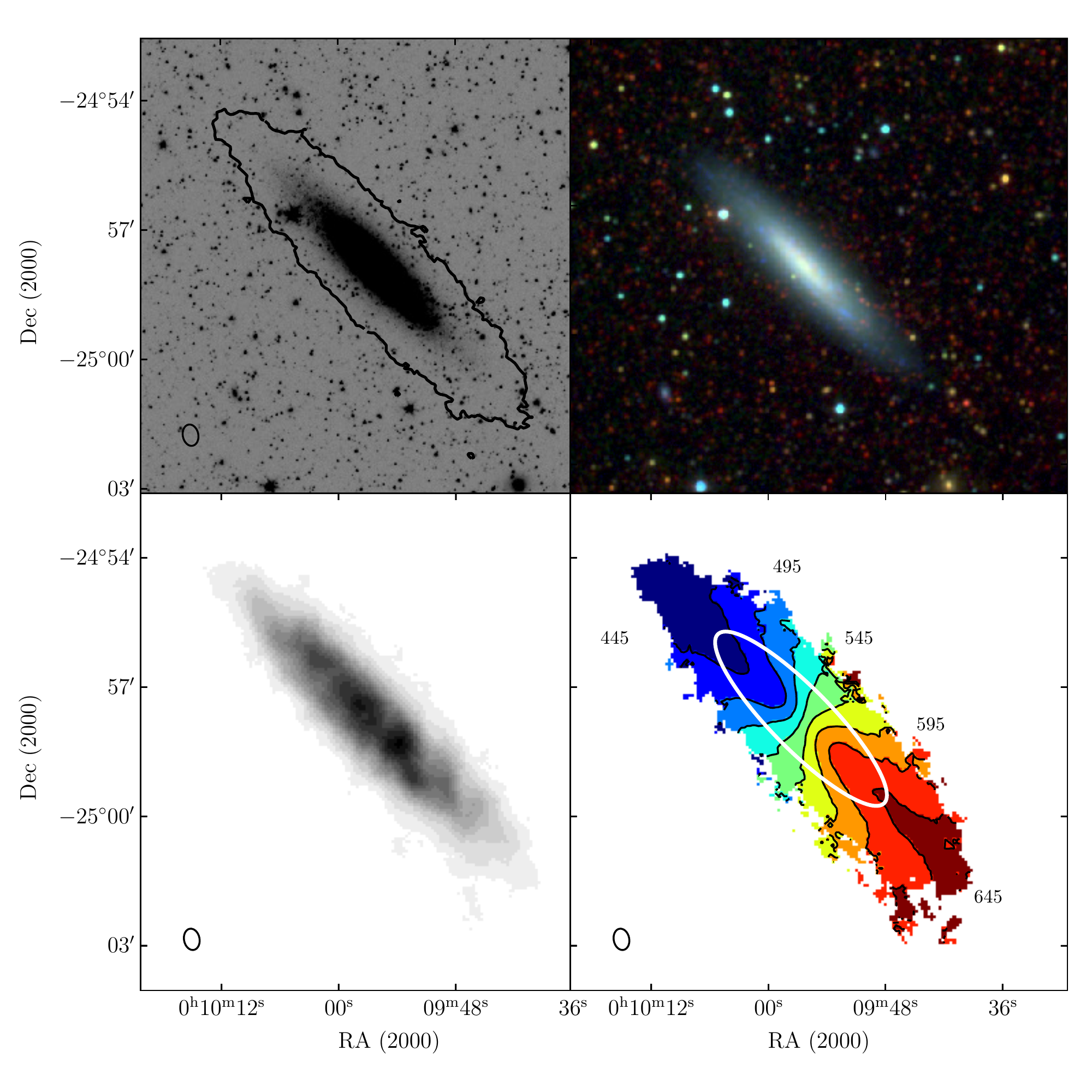}
	\caption{  NGC 0024. Top left panel shows the extent of the HI distribution (contour) in relation to the underlying stellar disc (3.6 $\mu$m image). Top left panel shows an RGB image where red is 3.6$\mu$m, green is R from WIYN, and blue is B + FUV from WIYN/Galex. Bottom left panel shows the HI surface density contours starting at 0.5 $\times$ 10$^{20}$ atoms cm$^{-2}$. The bottom right panel shows the HI velocity field.}
	\label{fig:ngc0024_4panel}
\end{figure*}

\clearpage

  \begin{figure}
     \centering
     \includegraphics[width=\columnwidth]{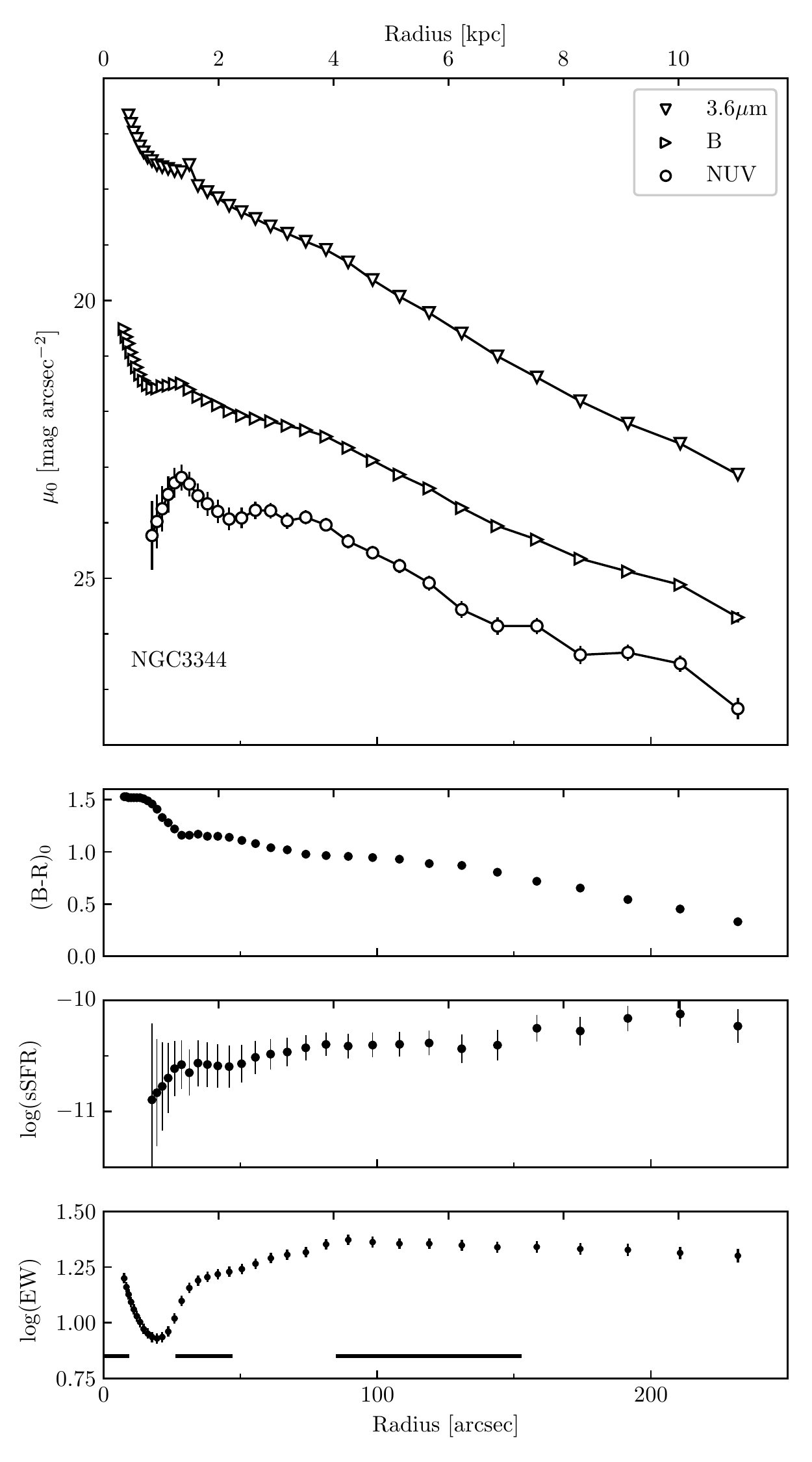}
     \caption{ \textbf{NGC 3344:}   Panels same as Figure \ref{fig:ngc0024_fulldata}. }
     \label{fig:ngc3344_fulldata}
 \end{figure}

 \begin{figure}
     \centering
     \includegraphics[width=\columnwidth]{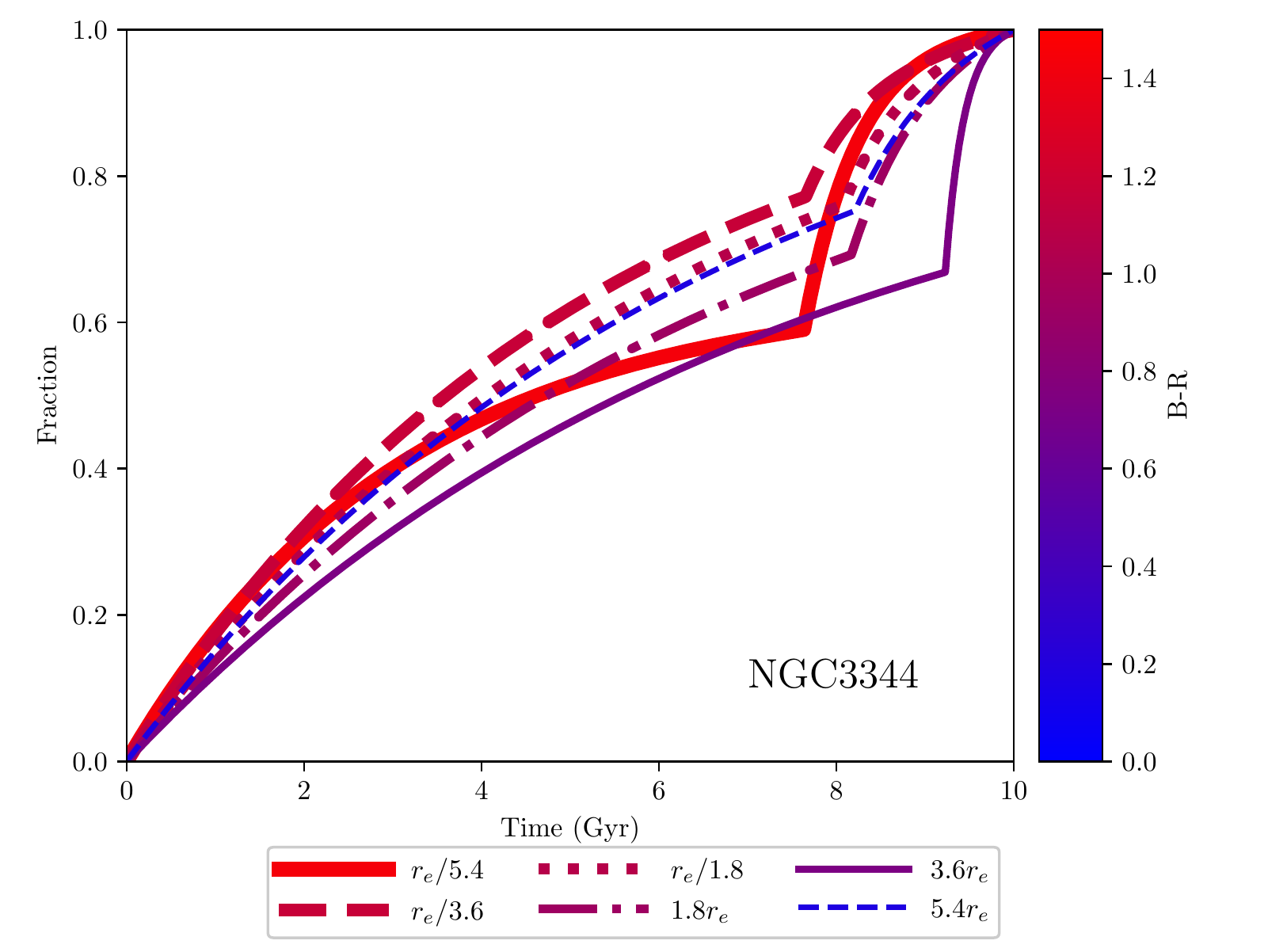}
     \caption{    NGC 3344 : Plot same as Figure \ref{fig:ngc0024_sfh}.}
     \label{fig:ngc3344_sfh}
 \end{figure}
 
\begin{figure*}
	\centering
	\includegraphics[width=2\columnwidth]{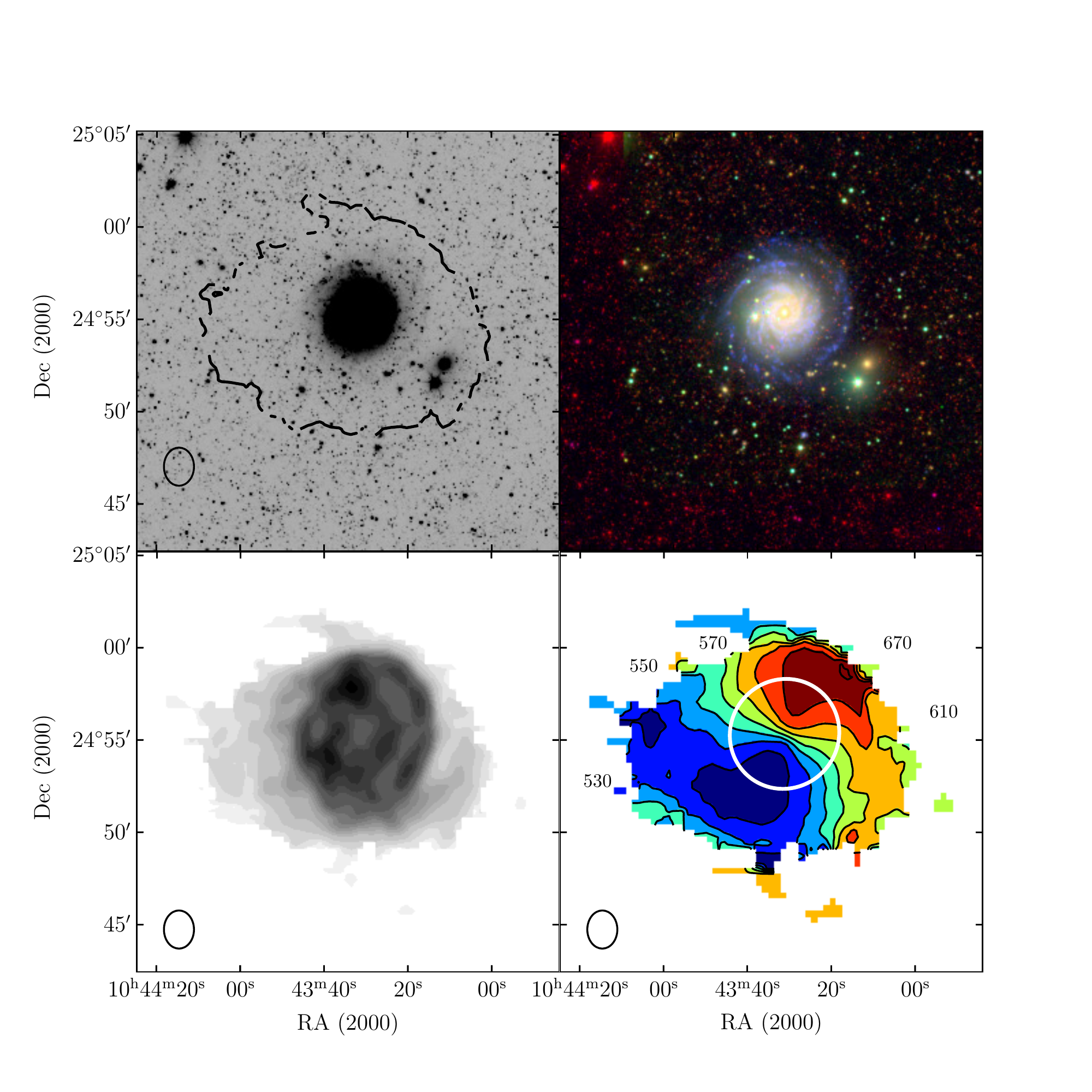}
	\caption{  NGC 3344:  Panels same as Figure \ref{fig:ngc0024_4panel}.}
	\label{fig:ngc3344_4panel}
\end{figure*}
 
\clearpage

  \begin{figure}
     \centering
     \includegraphics[width=\columnwidth]{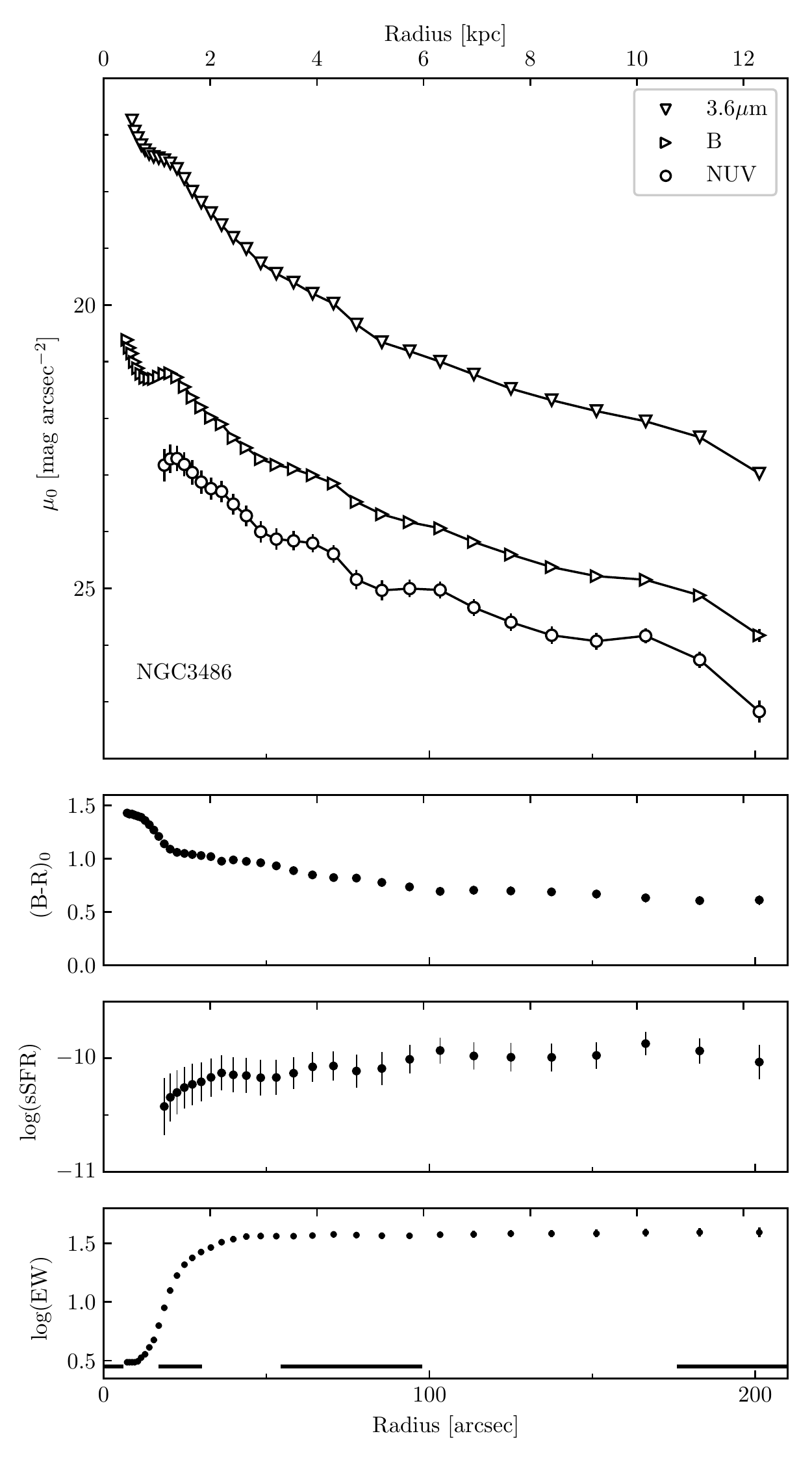}
     \caption{     NGC 3486: Panels same as Figure \ref{fig:ngc0024_fulldata}. }
     \label{fig:ngc3486_fulldata}
 \end{figure}
 
 \begin{figure}
     \centering
     \includegraphics[width=\columnwidth]{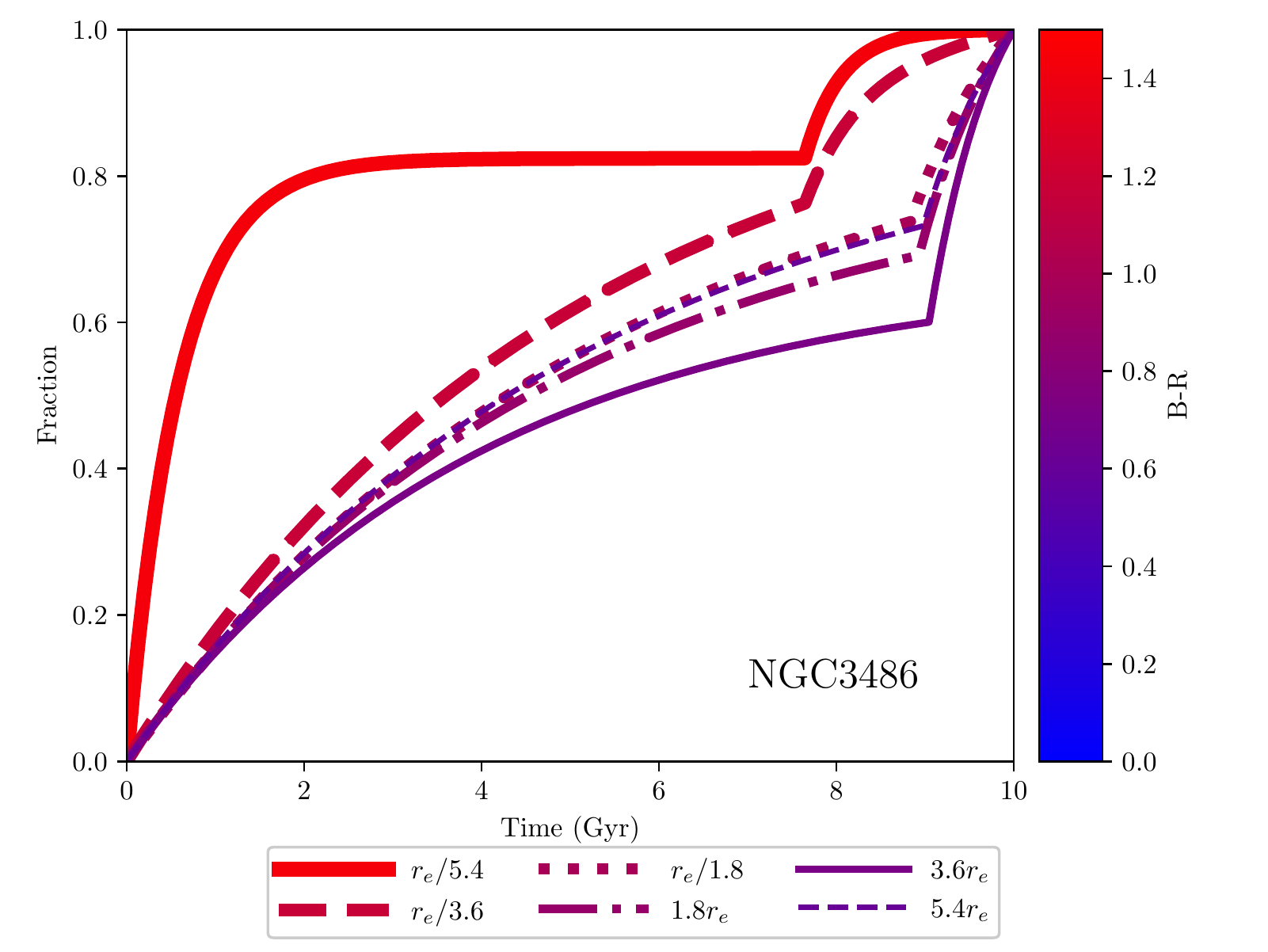}
     \caption{    NGC 3486 : Plot same as Figure \ref{fig:ngc0024_sfh}.}
     \label{fig:ngc3486_sfh}
 \end{figure}

\begin{figure*}
	\centering
	\includegraphics[width=2\columnwidth]{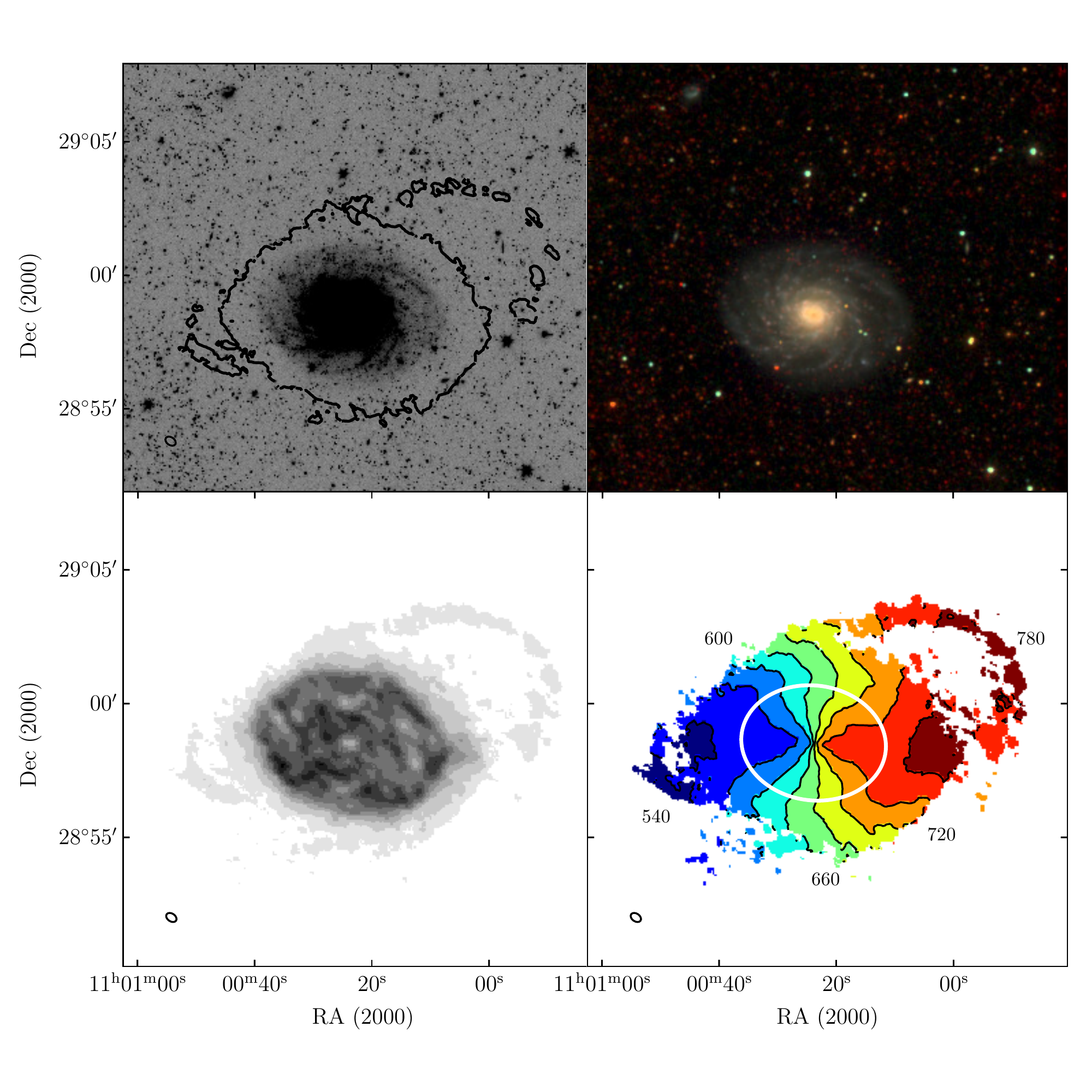}
	\caption{  NGC 3486: Panels same as Figure \ref{fig:ngc0024_4panel}.}
	\label{fig:ngc3486_4panel}
\end{figure*}

\clearpage

 \begin{figure}
     \centering
     \includegraphics[width=\columnwidth]{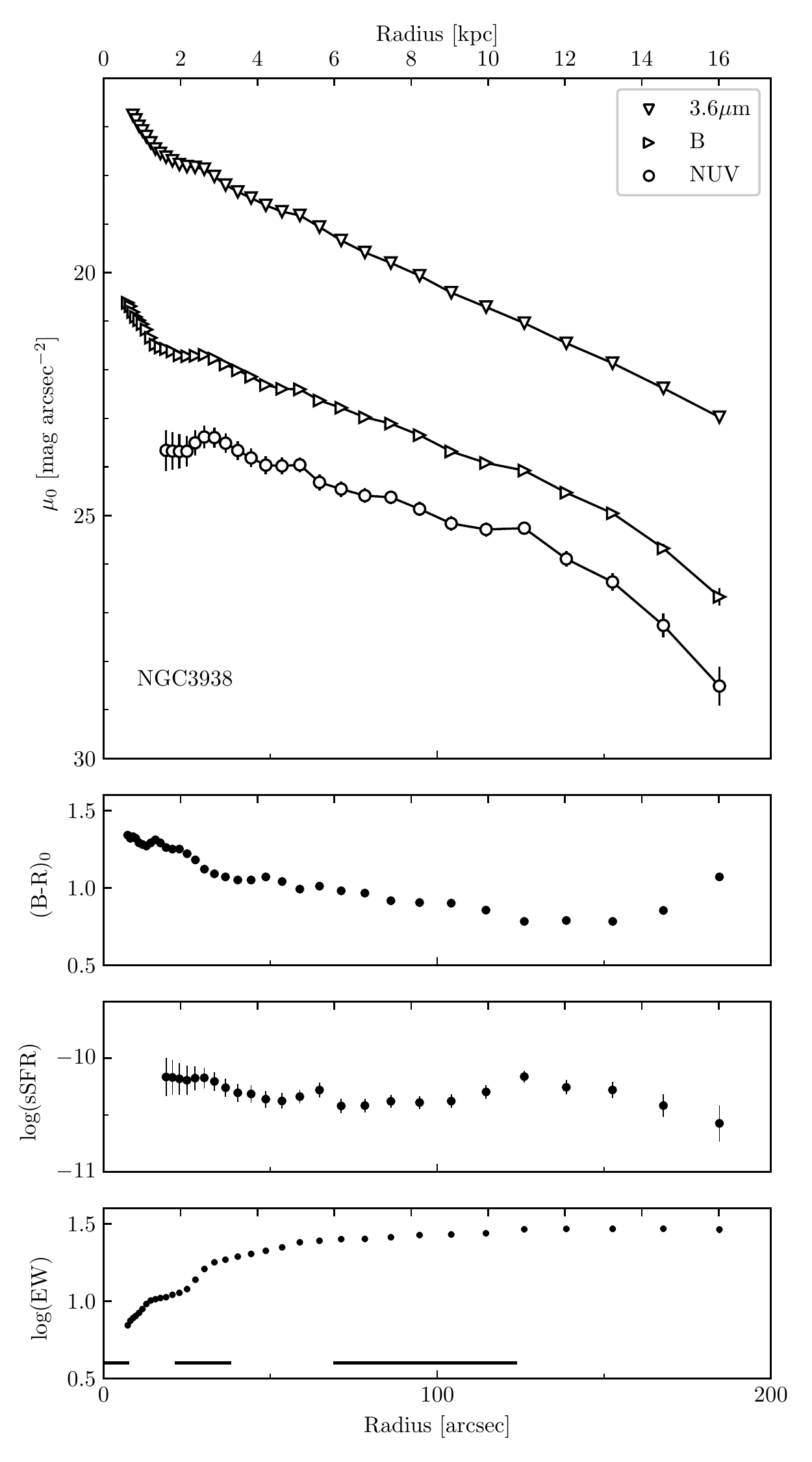}
     \caption{     NGC 3938 : Panels same as Figure \ref{fig:ngc0024_fulldata}. }
     \label{fig:ngc3938_fulldata}
 \end{figure}
 
 \begin{figure}
     \centering
     \includegraphics[width=\columnwidth]{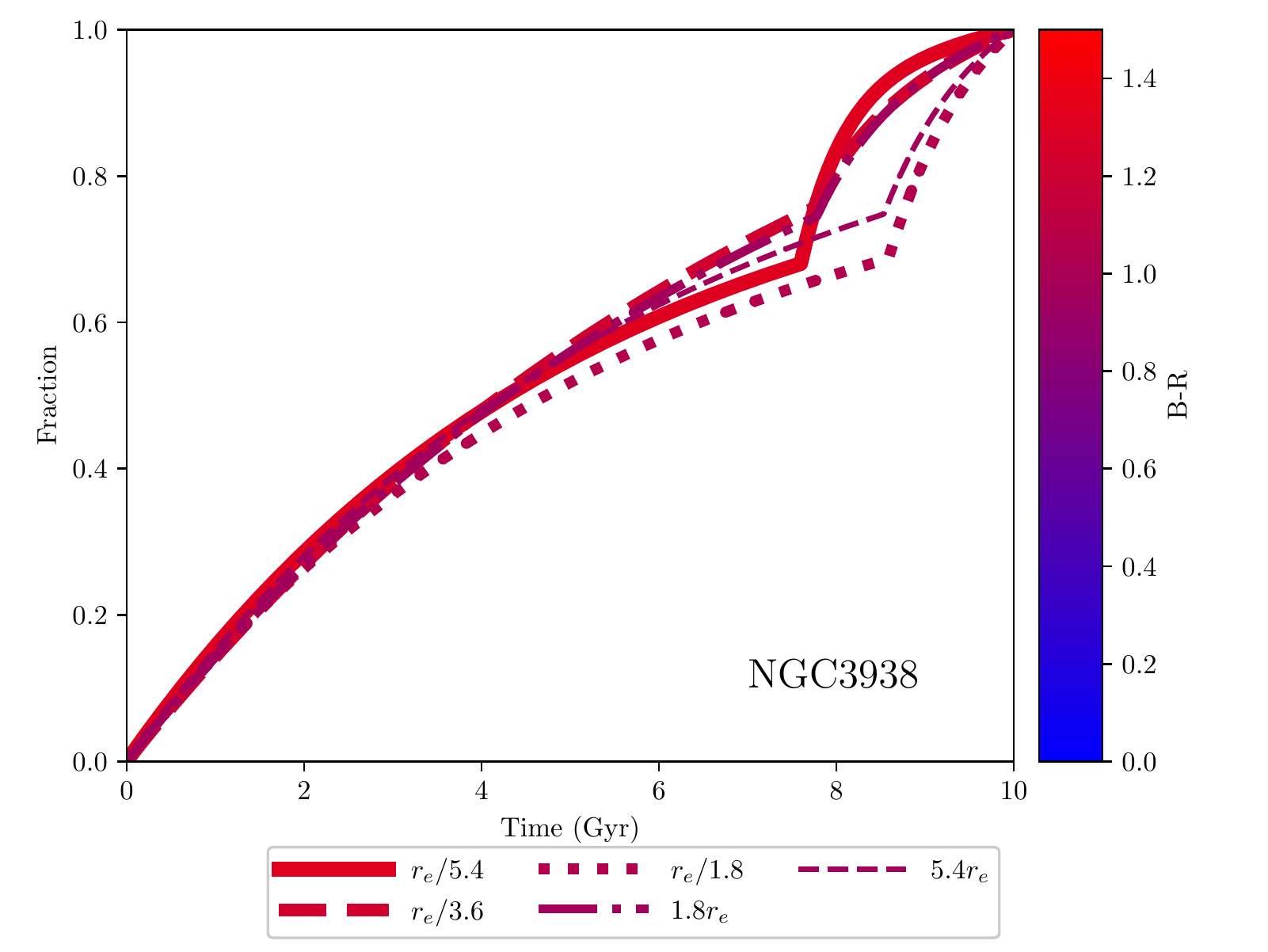}
     \caption{   NGC 3938 : Plot same as Figure \ref{fig:ngc0024_sfh}. }
     \label{fig:ngc3938_sfh}
 \end{figure}

\begin{figure*}
	\centering
	\includegraphics[width=2\columnwidth]{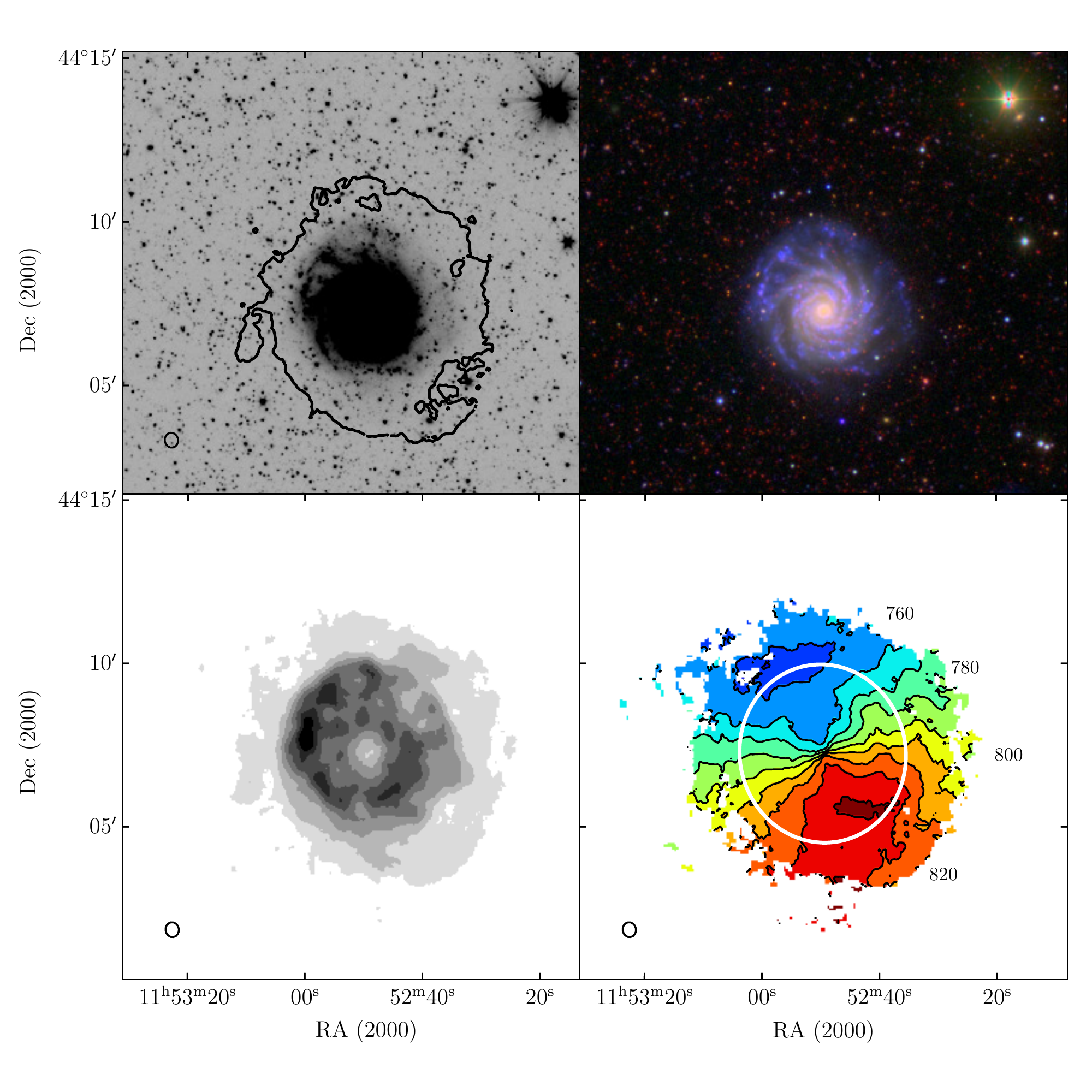}
	\caption{NGC 3938 : Panels same as Figure \ref{fig:ngc0024_4panel}.}
	\label{fig:ngc3938_4panel}
\end{figure*}

\clearpage

 \begin{figure}
     \centering
     \includegraphics[width=\columnwidth]{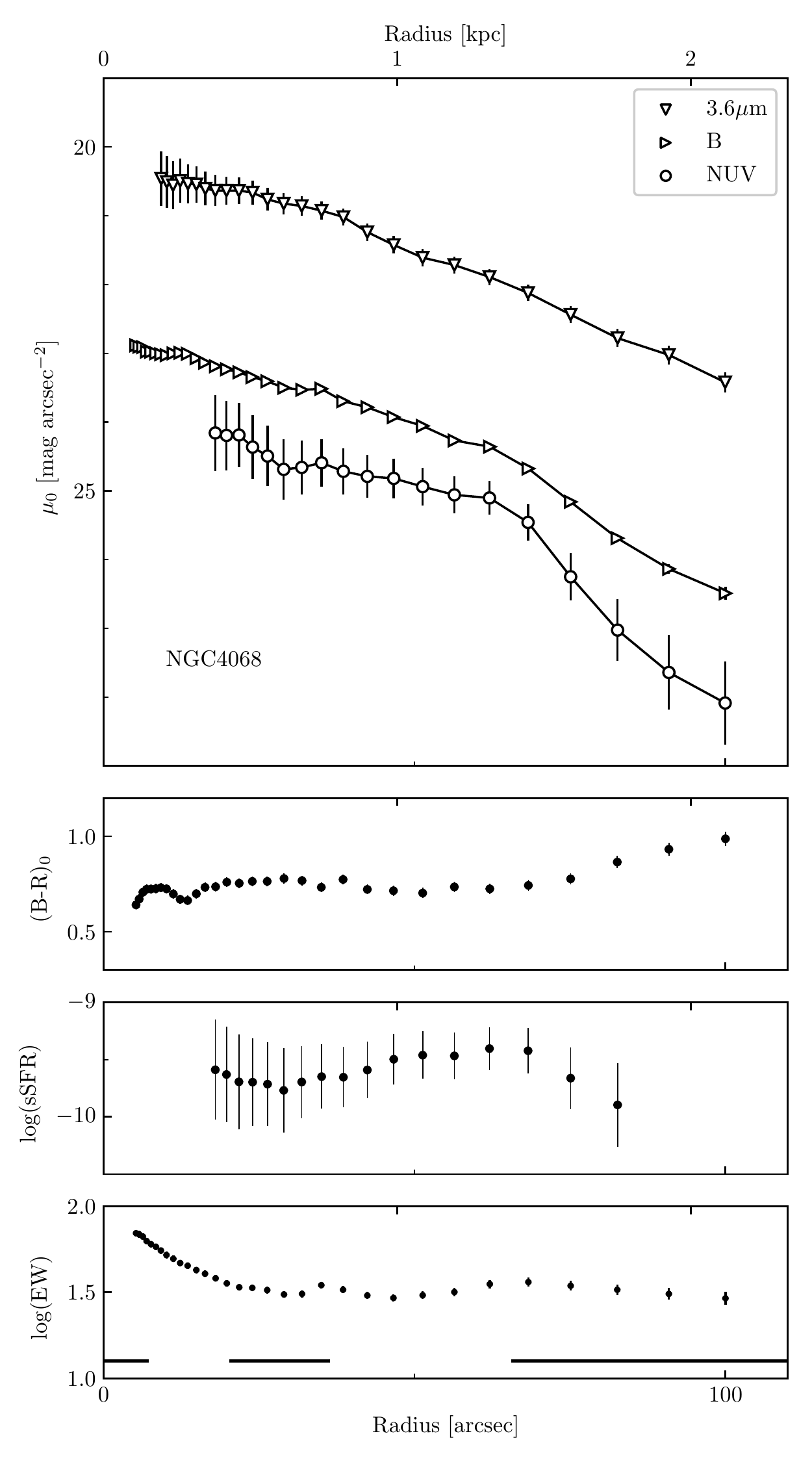}
     \caption{ \textbf{NGC 4068:}   Panels same as Figure \ref{fig:ngc0024_fulldata}. }
     \label{fig:ngc4068_fulldata}
 \end{figure}

\begin{figure}
    \centering
    \includegraphics[width=\columnwidth]{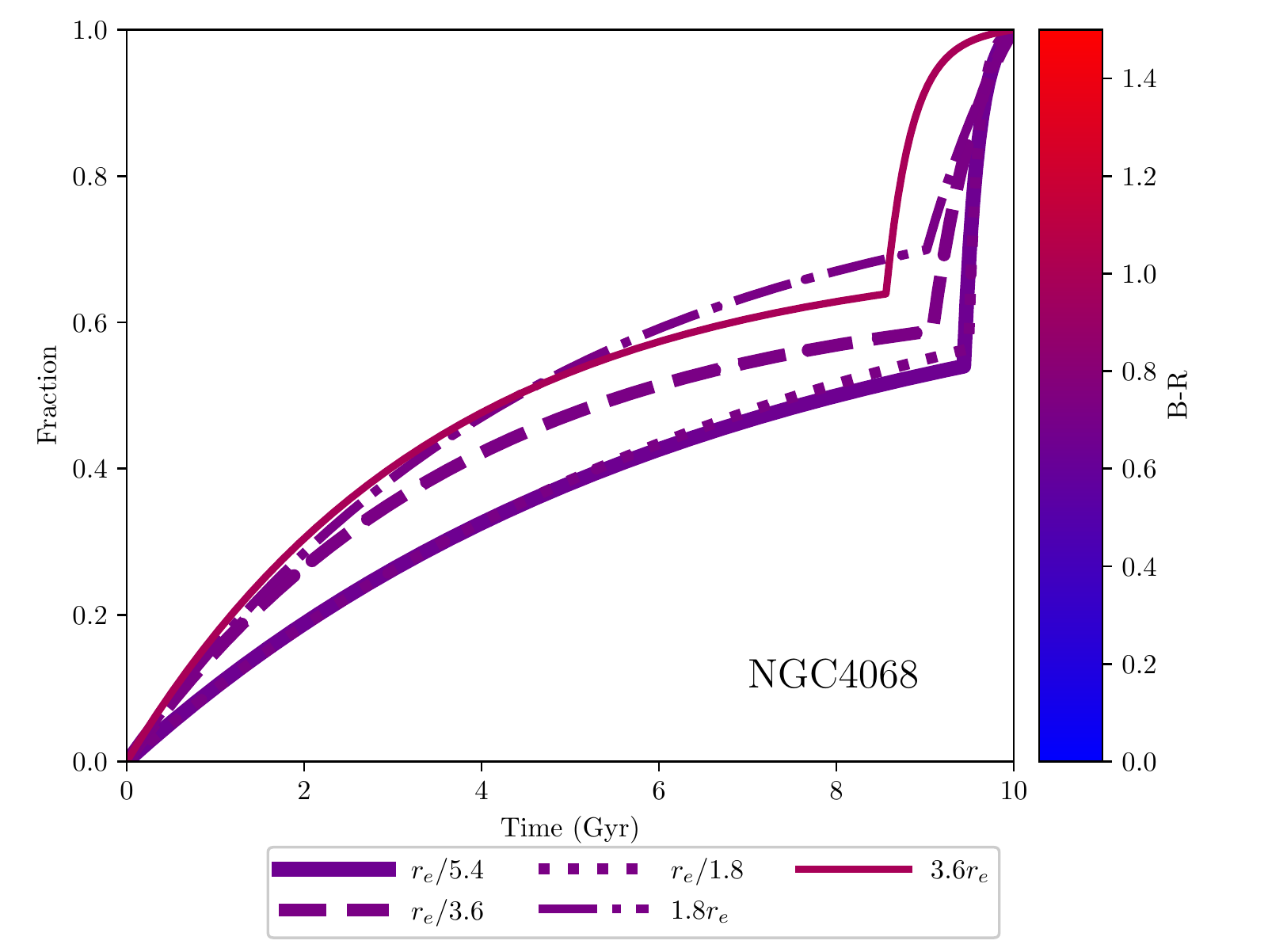}
    \caption{     NGC 4068 : Plot same as Figure \ref{fig:ngc0024_sfh}.}
    \label{fig:ngc4068_sfh}
\end{figure}

\begin{figure*}
	\centering
	\includegraphics[width=2\columnwidth]{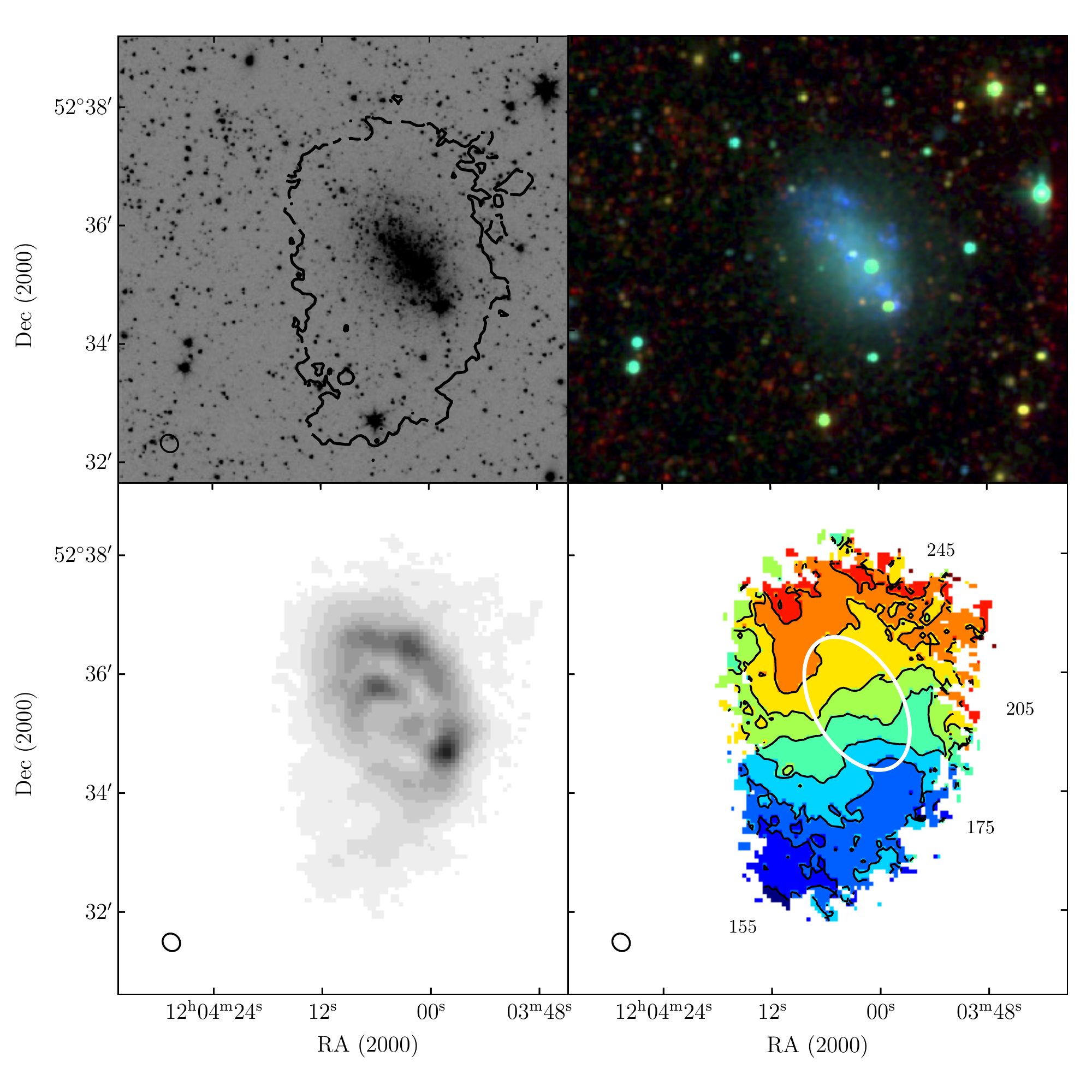}
	\caption{  NGC 4068: Panels same as Figure \ref{fig:ngc0024_4panel}.}
	\label{fig:ngc4068_4panel}
\end{figure*}

\clearpage

  \begin{figure}
     \centering
     \includegraphics[width=\columnwidth]{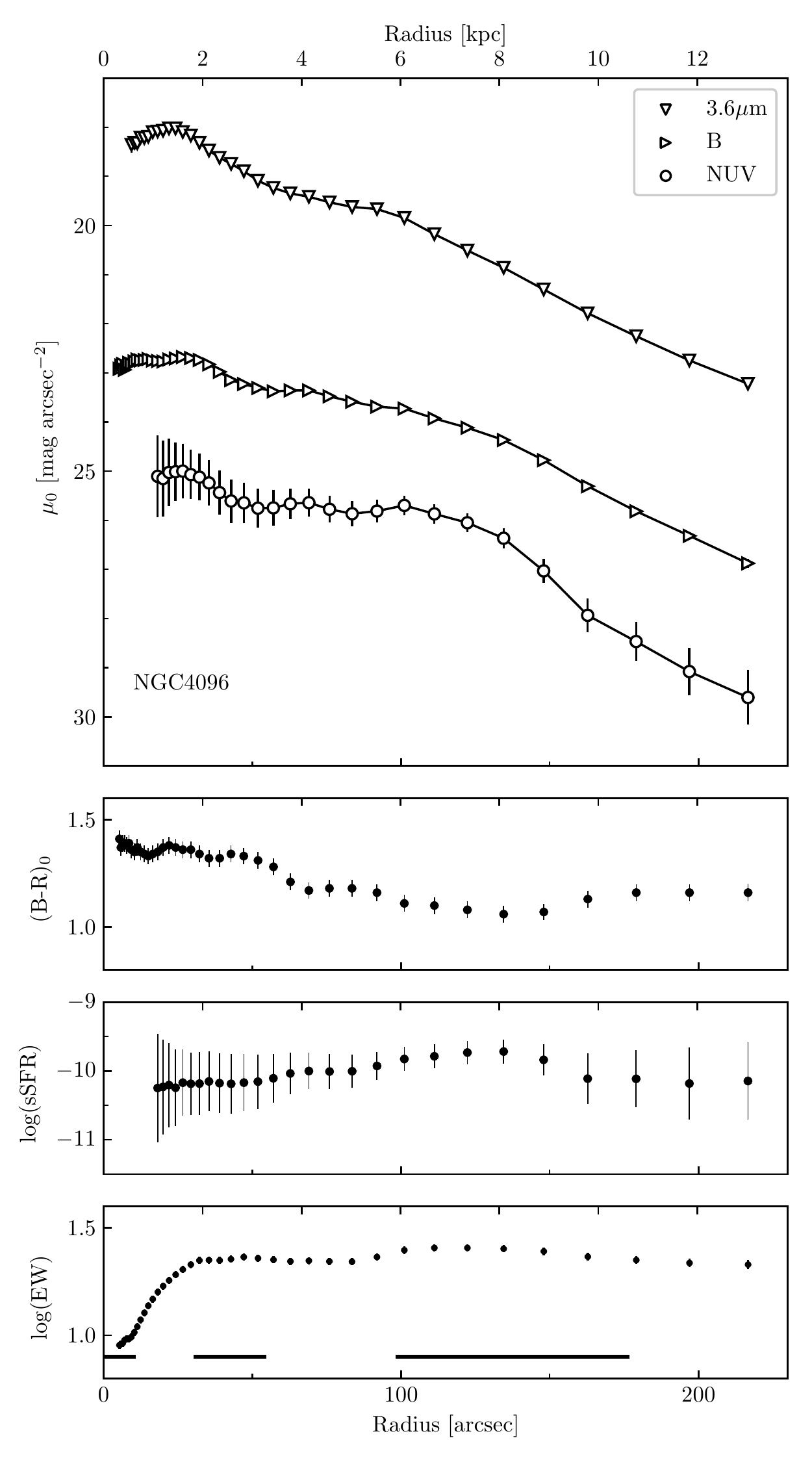}
     \caption{ \textbf{NGC 4096:}   Panels same as Figure \ref{fig:ngc0024_fulldata}. }
     \label{fig:ngc4096_fulldata}
 \end{figure}
 
 \begin{figure}
     \centering
     \includegraphics[width=\columnwidth]{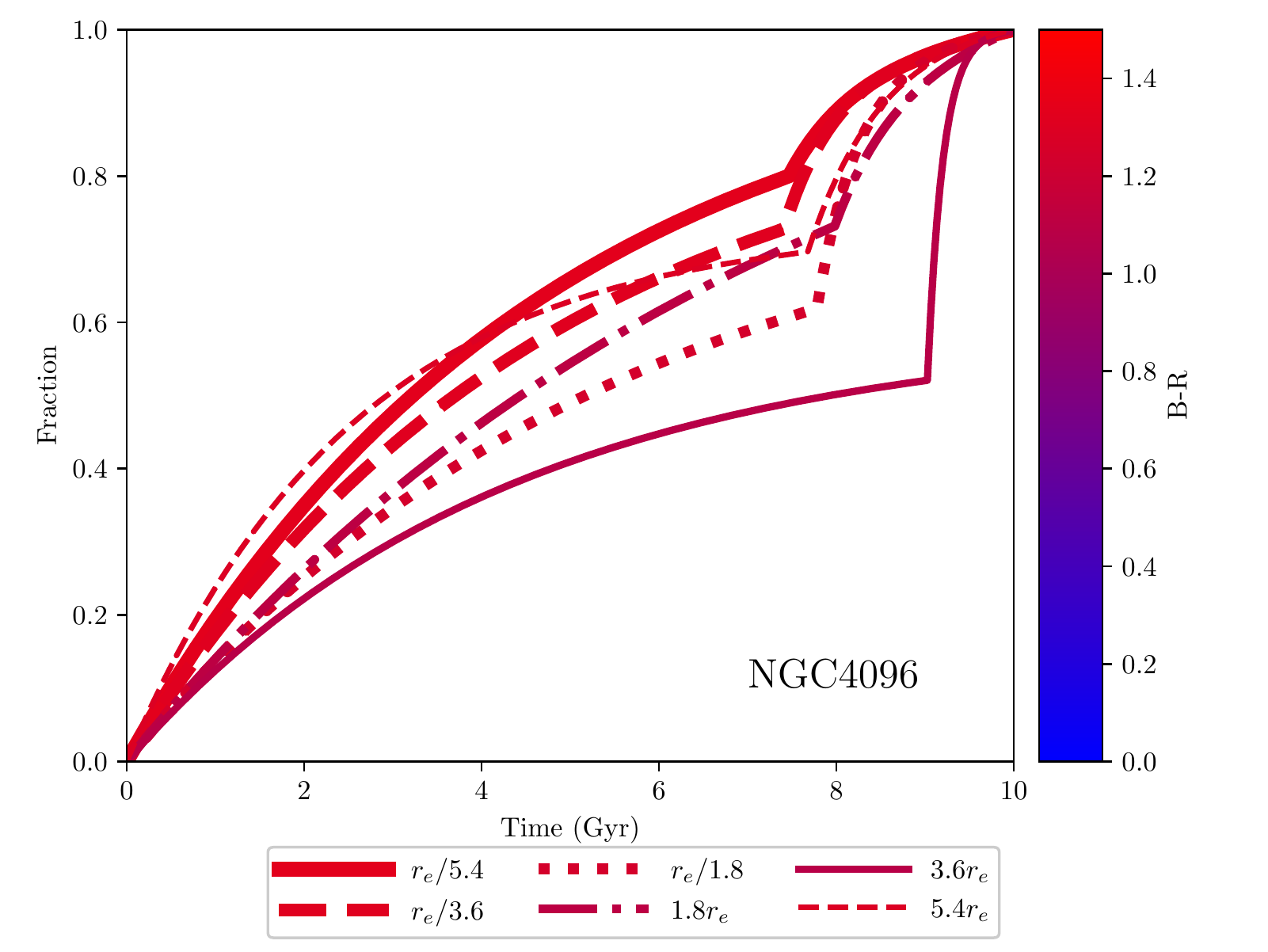}
     \caption{   NGC 4096 : Plot same as Figure \ref{fig:ngc0024_sfh}. }
     \label{fig:ngc4096_sfh}
 \end{figure}
 
 \begin{figure*}
	\centering
	\includegraphics[width=2\columnwidth]{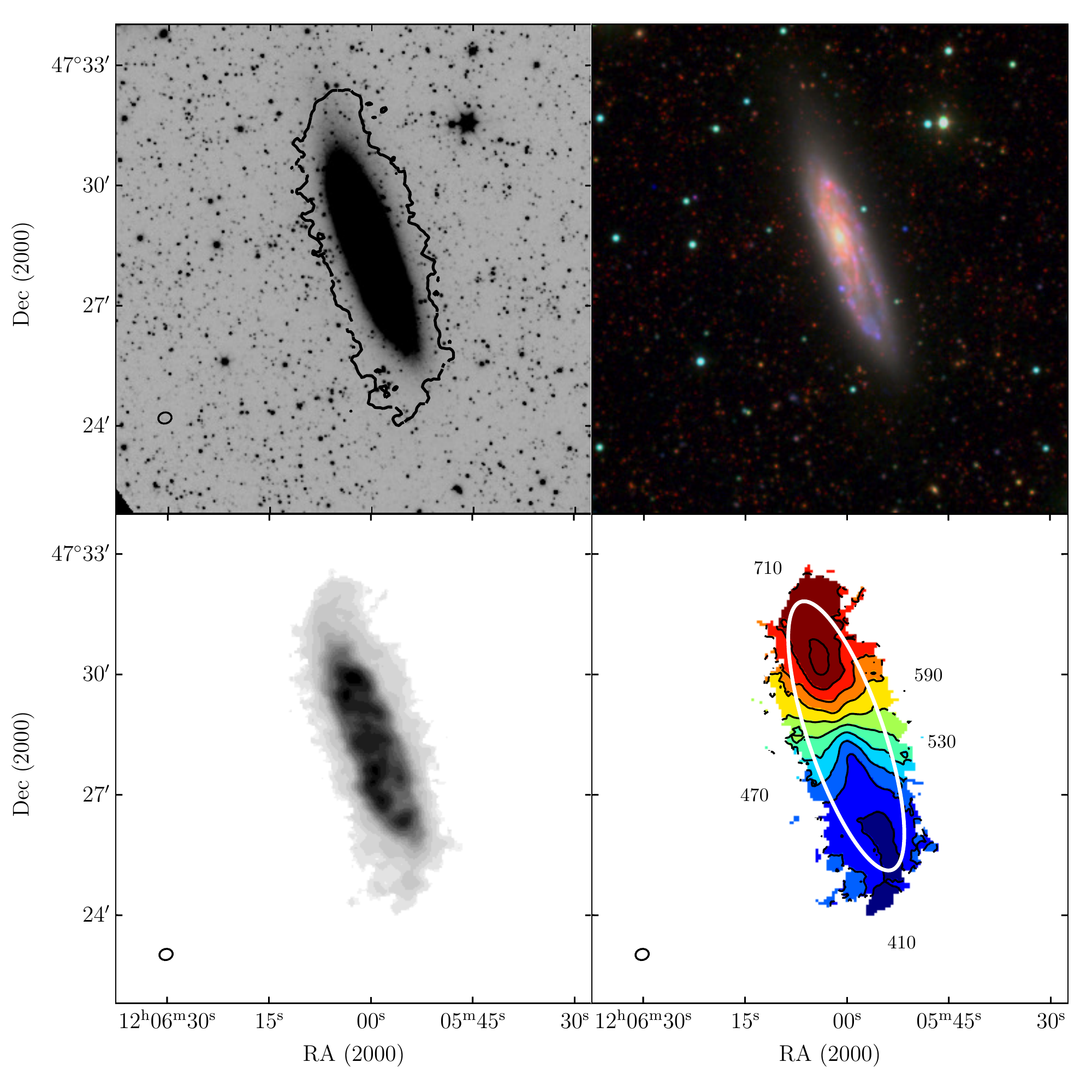}
	\caption{NGC 4096: Panels same as Figure \ref{fig:ngc0024_4panel}.}
	\label{fig:ngc4096_4panel}
\end{figure*}
 
\clearpage

 \begin{figure}
     \centering
     \includegraphics[width=\columnwidth]{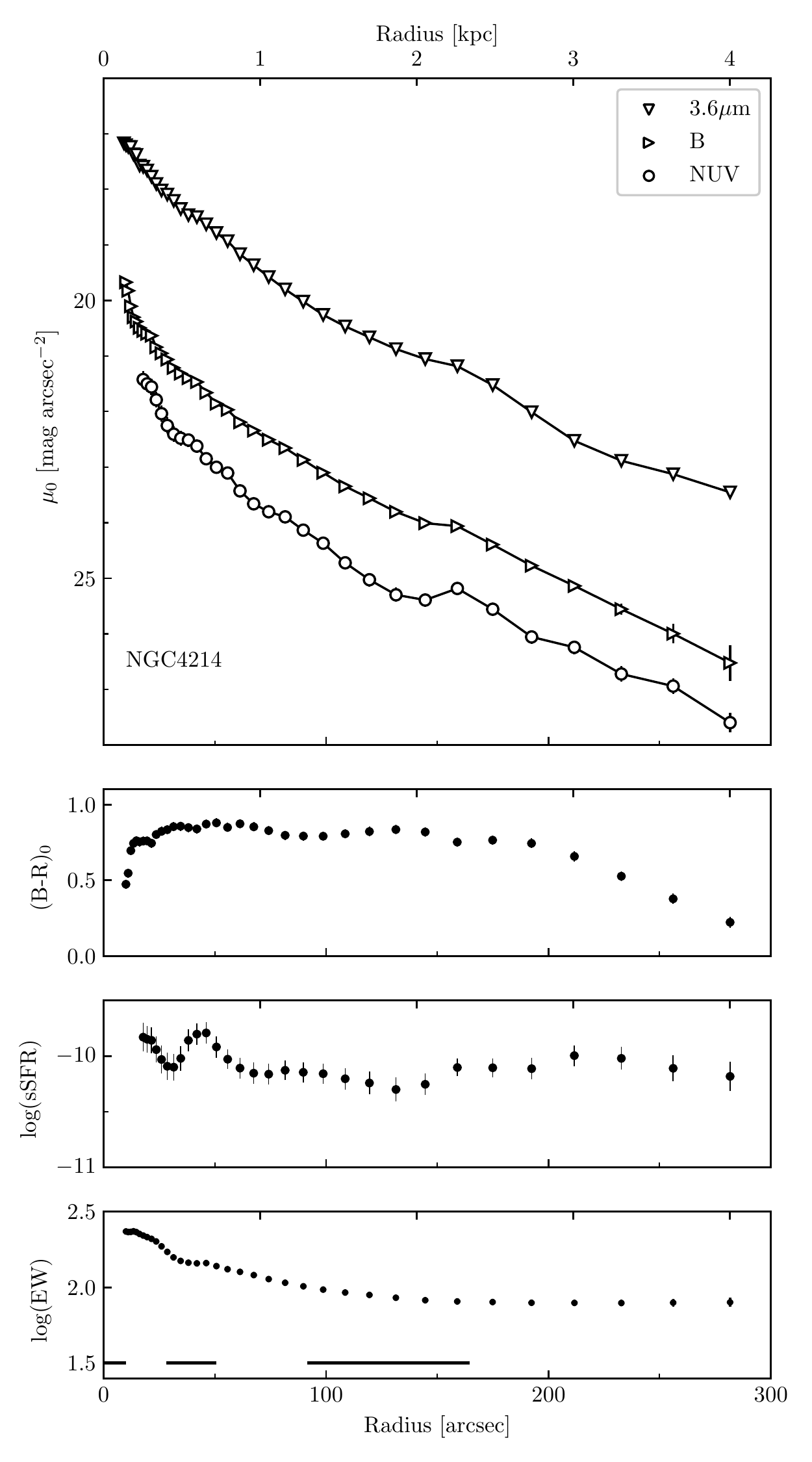}
     \caption{\textbf{NGC 4214:}   Panels same as Figure \ref{fig:ngc0024_fulldata} }
     \label{fig:ngc4214_fulldata}
 \end{figure}
 
\begin{figure}
    \centering
    \includegraphics[width=\columnwidth]{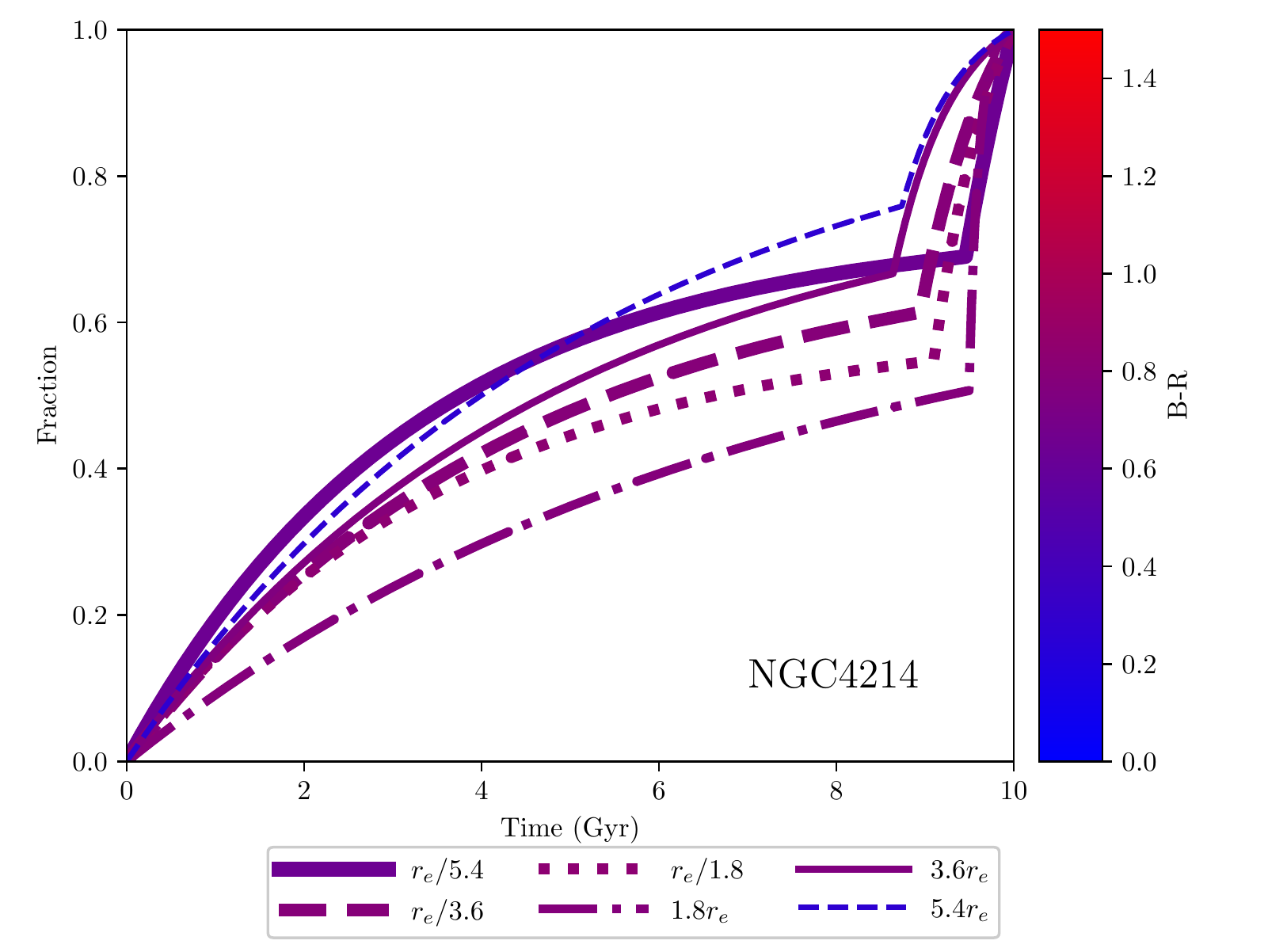}
    \caption{     NGC 4214 : Plot same as Figure \ref{fig:ngc0024_sfh}.}
    \label{fig:ngc4214_sfh}
\end{figure}

\begin{figure*}
	\centering
	\includegraphics[width=2\columnwidth]{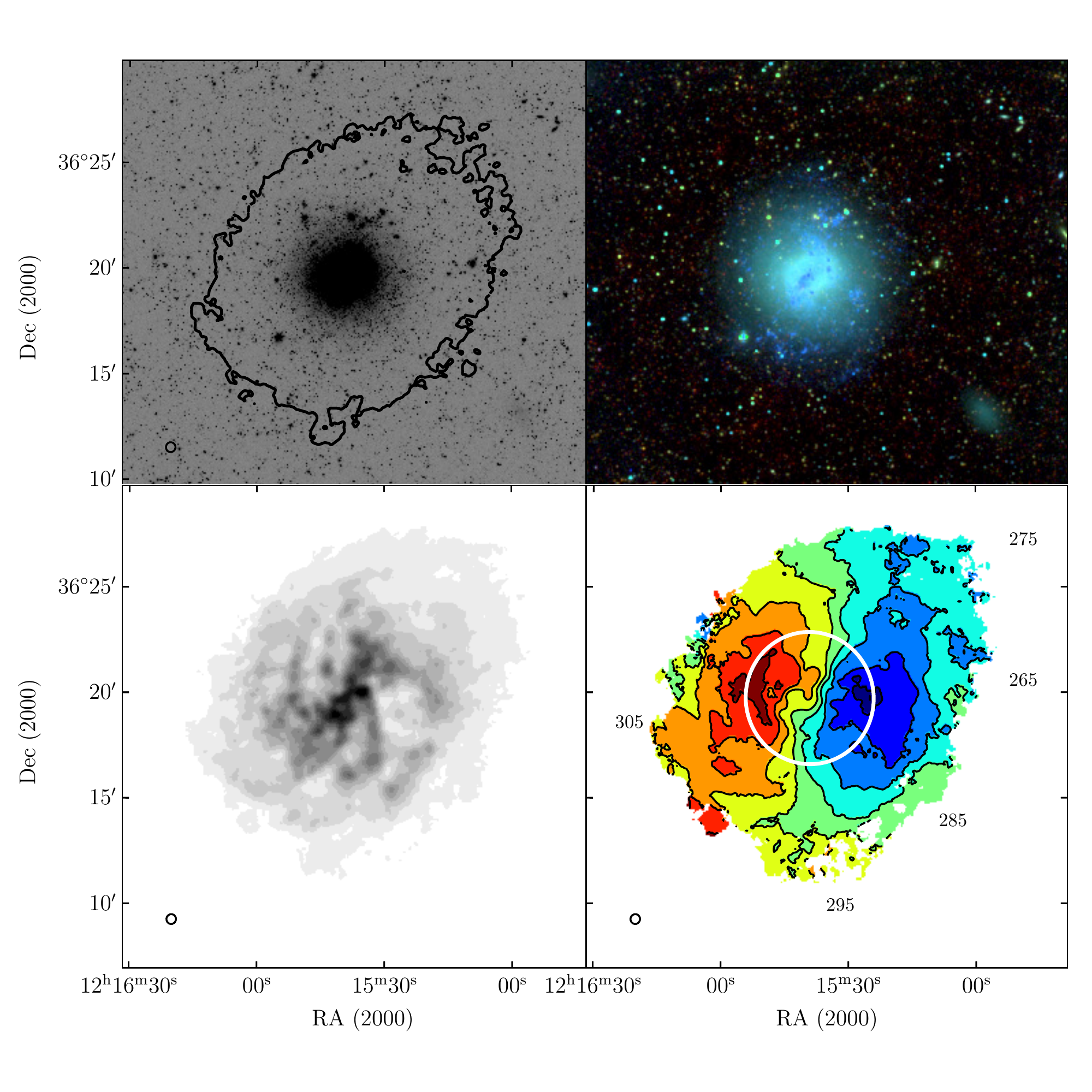}
	\caption{  NGC 4214: Panels same as Figure \ref{fig:ngc0024_4panel}.}
	\label{fig:ngc4214_4panel}
\end{figure*}

\clearpage

\begin{figure*}
	\centering
	\includegraphics[width=2\columnwidth]{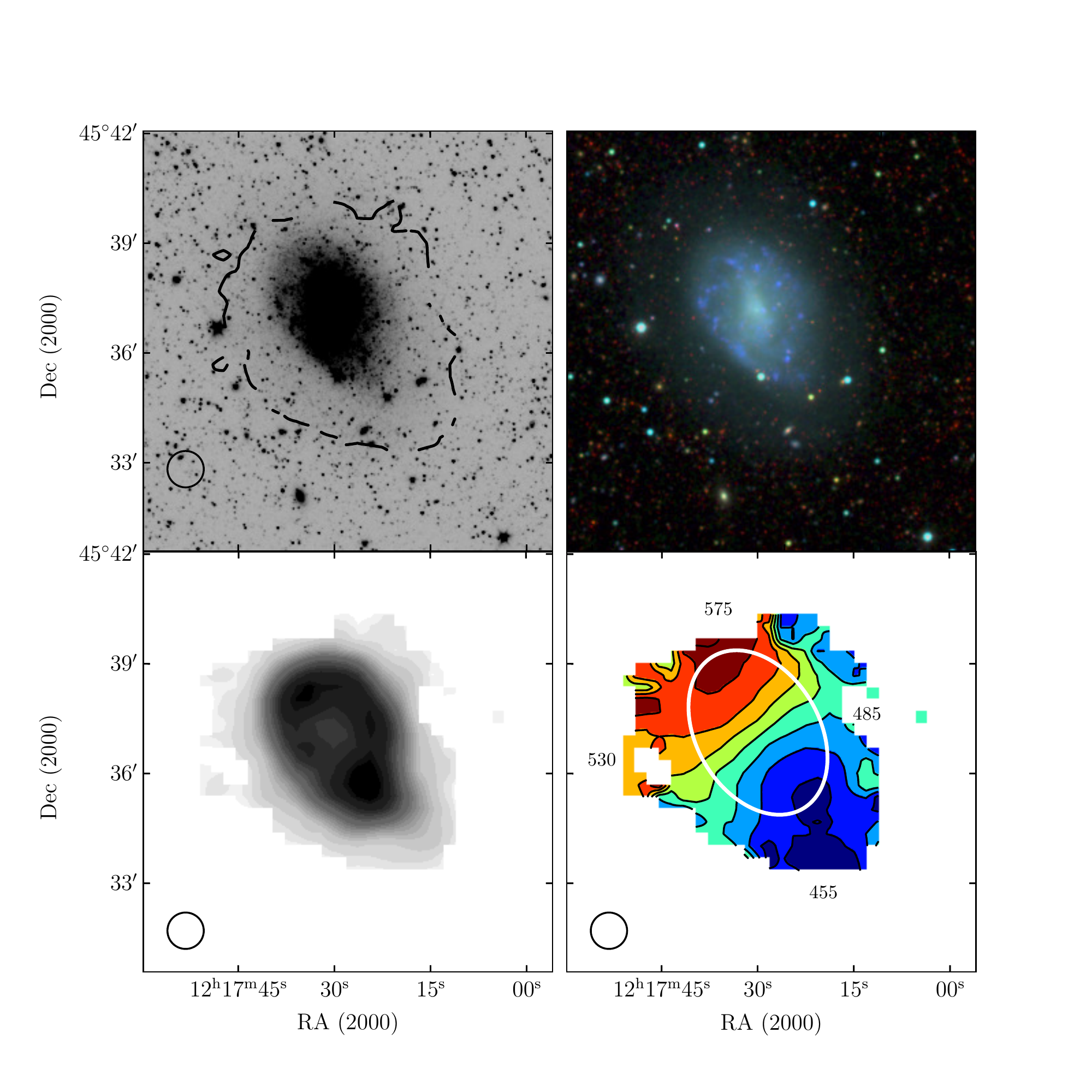}
	\caption{  NGC 4242: Panels same as Figure \ref{fig:ngc0024_4panel}.}
	\label{fig:ngc4242_4panel}
\end{figure*}

\clearpage

\begin{figure}
     \centering
     \includegraphics[width=\columnwidth]{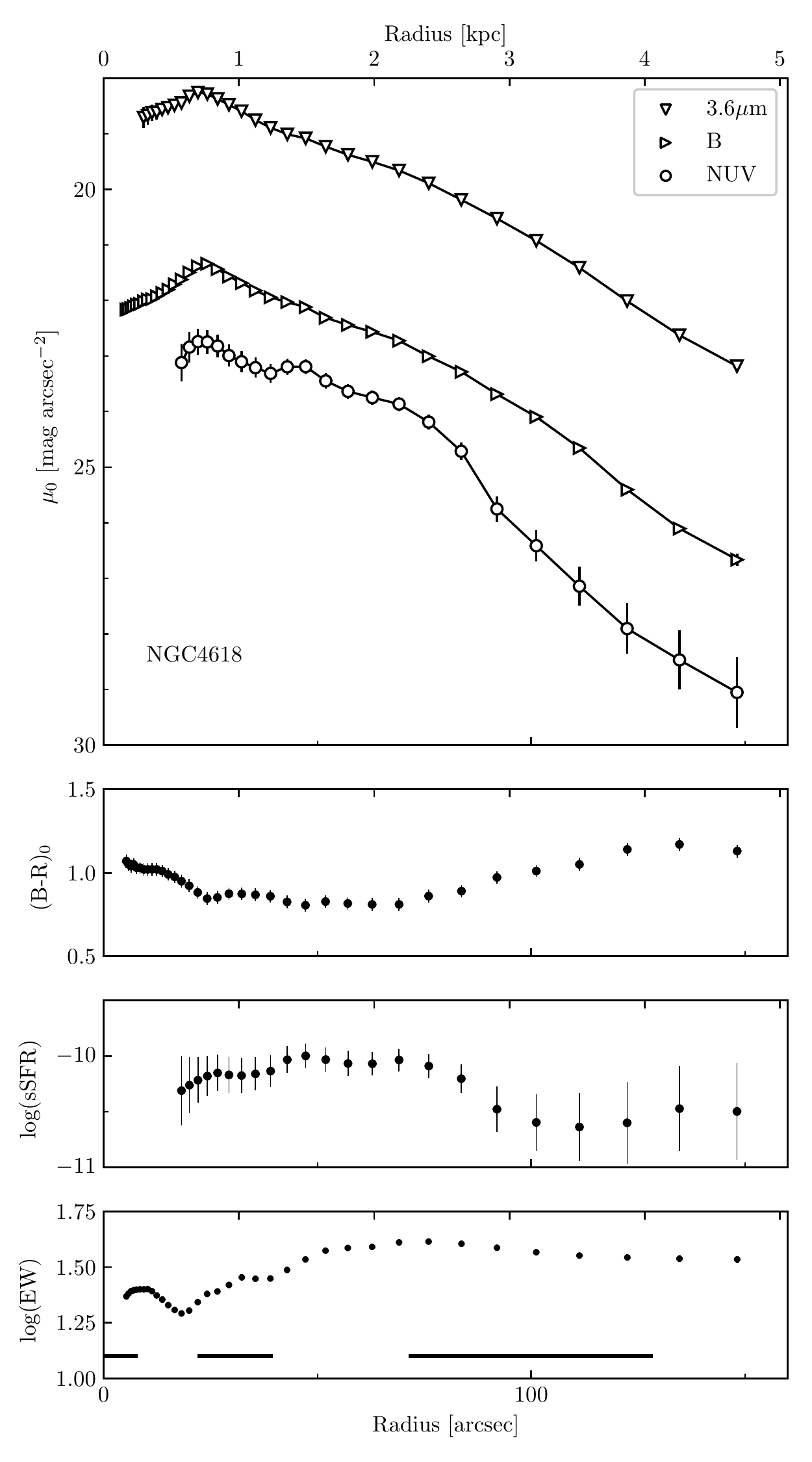}
     \caption{ \textbf{NGC 4618:}   Panels same as Figure \ref{fig:ngc0024_fulldata}. }
     \label{fig:ngc4618_fulldata}
 \end{figure}
 
 \begin{figure}
     \centering
     \includegraphics[width=\columnwidth]{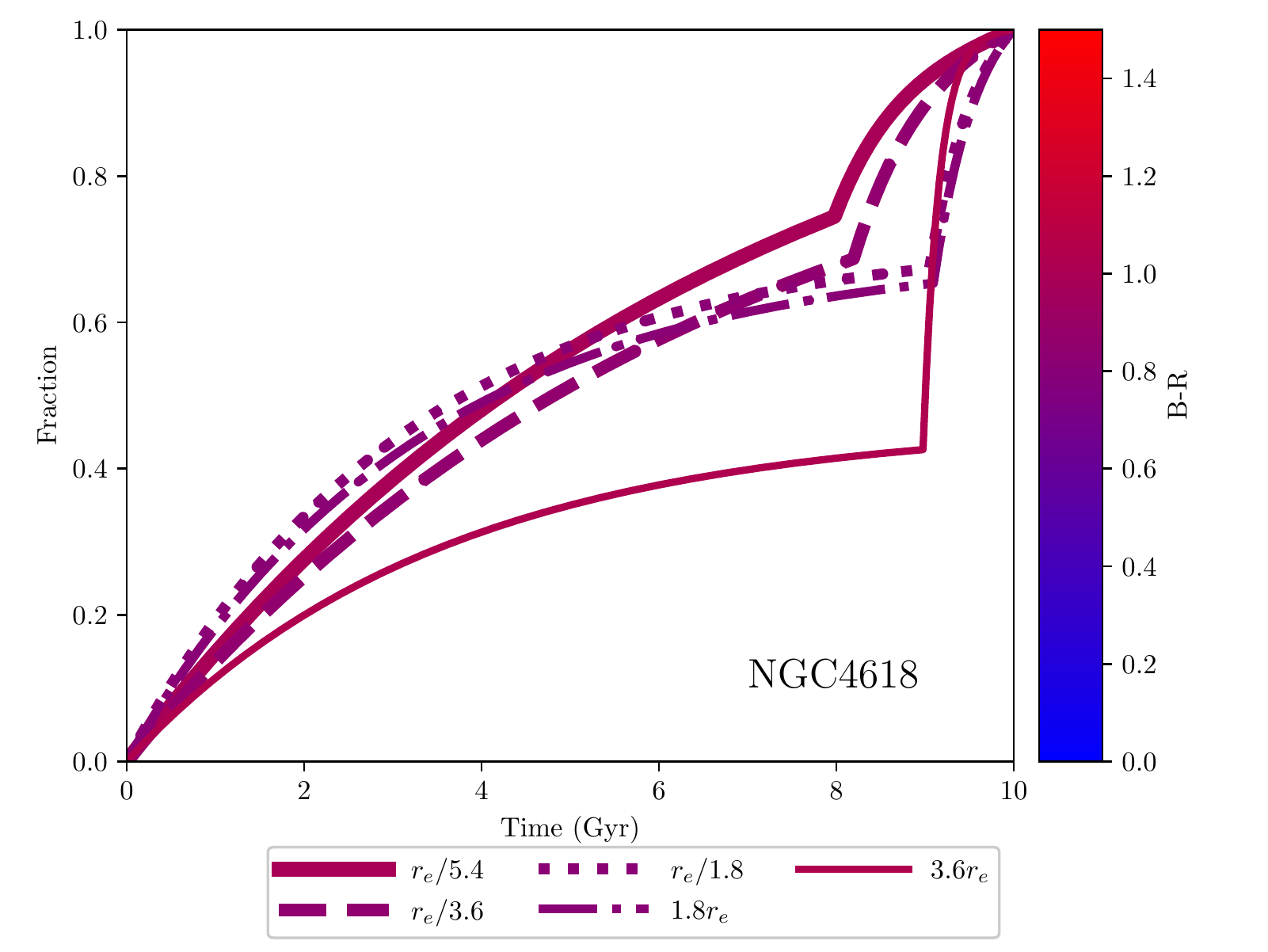}
     \caption{     NGC 4618 : Plot same as Figure \ref{fig:ngc0024_sfh}. }
     \label{fig:ngc4618_sfh}
 \end{figure}
 
 \begin{figure*}
	\centering
	\includegraphics[width=2\columnwidth]{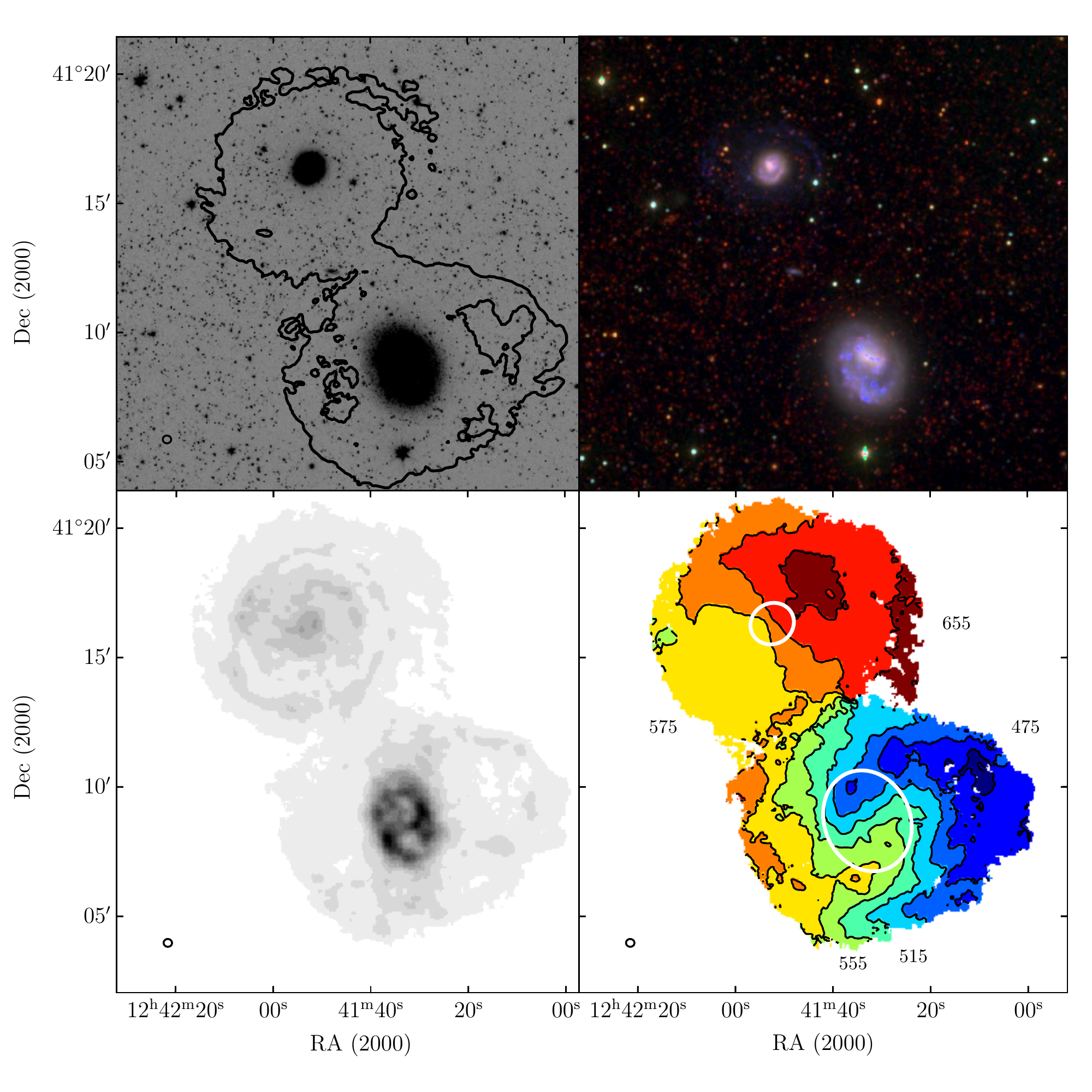}
	\caption{   NGC 4618 \& NGC4625: Panels same as Figure \ref{fig:ngc0024_4panel}.}
	\label{fig:ngc4618_4panel}
\end{figure*}

\clearpage

  \begin{figure}
     \centering
     \includegraphics[width=\columnwidth]{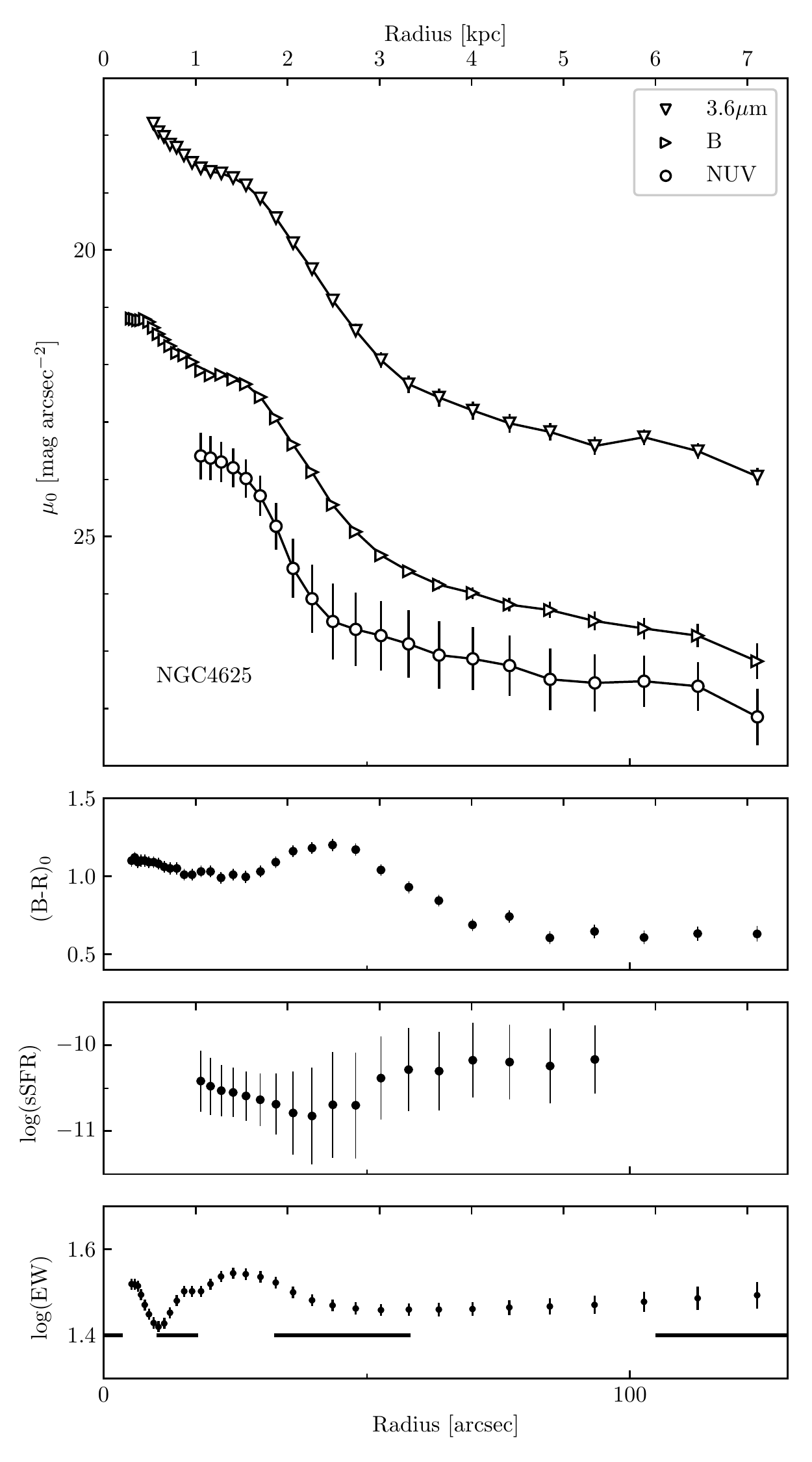}
     \caption{     NGC 4625: Panels same as Figure \ref{fig:ngc0024_fulldata}.}
     \label{fig:ngc4625_fulldata}
 \end{figure}
 
 \begin{figure}
     \centering
     \includegraphics[width=\columnwidth]{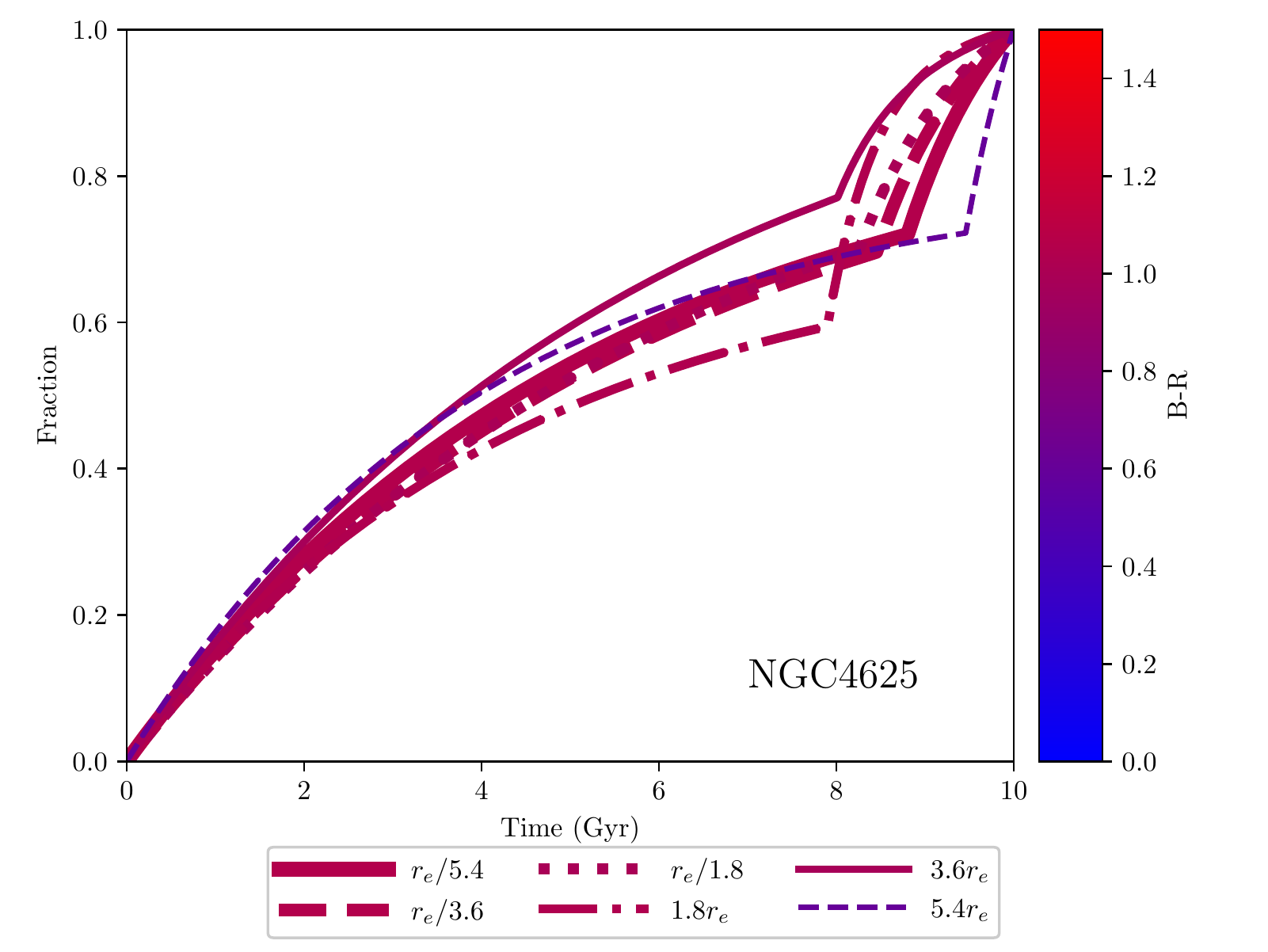}
     \caption{    NGC 4625 : Plot same as Figure \ref{fig:ngc0024_sfh}.}
     \label{fig:ngc4625_sfh}
 \end{figure}

\clearpage

\begin{figure}
    \centering
    \includegraphics[width=\columnwidth]{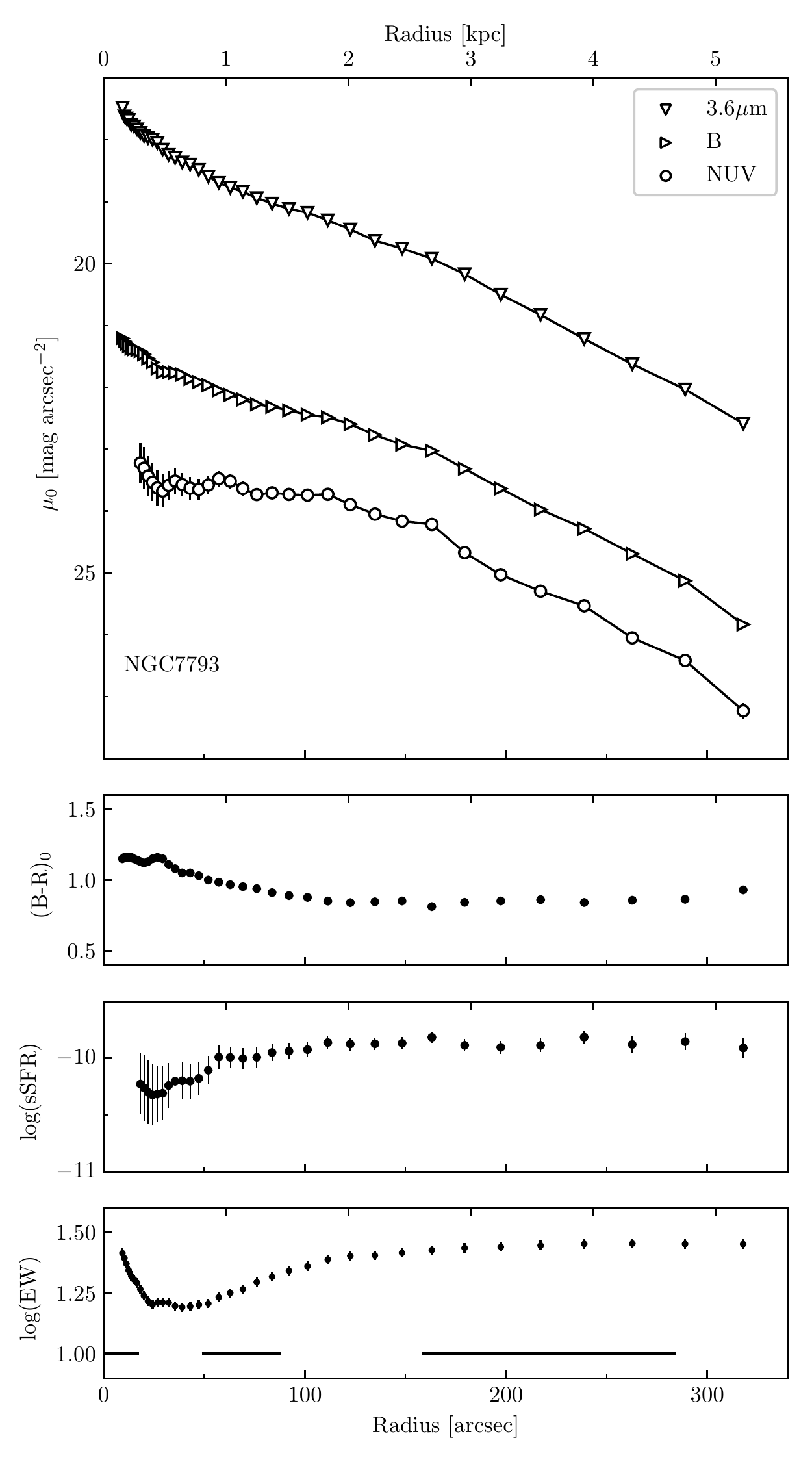}
    \caption{     NGC 7793: Panels same as Figure \ref{fig:ngc0024_fulldata}.}
    \label{fig:ngc7793_fulldata}
\end{figure}

 \begin{figure}
     \centering
     \includegraphics[width=\columnwidth]{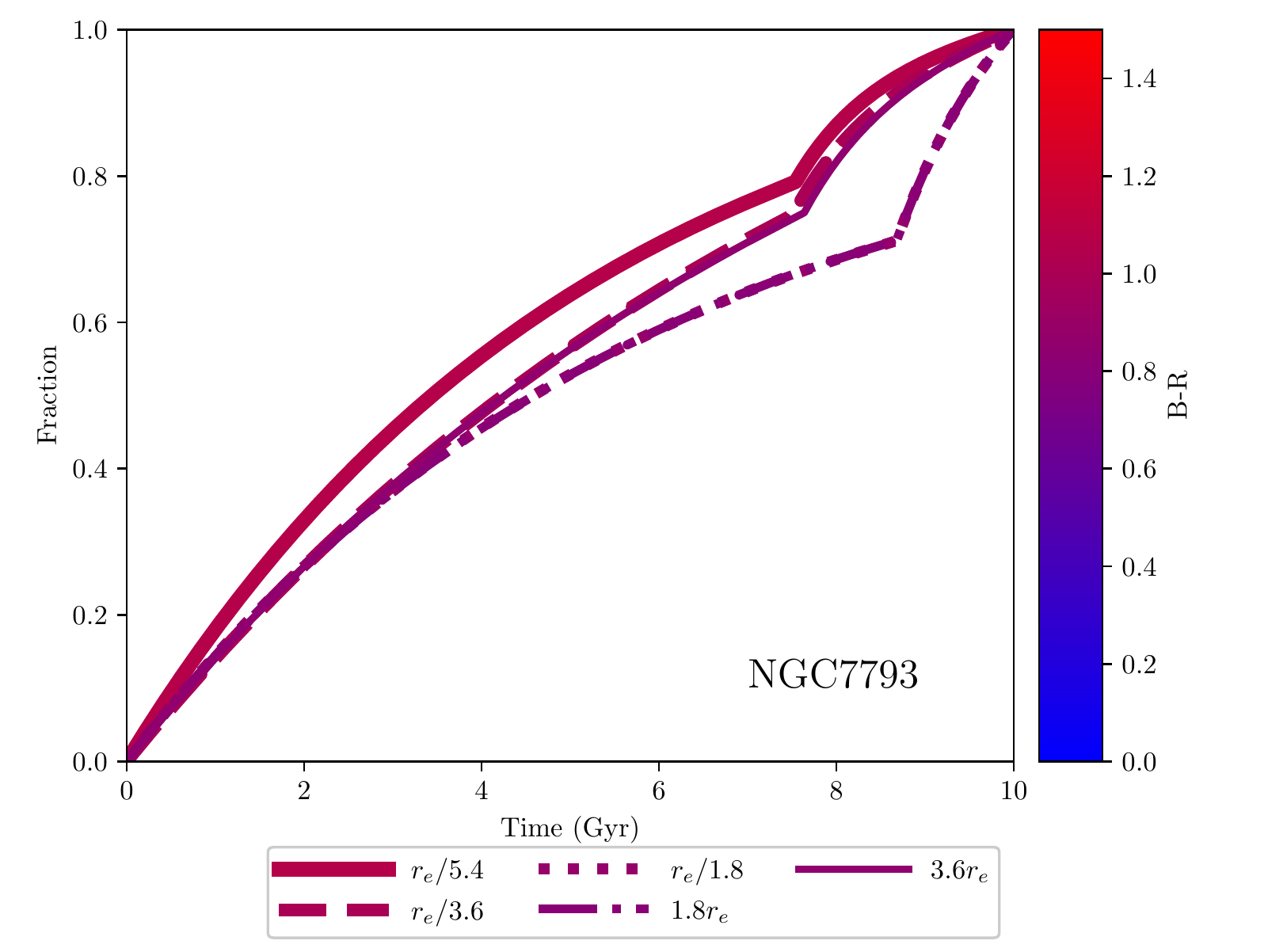}
     \caption{    NGC 7793 : Plot same as Figure \ref{fig:ngc0024_sfh}. The results of NGC 7793 are compared to the CMD-derived SFH in Section \ref{ssec:cmd sfh}.}
     \label{fig:ngc7793_sfh}
 \end{figure}
 
 \begin{figure*}
	\centering
	\includegraphics[width=2\columnwidth]{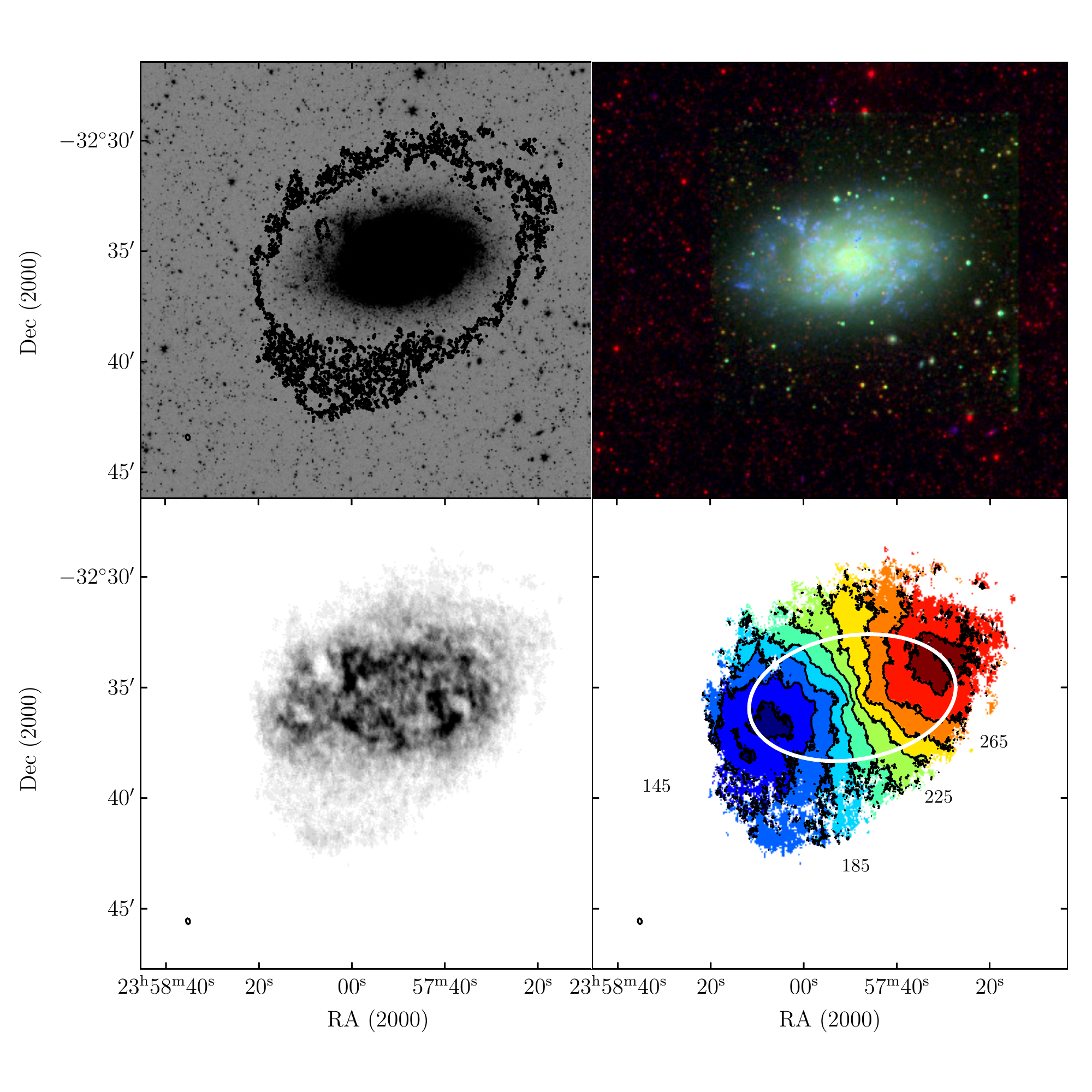}
	\caption{   NGC 7793 : Panels same as Figure \ref{fig:ngc0024_4panel}.}
	\label{fig:ngc7793_4panel}
\end{figure*}

\clearpage

  \begin{figure}
     \centering
     \includegraphics[width=\columnwidth]{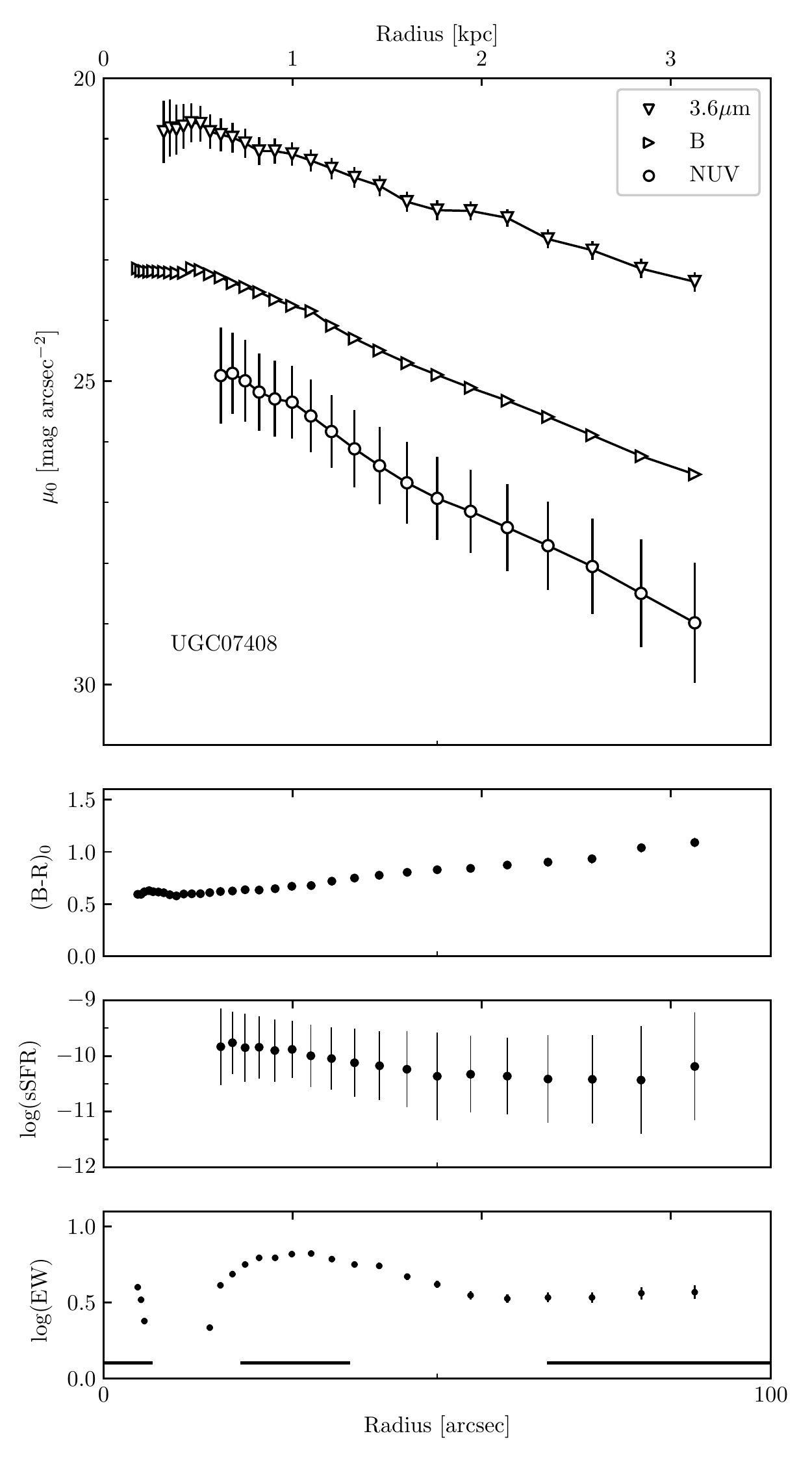}
     \caption{     UGC 07408 : Panels same as Figure \ref{fig:ngc0024_fulldata}.}
     \label{fig:ugc07408_fulldata}
 \end{figure}
 
 \begin{figure}
     \centering
     \includegraphics[width=\columnwidth]{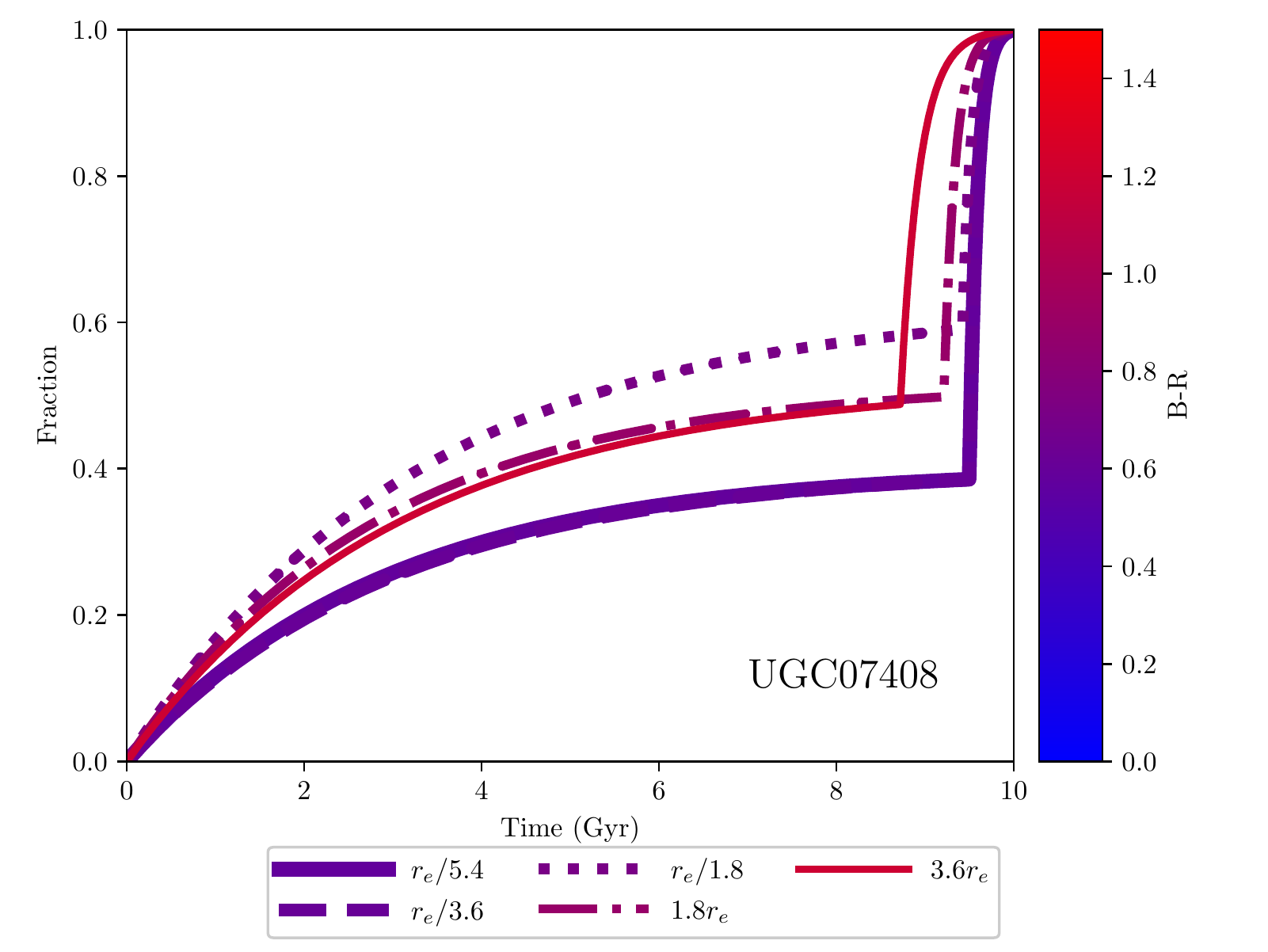}
     \caption{    UGC 07408 : Plot same as Figure \ref{fig:ngc0024_sfh}.}
     \label{fig:ugc07408_sfh}
 \end{figure}
 
 \begin{figure*}
 	\centering
	\includegraphics[width=2\columnwidth]{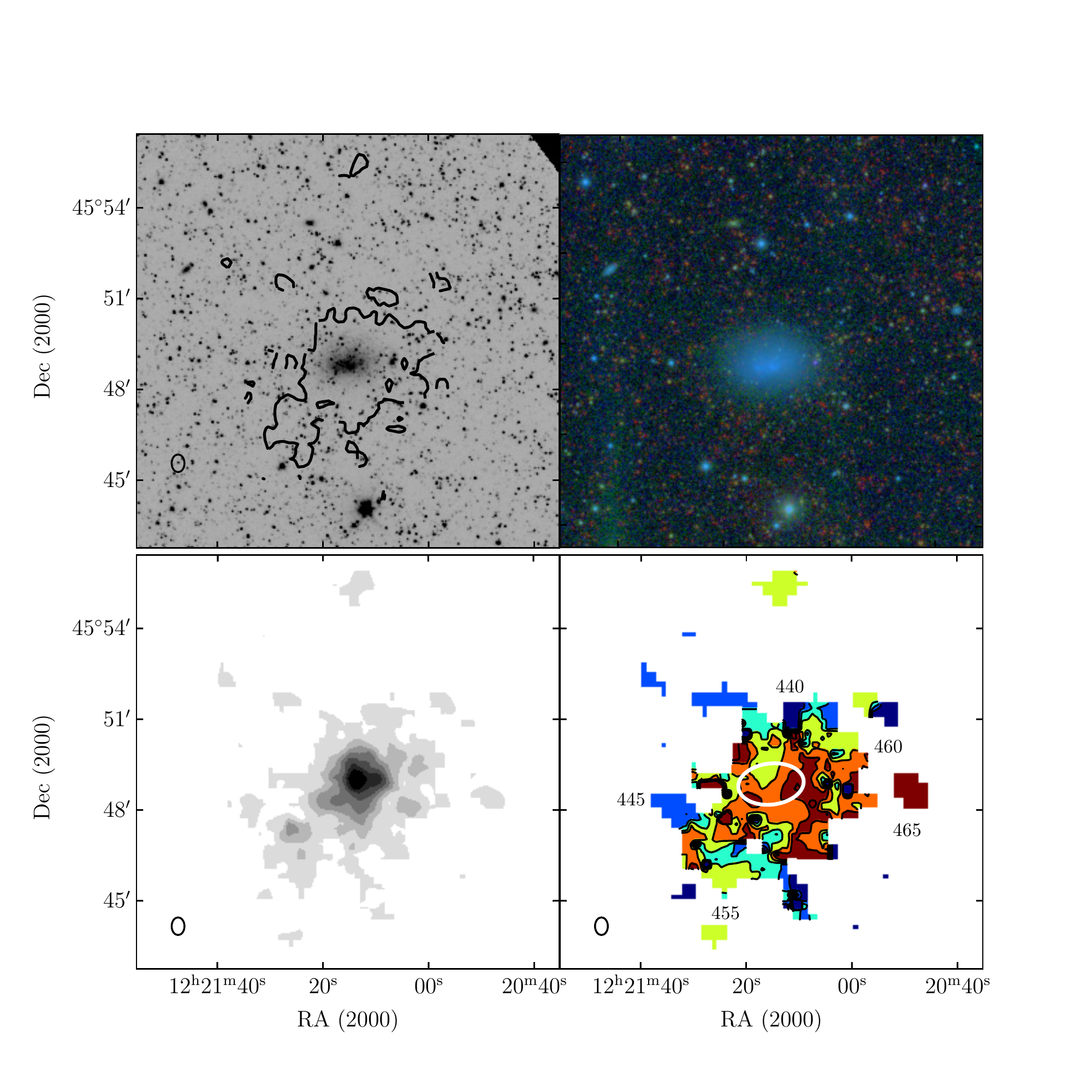}
	\caption{UGC 07408 : Plot same as Figure \ref{fig:ngc0024_4panel}.}
	\label{fig:ugc07408_4panel}
 \end{figure*}

\clearpage

  \begin{figure}
     \centering
     \includegraphics[width=\columnwidth]{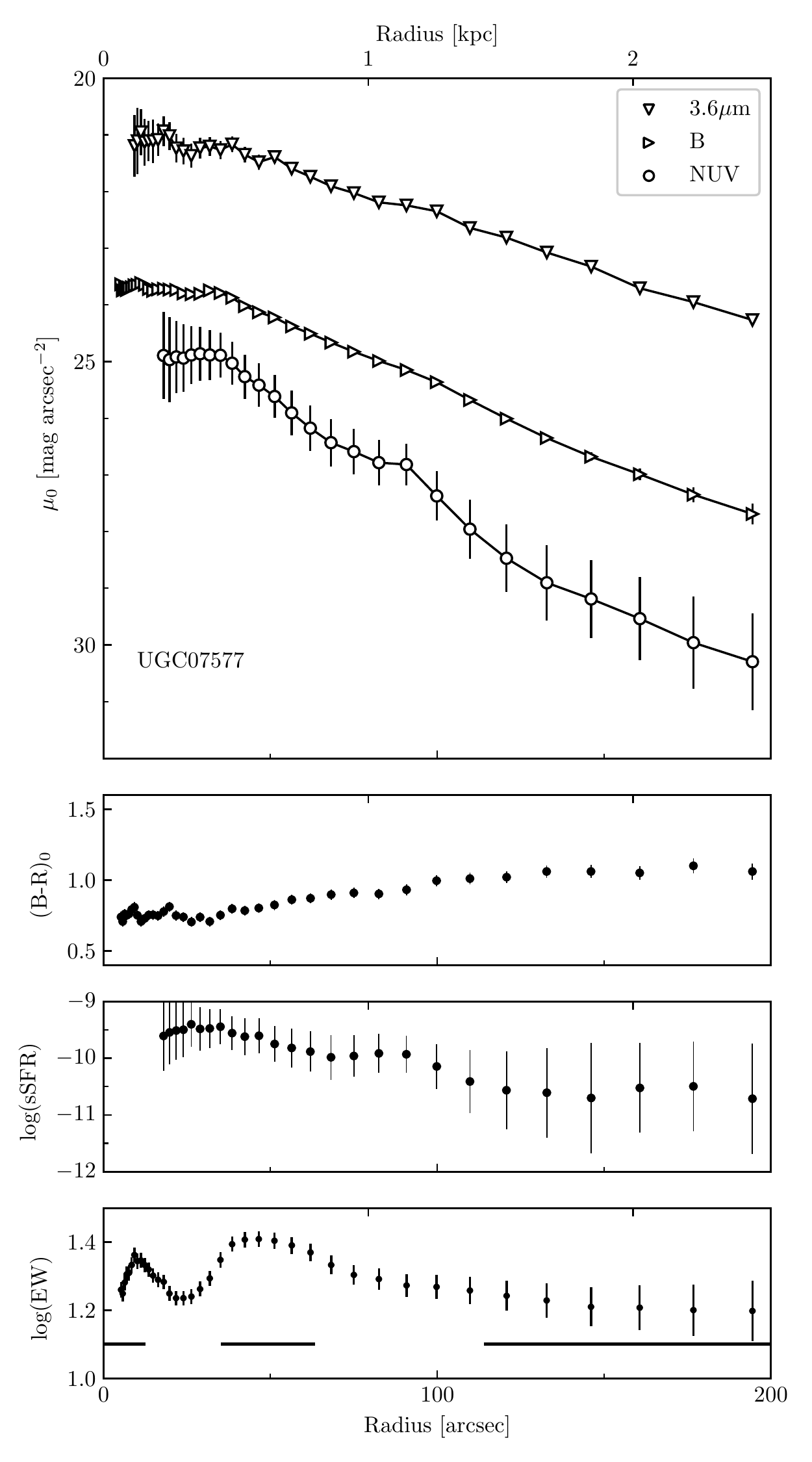}
     \caption{     UGC 07577 : Panels same as Figure \ref{fig:ngc0024_fulldata}.}
     \label{fig:ugc07577_fulldata}
 \end{figure}
 
 \begin{figure}
     \centering
     \includegraphics[width=\columnwidth]{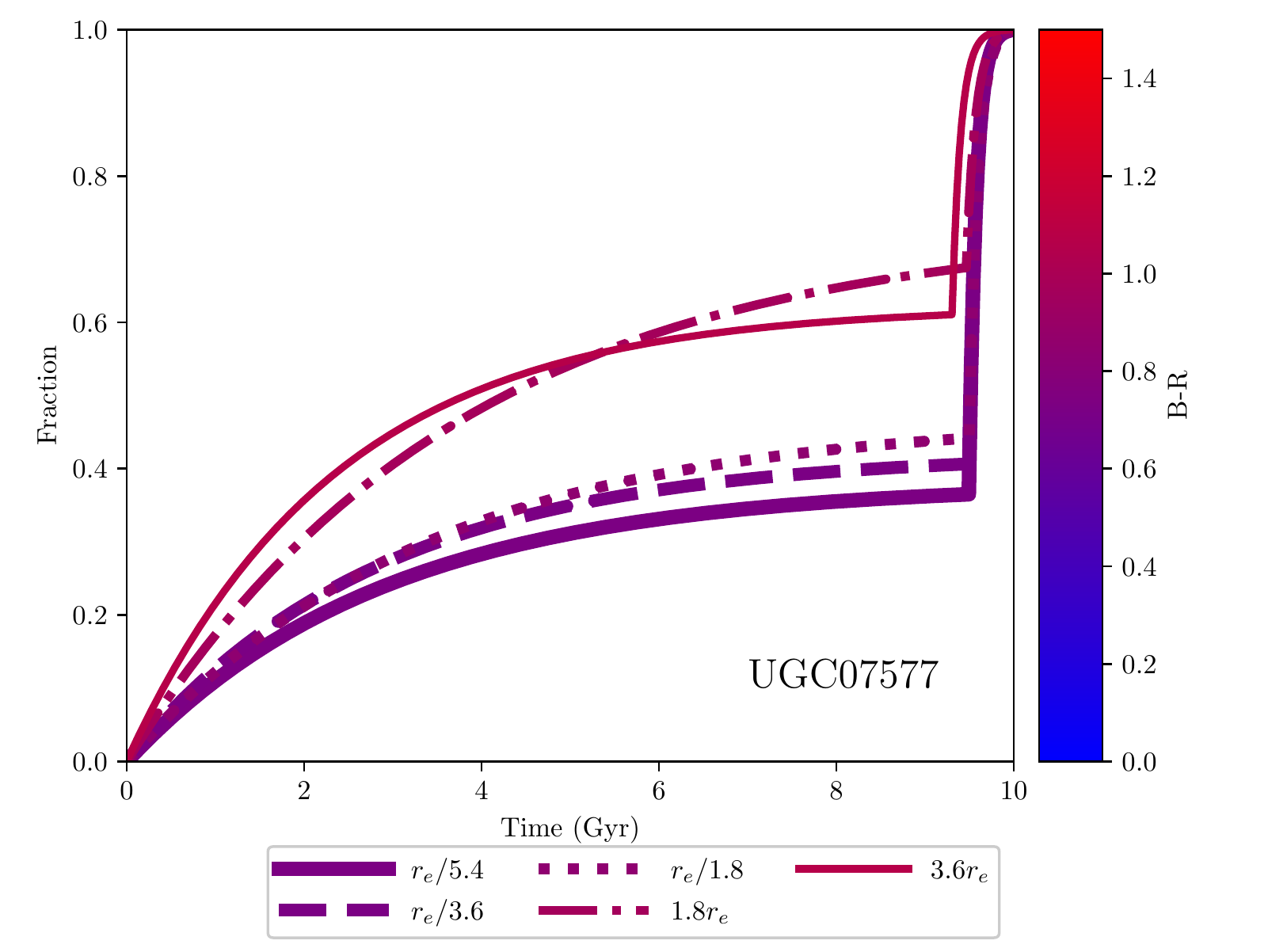}
     \caption{    UGC 07577 : Plot same as Figure \ref{fig:ngc0024_sfh}.}
     \label{fig:ugc07577_sfh}
 \end{figure}
 
\begin{figure*}
	\centering
	\includegraphics[width=2\columnwidth]{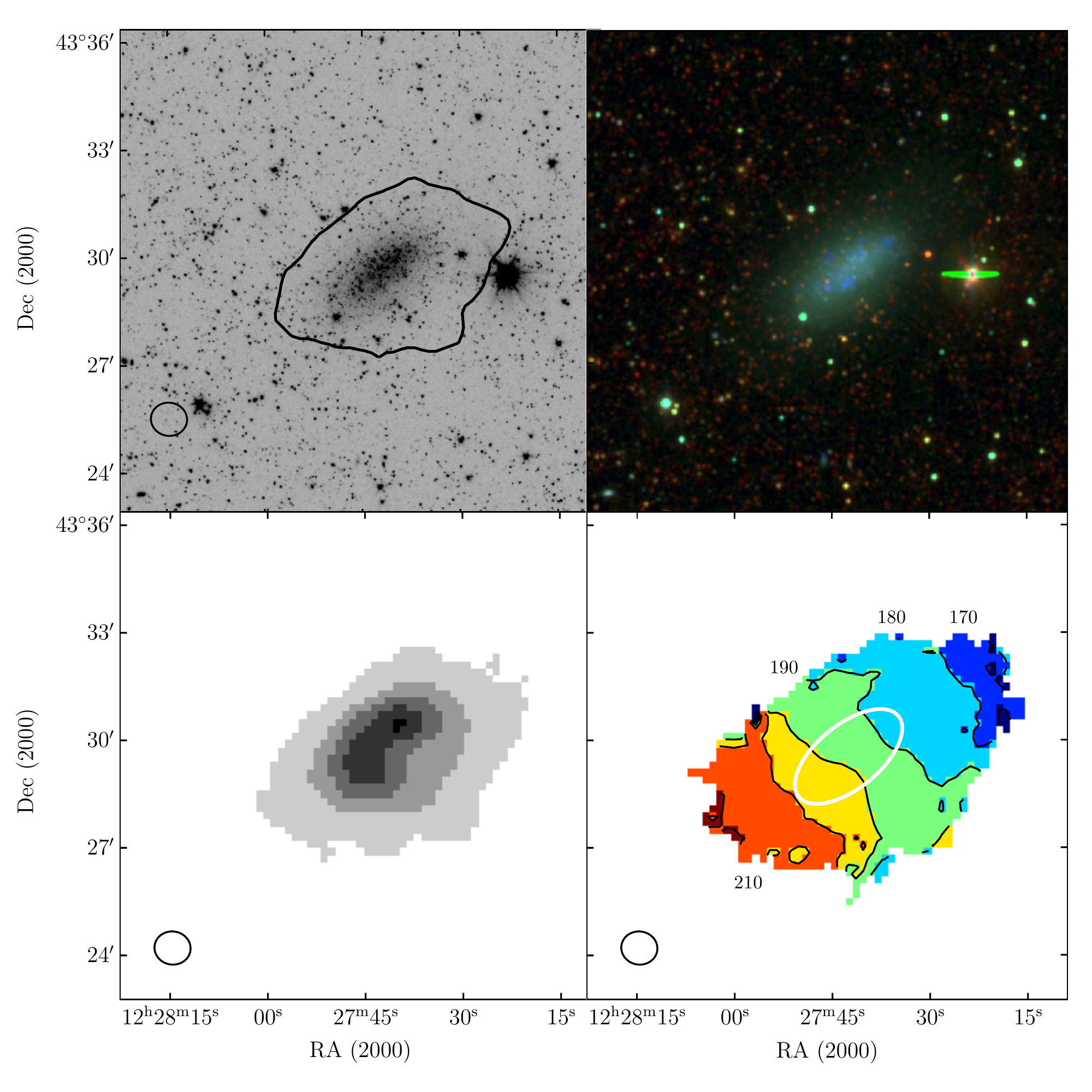}
	\caption{  UGC 07577 : Panels same as Figure \ref{fig:ngc0024_4panel}.}
	\label{fig:ugc07577_4panel}
\end{figure*}

\clearpage

  \begin{figure}
     \centering
     \includegraphics[width=\columnwidth]{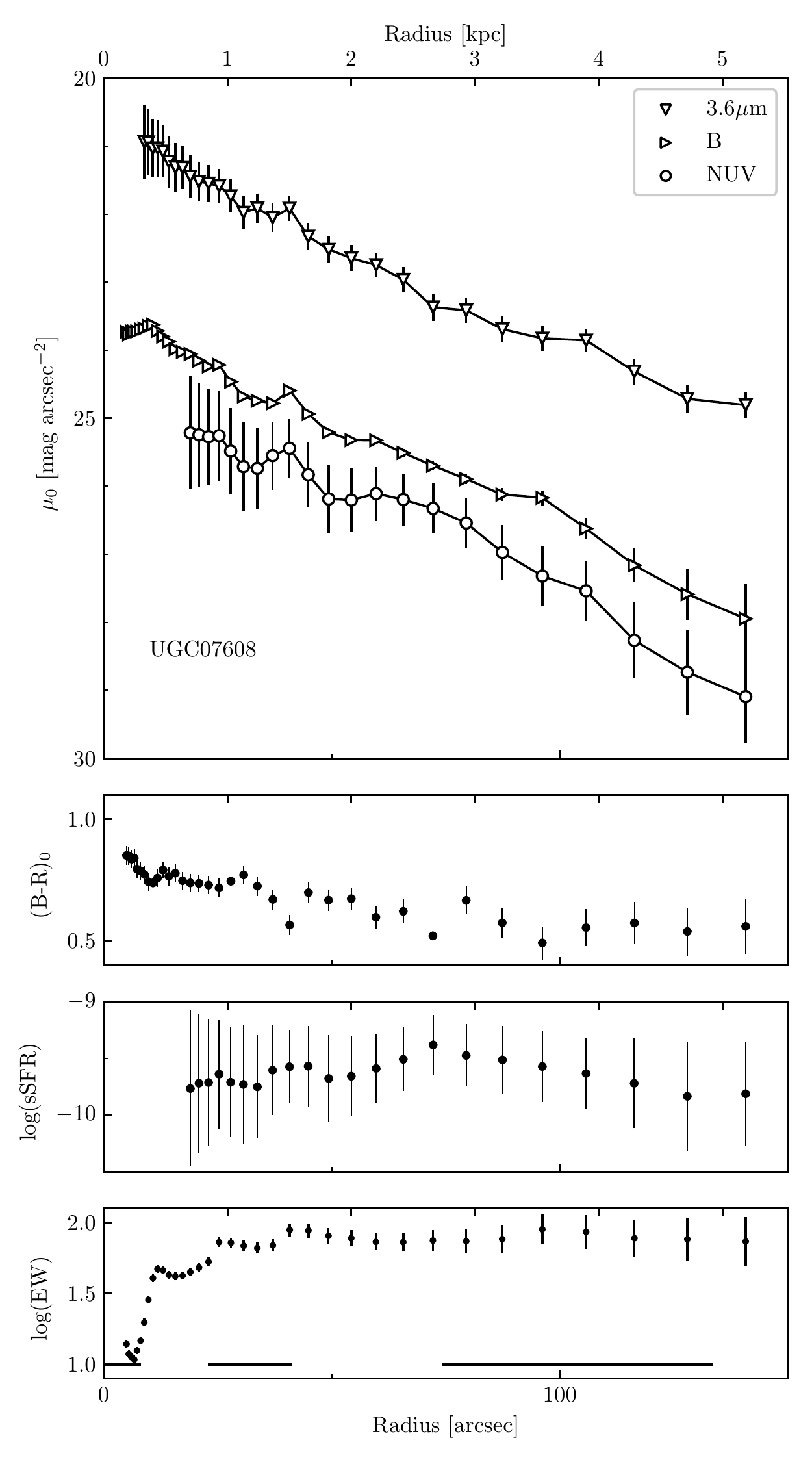}
     \caption{     UGC 07608 : Panels same as Figure \ref{fig:ngc0024_fulldata}.}
     \label{fig:ugc07608_fulldata}
 \end{figure}
 
 \begin{figure}
     \centering
     \includegraphics[width=\columnwidth]{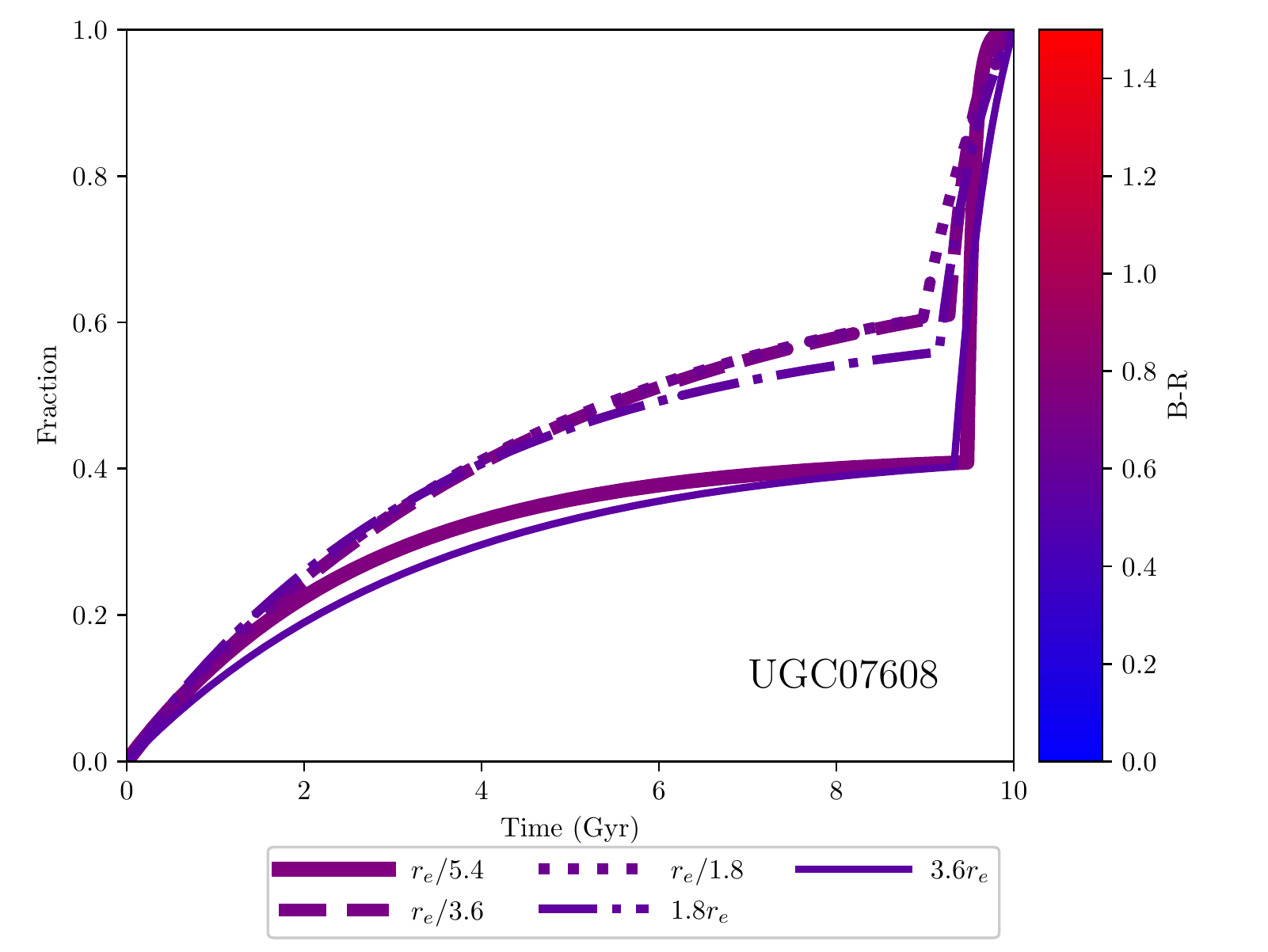}
     \caption{    UGC 07608 : Plot same as Figure \ref{fig:ngc0024_sfh}.}
     \label{fig:ugc07608_sfh}
 \end{figure}
 
\begin{figure*}
	\centering
	\includegraphics[width=2\columnwidth]{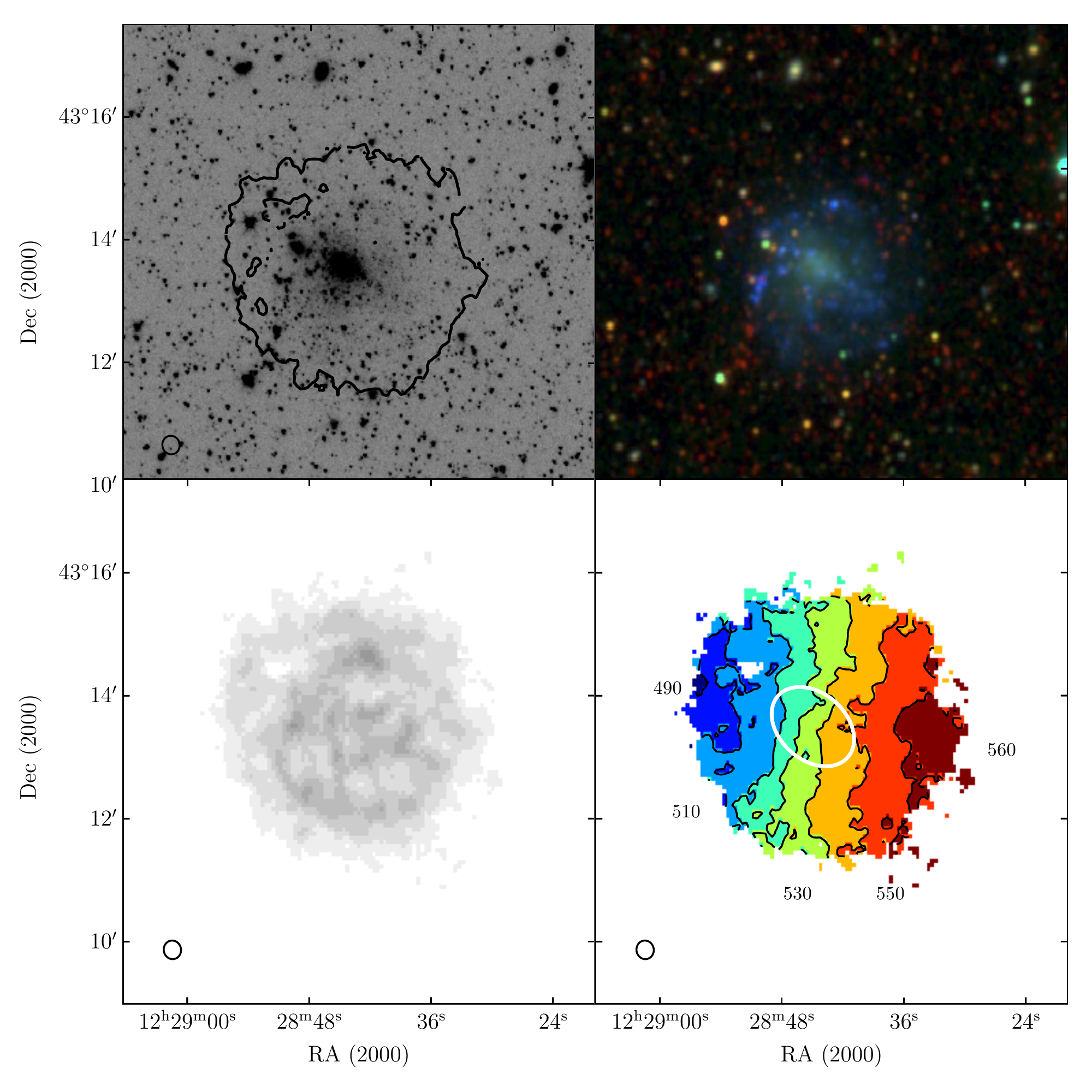}
	\caption{   UGC 07608 : Panels same as Figure \ref{fig:ngc0024_4panel}.}
	\label{fig:ugc07608_4panel}
\end{figure*}

\clearpage

 \begin{figure}
     \centering
     \includegraphics[width=\columnwidth]{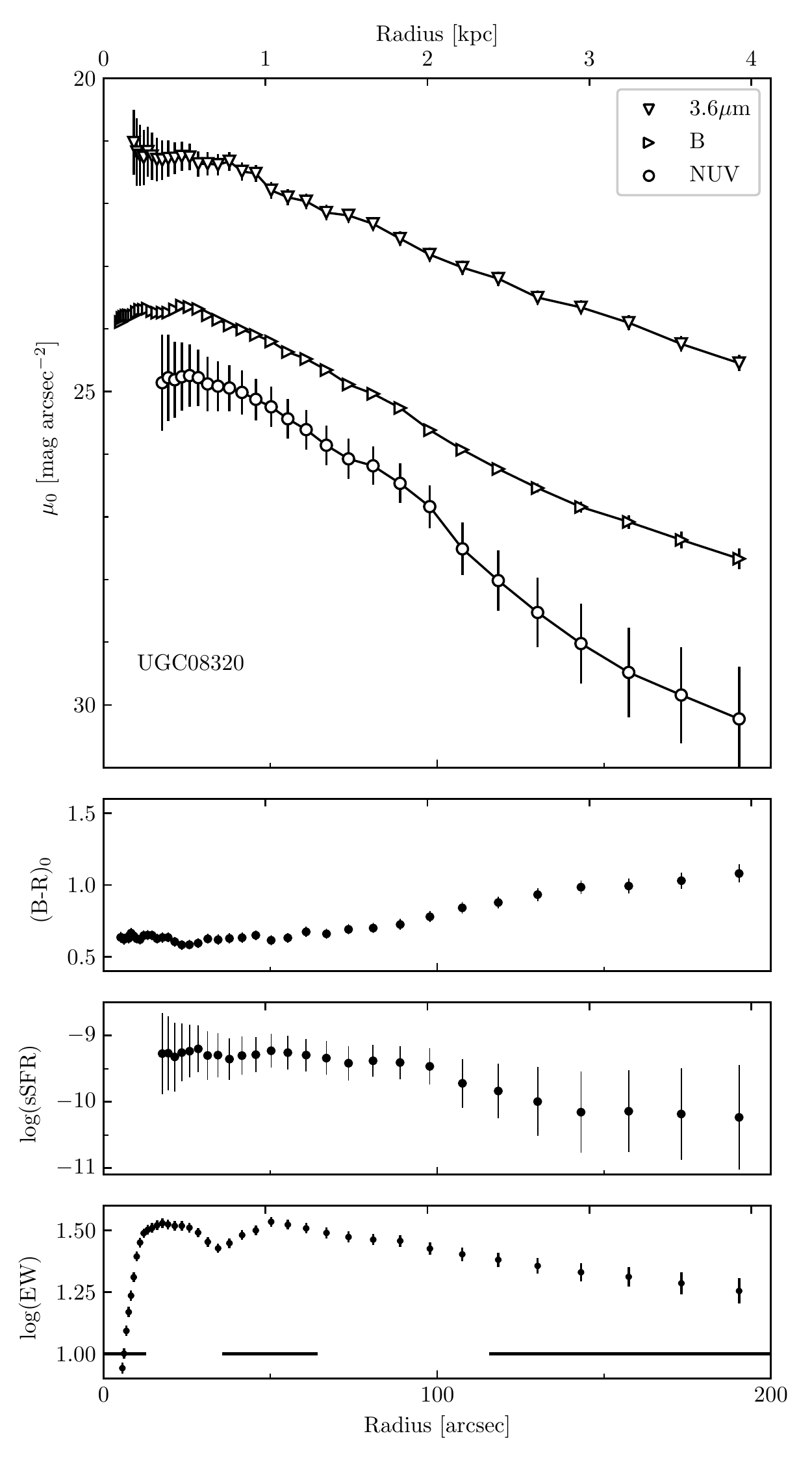}
     \caption{     UGC 08320 : Panels same as Figure \ref{fig:ngc0024_fulldata}.}
     \label{fig:ugc08320_fulldata}
 \end{figure}
 
 \begin{figure}
     \centering
     \includegraphics[width=\columnwidth]{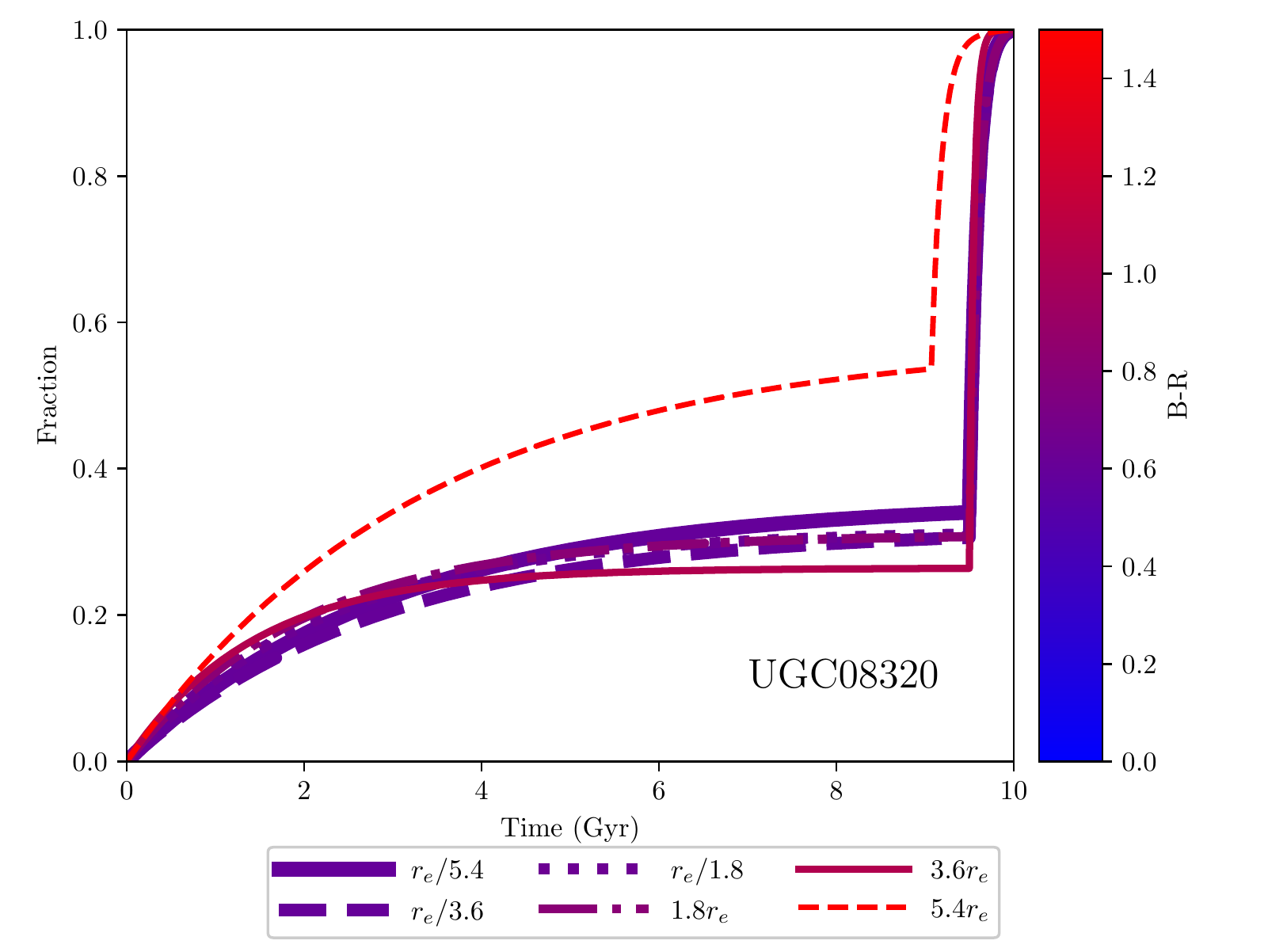}
     \caption{     UGC 08320 : Plot same as Figure \ref{fig:ngc0024_sfh}. }
     \label{fig:ugc08320_sfh}
 \end{figure}
 
\clearpage

 \begin{figure*}
	\centering
	\includegraphics[width=2\columnwidth]{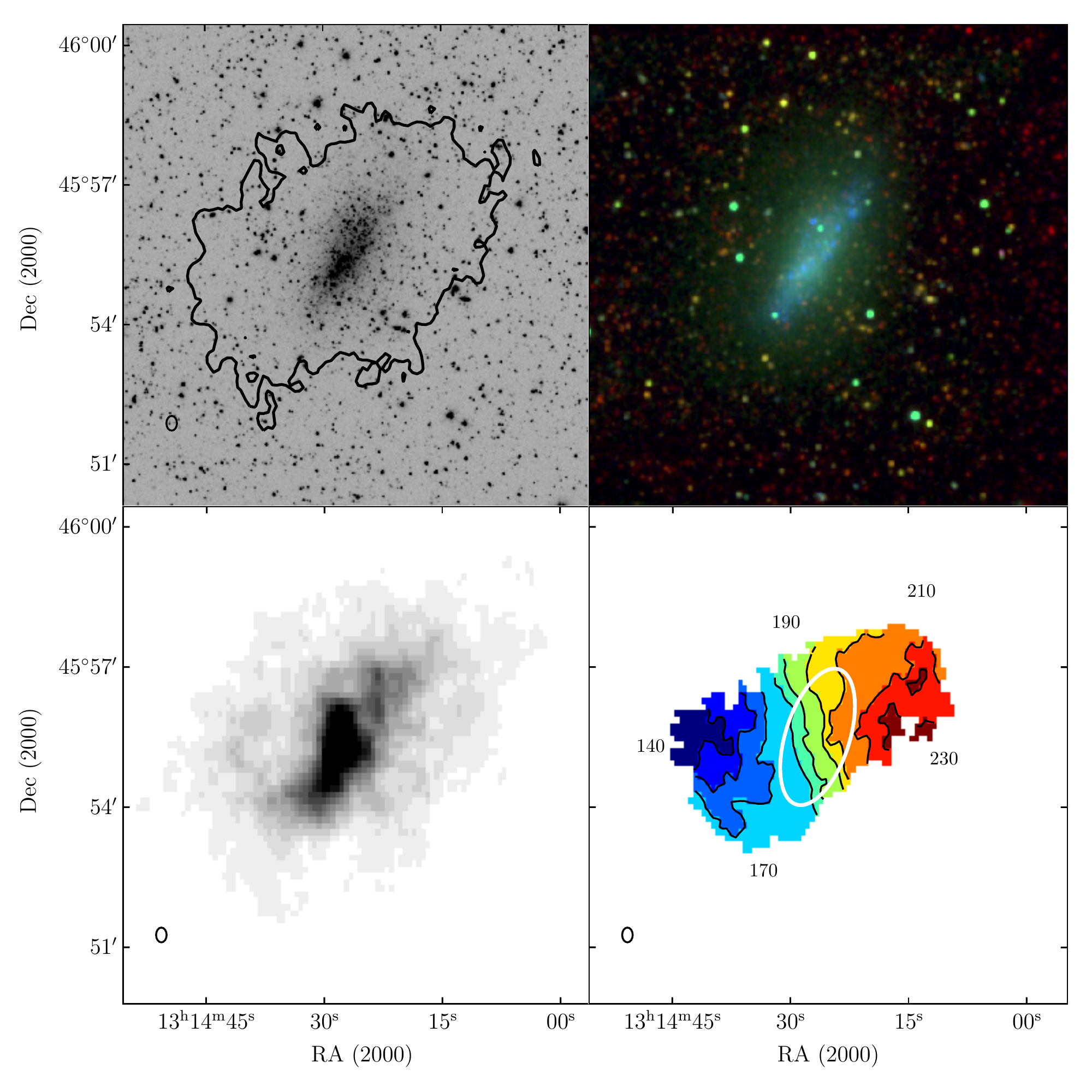}
	\caption{   UGC 08320 : Panels same as Figure \ref{fig:ngc0024_4panel}.}
	\label{fig:ugc08320_4panel}
\end{figure*}

% If you want to present additional material which would interrupt the flow of the main paper,
% it can be placed in an Appendix which appears after the list of references.

%%%%%%%%%%%%%%%%%%%%%%%%%%%%%%%%%%%%%%%%%%%%%8%%%%%

% Don't change these lines
\bsp	% typesetting comment
\label{lastpage}
\end{document}